\documentclass[pre,twocolumn,showpacs,floatfix]{revtex4}
\usepackage{graphicx}

\def\t2#1{\underline{\underline{#1}}}
\newcommand{\be}{\begin{equation}}
\newcommand{\ee}{\end{equation}}
\newcommand{\ppp}{{\mathcal P}}
\newcommand{\fff}{{\mathcal F}}
\newcommand{\sss}{{\mathcal S}}
\newcommand{\ccc}{{\mathcal C}}
\newcommand{\vvv}{{\mathcal V}}
\newcommand{\eee}{{\mathcal E}}
\newcommand{\ddd}{{\mathcal D}}
\def\R{\mbox{I\hspace{-.15em}R}} 

\begin{document}
\title{Geometric origin of mechanical properties of granular materials}
\author{Jean-No\"el Roux}
\email{jean-noel.roux@lcpc.fr}
\affiliation{Laboratoire Central des Ponts et Chauss\'ees,
58, boulevard Lef\`ebvre,75732 Paris cedex 15,
France}
\begin{abstract}
Model granular assemblies, in which grains are assumed rigid and
frictionless, at equilibrium under some prescribed external load, are
shown to possess, under generic conditions, several remarkable mechanical
properties,
related to isostaticity and potential energy minimization.
Isostaticity -- the uniqueness of the contact forces, once the list of
contacts is known-- is established in a quite general context,
and the important distinction between isostatic
\emph{problems} under given external loads and isostatic
(rigid) \emph{structures} is presented.
Complete rigidity is only guaranteed, on stability grounds,
in the case of spherical cohesionless grains. Otherwise,
the network of contacts might
deform elastically in response to small load increments,
even though grains are perfectly rigid.
In general, one gets an upper bound on the contact coordination number.
The approximation of small displacements, that is introduced 
and discussed, allows to draw
analogies with other model systems studied in statistical mechanics, such as 
minimum paths on a lattice. It also entails the uniqueness of the
equilibrium state (the list of contacts itself is geometrically determined)
for cohesionless grains,
and thus the absence of plastic dissipation in rearrangements
of the network of contacts. Plasticity and hysteresis are related to the lack
of such uniqueness, which can be traced back,
apart from intergranular friction, to non-reversible rearrangements
of small but finite extent,
in which the system jumps between two distinct potential energy minima in
configuration space, or to bounded tensile forces, deriving from a non-convex
potential, in the contacts. Properties of response functions to load
increments are discussed. On the basis of past numerical studies, it is argued
that, provided the approximation of small displacements is valid,
displacements due to the rearrangements of the rigid grains in response to
small load increments,  once averaged on the macroscopic scale, are solutions
to elliptic boundary value problems (similar to the Stokes problem for viscous
incompressible flow).   \end{abstract}
\maketitle
\section{Introduction}
\subsection{Motivations}
A large research effort,
both in the statistical physics and the mechanics and 
civil engineering communities,
is currently being devoted to granular materials,
aiming in particular at a better understanding of the relationships between
grain-level micromechanics (intergranular contact laws) and macroscopic
behaviours (global equilibrium conditions, constitutive
relations)~\cite{BJ97,WG97,HHL98}

This aim --the traditional program of Statistical Mechanics -- is far from
fully
achieved in dense granular systems near equilibrium, for one is facing at
least two fundamental difficulties.

Firstly, the non-smooth character of contact laws, 
that involve unilaterality and, possibly, dry friction, is a common feature of
granular assemblies that endows them with a high level of disorder and a high
sensitivity to perturbations. 
Tiny motions might
significantly affect the way forces are transmitted, since contacts between
neighbouring grains might open or close
(and the sliding or non-sliding status of closed ones might change).
Hence the characteristically heterogeneous aspect of
force transport in dense granulates: large forces are carried by a network of
preferred paths (the ``force chains'') while some grains or sets of grains
carry but vanishing efforts (``arching effect''). The histogram of contact
forces spans a wide range. These phenomena have been experimentally observed
thanks to techniques like photoelastic stress visualization~\cite{DA57,JDJV69}
and carbon paper print analysis~\cite{DDL90,MJN98}.
They have also been studied in numerical
simulations~\cite{RJMR96,OR97b}, and
some attempts of theoretical descriptions have been proposed~\cite{CLMNW96}.
Such peculiar aspects of granular systems render more difficult the reference
to existing models from other fields. Indeed, a recent trend in the
physics literature on static
granular systems~\cite{BCC95,WCC97,CWBC98,Claudin} 
insists on their difference with ordinary, elastic solids, and suggests, 
instead of resorting to macroscopic displacement or strain variables, 
to search for direct relations betwen the components of the stress tensor. 

The second basic difficulty stems from the incomplete knowledge of the
mechanical properties of granular systems, especially those ruling the
dynamics. When a granular sample is submitted to some prescribed external
actions that are sufficiently slowly changing in time, its evolution is
customarily described as an ordered set of equilibrium states that are
successively reached, with little or no dependence on physical time.
The physical processes by which kinetic energy is dissipated are, however,
most often somewhat mysterious or poorly characterised. They are, in the
framework of the \emph{quasi-static} description we have just mentioned,
implicitly regarded as irrelevant. One might wish to assess the
validity of such an assumption. Numerical simulations, that have
to adopt some rule to move the grains, could in principle allow useful
investigations of the influence of the dynamics. However, in view of the
practical difficulty to obtain representative configurations close enough to
equilibrium within a reasonable computation time, they
sometimes resort to non-physical parameters, and pick up the dynamical rule
among the restricted range of those that allow tractable calculations. 

This paper addresses both those basic concerns, in the following way.
Simplifying assumptions are introduced (we consider, \emph{e.g.}, rigid
frictionless grains), thus restricting our attention to a certain class of
model systems, that are however argued to exhibit the same qualitative
behaviours as more realistic ones. Those systems are suitable candidates to
test, most easily by numerical means, some recently proposed models and
speculations, at the expense of rather extensive numerical computations. 
The purpose of the present article is not, however, to present new results of
numerical simulations. 
We shall state
and establish, rather, with a fair level of generality, some basic
properties of such systems, and study their qualitative
consequences in terms of macroscopic mechanical behaviour. This analysis
will shed some light on some analogies and differences with other previously
studied problems in statistical mechanics, such as directed `polymers' in
random environments and percolation models. It will also,
along with the exploitation of 
past numerical results on a simplified
model~\cite{OR97a,OR97b,JNR97a,JNR97b,Sofiane}, allow
us to investigate the possible origins of
some macroscopic features of granular mechanics,
that are classically modelled with elastoplastic constitutive
laws~\cite{HHL98,Muirwood}, and
to discuss other recently proposed
approaches~\cite{BCC95,WCC97,CWBC98,Claudin}.

We will show that mechanics is to a
large extent determined by geometrical aspects (steric exclusion), thus
partially answering concerns about the role of dynamical parameters. Finally
we will discuss the status of displacement and strain variables
in quasi-static
assemblies of rigid grains, and give perpectives for future investigations.

\subsection{Synopsis.}
The paper is composed of two main parts. 

First, sections II to V introduce useful definitions and
state basic properties that are necessary for the derivation
of the main results. 

Thus, section II presents useful definitions and
mechanical properties of static granular systems, \emph{i.e.}, collections of
rigid bodies essentially interacting via point forces mutually exerted on
their surfaces. Those notions, that include the theorem of virtual power,
generalized forces and velocities for collective degrees of freedom, and the
degree of indeterminacy of forces and of velocities, are not always familiar
in the condensed matter physics community. Section III introduces the
potential energy minimization problems for various simple frictionless contact
laws. Section IV defines the approximation of small displacements, a modelling
step of both technical and conceptual importance, since it allows, in
particular, an analogy with problems of scalar transport on discrete networks,
as explained in section V.

Once those essential ingredients made available, the second part of the paper
(sections VI to IX) establishes the main results and discusses their
consequences, with reference to previous theoretical and numerical work, and
to known aspects of the mechanical behaviour of granular materials. 

Section VI is devoted to the \emph{generic isostaticity property} of
equilibrium states in systems of rigid grains that may only exert normal
contact forces on one another.  We then prove and discuss (section VII) the
\emph{uniqueness} of the equilibrium state in cohesionless systems within the
approximation of small displacements, and compare the determinatin of
equilibrium states of such systems with other mechanical or scalar transport
problems. Section VIII introduces the additional requirement of stability,
outside the approximation, which is dealt with, in the absence of friction,
in terms of potential energy minimization. In some restricted models, this
allows to conclude to the isostaticity of the structure, a stronger property
than mere isostaicity of the problem under a given load. It is then possible
to discuss the possible origins of plasticity in systems of frictionless
grains and the form of the mechanical response to small load increments. The
paper ends with concluding remarks (X) on the role of displacements and
strains in granular materials and suggestions for future research.

\section{Basic definitions and properties.} 
We are interested in the
modelling of large packings of solid bodies (grains),  in equilibrium under
some prescribed external forces. Grains are assumed to interact \emph{via}
point forces mutually exerted on their surfaces, which means that the
distribution of stress on their areas of contact or of influence can
effectively be viewed as localized at a point, on the scale of the whole
grain. Apart from this reservation, that excludes flat or conforming
surfaces \footnote{Our considerations do apply, in fact, to flat surfaces,
provided face to face contacts are counted $d$ times in $d$ dimensions, as
they transmit one force and $d-1$ torques.}, grains might have arbitrary
shapes, and our considerations apply to spatial dimension $d$ equal to 2 or 3,
although most examples will be taken with two-dimensional systems of discs.
Note that we do not require  interacting grains to touch one another at this
stage. We mostly restrict our attention here to \emph{frictionless} bodies,
\emph{i.e.}, such that contact forces are normal to the grain surfaces. This
might look like a severe limitation, but we shall argue that such simplified
systems do possess the generic properties of granular media. We shall also
assume, unless otherwise specified, that the grains behave as rigid
undeformable objects.

\subsection{System, external forces} 
We consider a set of $n$ grains, labelled with indices $i$, with $1\le i\le
n$.  In each of them we arbitrarily choose a `center', which might
\emph{e.g.,} coincide with its center of mass. In the case of spherical grains
it is of course convenient to take the geometrical center of the sphere. The
($d$-dimensional) velocities of those centers,  $({\bf V}_i)_{1\le i\le n}$,
together with the $d'$-dimensional (with $d'={d(d-1)\over 2}$) angular
velocities $({\bf \Omega}_i)_{1\le i\le n}$,  make up
the kinematic degrees of freedom of the whole system, thus labelled by couples
of indices $(i,\alpha)$, with $1\le i\le n$ and $1\le \alpha \le
d+d'={d(d+1)\over 2}$. We denote as $I$ the set of such couples. If $\alpha >
d$, $V_{i,\alpha}$ is now a notation for $\Omega_{i,\alpha -d}$. Boundary
conditions are often enforced by prescribing the motion, or the absence of
motion, of walls. Those might be regarded as solid bodies, or particular
`grains' themselves. In the following we shall sometimes write down large
`velocity vectors' that gather all $N_f$ kinematic degrees of freedom of the
system, then denoted, with a single index, as $(v_\mu)_{1\le\mu\le N_f}$.

It might also be convenient to keep some grain coordinates fixed (thus
choosing one particular Galilean frame), \emph{i.e.,} to impose, for all
couples $i,\alpha$ belonging to some subset $I_0$ of $I$,
$V_{i,\alpha}=0$. Indices $\mu$ are then renumbered, and $N_f$ is reduced
accordingly, to label and to count the free kinematic parameters. Another
classical way to impose some boundary conditions is to require, for all
$i,\alpha$ in some subset $I_1$ of $I$,  $V_{i,\alpha}$ to depend linearly on
one or several parameters, \emph{e.g.}: \be (i,\alpha)\in I_1 \ \ \ 
V_{i,\alpha}=A_{i,\alpha}\lambda_1, \label{eqn:deflambda1}
\ee
introducing some collective `generalized velocity' $\lambda_1$.
Once again, in such a case, $N_f$ is reduced to count elements of
$I\setminus (I_0\cup I_1)$, plus $\lambda_1$.

At least locally, it is possible to regard velocities and generalized
kinematic parameters (like $\lambda_1$ in eqn.~\ref{eqn:deflambda1}) as time
derivatives of spatial coordinates, which we shall do in the following, thus
writing, \emph{e.g.,} ${\displaystyle V_{i,\alpha}=\frac{d X_{i,\alpha}}{dt}}$.
As we are only interested in those properties that do not depend on dynamics,
grain trajectories might as well be described by any parameter, not
necessarily by physical time. In the case of kinematic constraints of type
\ref{eqn:deflambda1}, parameters $A_{i,\alpha}$ will be regarded as fixed,
although positions of the grains and the walls change. One then defines a
generalized coordinate $\Lambda_1$, such that ${d \Lambda_1\over
dt}=\lambda_1$. Just like for velocities, the compact notation $(x_\mu)_{1
\le \mu \le N_f}$ refers to the whole set of positional coordinates.

External forces and torques may at will be exerted on the grains that are
free of kinematic constraints. We shall use the same notations as for
velocities, writing down large $N_f$-vectors of `external forces' (some of
their coordinates standing, actually, for torques), as $(F^{ext}_\mu)_{1 \le
\mu \le N_f}$. At equilibrium, they are of course to be balanced by internal
forces $(F^{int}_\mu)_{1 \le \mu \le N_f}$:
\be (1 \le \mu \le N_f)\ \
F^{ext}_\mu + F^{int}_\mu=0. \label{eqn:eqfor}
\ee
In order to enforce constraints of type~\ref{eqn:deflambda1}, some external
efforts have to be exerted on the concerned bodies. On requiring the power of
such efforts to be balanced by that of internal forces $(F^{int}_\mu)_{1 \le
\mu \le N_f}$, one identifies the generalized force conjugate to $\lambda_1$
as
\be
Q_1 = -\sum_{(i,\alpha)\in I_1}
F^{int}_{(i,\alpha)}A_{(i,\alpha)}.\label{eqn:defQ1}
\ee
We just used the
power to find generalized forces: this is a manifestation of the
\emph{duality} between forces and displacements or velocities, which will be
repeatedly exploited in the sequel. The $N_f$-dimensional vector space $\fff$
of external forces, is, by construction, to be regarded as the dual space, in
the ordinary sense of linear algebra, of the $N_f$-dimensional space $\vvv$
of kinematic degrees of freedom. 

In general, it should be appreciated that the appropriate mathematical
description of configuration space is not $\R^{ N_f}$ with its Euclidean
structure, but, due to rotational degrees of freedom, an $N_f$-dimensional
manifold, on which $(x_\mu)_{1\le \mu \le N_f}$ is a set of (local)
curvilinear coordinates. $\vvv$ and $\fff$ are respectively the tangent and
cotangent vector space at a given point, and depend on that point. Thus the
definition of `constant velocities', or of `constant forces' requires some
care. However, these difficulties are inessential in our subsequent treatment,
and we shall assume `constant external forces' are applied, and derive from a
potential energy: \be W = -\sum_{\mu=1}^{N_f}F_{\mu}^{ext}x_{\mu}.
\label{eqn:defW}
\ee
It is easily checked that such a definition is devoid of ambiguity in
the following important cases.
\begin{itemize}
\item
The set of grain center positions, as opposed to grain orientations, define a
`flat' space, on which constant vectors and covectors are unambiguous.
Whenever external efforts are not sensible to orientational coordinates, as in
the case of gravity (if the grain `centers' are their centers of mass), one
may therefore `apply constant forces'. \item Anticipating on part IV, the
approximation of small displacements assumes that the manifold might locally
be replaced by its flat tangent space. \end{itemize}

The complete $N_f$-vector of external forces is referred to as the
\emph{load}. Sometimes, it is convenient to deal with parametrized sets of
loads. When the direction of the load is fixed, while its intensity might
vary, one has a \emph{one-parameter loading mode}. In such a situation, all
external force components are kept proportional to a single loading parameter
$Q$, and a generalized velocity conjugate to $Q$, $\lambda$, can be identified
on equating the power of external forces with the product $Q\lambda$.
$\lambda$ is some linear combination of the kinematic degrees of freedom
$(v_\mu)_{1\le \mu \le N_f}$, and the time derivative of a generalized
coordinate $\Lambda$, equal to the same combination of coordinates
$(x_\mu)_{1\le \mu \le N_f}$.  The potential energy is then simply 
\be
W = -Q\Lambda  . \label{eqn:defWoplm}
\ee

\begin{figure}[htb]
\vbox{
\begin{center}
\includegraphics[width=8cm]{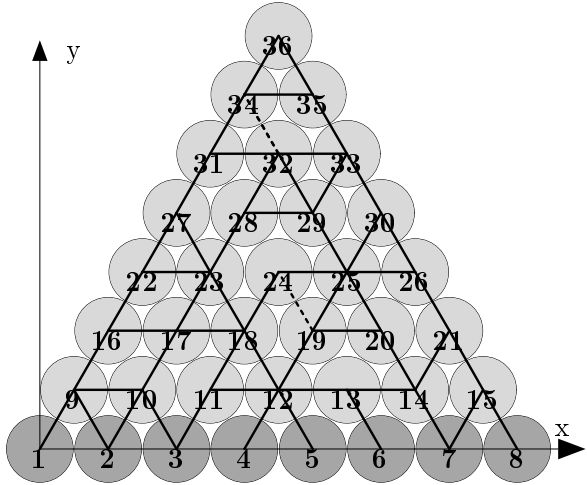}
\caption{System A : a pile under gravity. The bottom boundary conditions are
explained in the text
\label{fig:sysA}}
\end{center}
}
\end{figure}

Let us now illustrate those notions with simple examples,
that will be repeatedly used in the following. Systems A and B are packings of
discs that are placed on the sites of a regular triangular lattice. (Later
on, we shall allow for a slight polydispersity of the grains. They might move,
gain or lose contacts with their neighbours, and the lattice might be slightly
distorted). System A (fig.\ref{fig:sysA}) is a pile with slope inclined at
$60$ degrees with respect to the horizontal direction. 
Each disc is submitted to its own weight,
except those of the bottom row, which collectively set the boundary condition.
One might keep them fixed at regularly spaced positions, imposing, say
(numbering them as on the figure, and denoting as $a$ the lattice spacing)
\begin{equation}
(1\le i\le 8) \left\{
\begin{array}{ll}
x_i & = (i-1)(1-\Lambda_1)a \\
y_i & = 0 
\end{array}
\right.,
\label{eqn:cl1tas}
\end{equation}
allowing for a horizontal deformation parameter $\Lambda_1$. One may also
require them to stay on the horizontal axis $y=0$ and satisfy
\be
v_i^y=-\lambda_1 (i-1) a,
\label{eqn:cl2tas}
\ee
with a free kinematic parameter $\lambda_1$. According to
eqn.~\ref{eqn:defQ1}, the generalized force conjugate to $\lambda_1$ is
\be
Q_1=\sum _{i=1}^8 F_{i,x}^{int}(i-1)a. 
\label{eqn:defQ1BC2}
\ee
These two slightly different boundary conditions (BC) are
respectively abbreviated as BC1 and BC2 in the following.

\begin{figure}[htb]
\centering
\includegraphics[width=8cm]{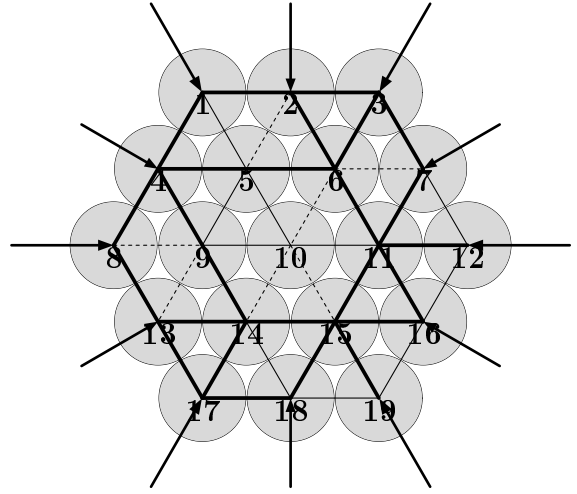}
\caption{System B : a hexagonal sample. Arrows depict external forces applied
on peripheral discs.} \label{fig:sysB}
\end{figure}

System B (fig.~\ref{fig:sysB}) is a hexagonal sample of the same material. It
is submitted to external forces on the periphery, which mimic hydrostatic
pressure.

\begin{figure}[tbh]
\centering
\includegraphics[width=8cm]{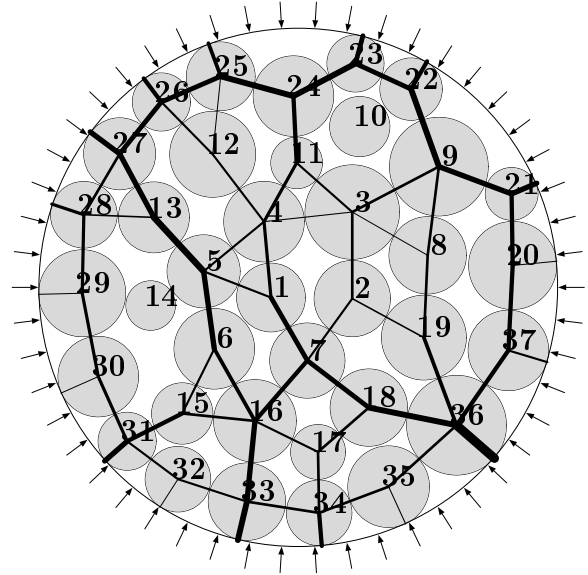}
\caption{System C : a disordered packing surrounded by a circular wall that
might uniformly expand or shrink, as indicated by the
small arrows. \label{fig:sysC} }
\end{figure}

System C (fig.~\ref{fig:sysC}) is a disordered collection of discs with a
larger polydispersity. It is embedded within a circular wall the radius $R$ of
which might change. One controls the generalized force conjugate to
$\lambda_1 = \frac{dR}{dt}$, \emph{viz.} 
\be
Q_1 = \sum _i f_{iw}, \label{eqn:Qcerc}
\ee
where the sum runs over all particles $i$ exerting forces $f_{iw}$
normally onto the wall.

\subsection{The structure: a set of bonds.}
The definitions we introduce here pertain to one specific configuration
of the grains, with the positions and orientations fixed.

We call `bonds' the pairs of neighbouring grains that \emph{may}
exert a force on one another. We
require this force to be concentrated at the point of each grain which is the
closest to the other one, and directed normally to the surface.\footnote{
This latter condition is not essential: the properties of Section II hold true
provided the direction of the force carried by a bond is fixed.} The more
general case of arbitrary bond forces will be briefly evoked later.
\begin{figure}[htb]
\centering
\includegraphics[width=8cm]{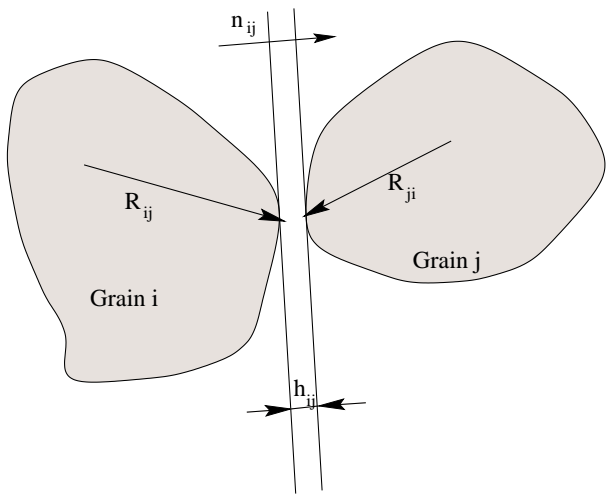}
\caption{Two grains $i$ and $j$ joined by a bond. $h_{ij}$ is the minimum
distance between their surfaces, measured where a common normal unit vector
is ${\bf n}_{ij}$. Vector ${\bf R}_{ij}$ (respectively ${\bf R}_{ji}$) points
from the center of $i$ (resp. of $j$) to the point of its surface that is
closest to $j$ (resp. to $i$) \label{fig:defbond} } 
\end{figure}
Note that we 
neither require the grains that are joined by a bond to be in contact, nor
impose any sign constraint on the force. We thus define, somewhat arbitrarily
at this stage, $N$ such bonds as depicted on figure~\ref{fig:defbond},
alternatively labelled with an index $l$, $1\le l\le N$, or with the pair of
labels of the two grains they join. If bond $l$ connects $i$ and $j$, ${\bf
n}_l$ or ${\bf n}_{ij}$ denotes the unit vector that points from $i$ to $j$,
normally to the surfaces of both grains where the distance between them ,
$h_{ij}$, is the smallest. ${\bf R}_{ij}$ is the vector
joining the center of grain $i$ (origin), to the point on its
surface that is closest to grain $j$ (extremity). This contact zone might
transmit a \emph{normal} force, along ${\bf n}_{ij}$, of magnitude $f_{ij}$
that will be counted positively when the grains repell each other. 
Once this set of bonds is defined, it is referred to as the \emph{structure}.
The set of bonds defined by intergranular contacts ($h_{ij}=0$) will
be called the \emph{contact structure}.

As a consequence of the definition of a structure, the form of internal forces
($({\bf F}_i^{int})_{1\le i\le n}$) and torques
($({\bf \Gamma}_i^{int})_{1\le i\le n}$)
in the system is specified: they linearly depend on bond forces $f_{ij}$, as
\be 
\begin{array}{ll}

{\bf F}_i^{int} &={\displaystyle -\sum_{j\ne i}f_{ij}{\bf n}_{ij}} \\
{\bf \Gamma}_i^{int}
&={\displaystyle -\sum_{j\ne i}f_{ij}{\bf R}_{ij}\wedge {\bf n}_{ij}}
\end{array}
.
\label{eqn:fint}
\ee

Given the load $(F^{ext}_\mu)_{1\le \mu\le N_f}$, equilibrium
requires, in view of eqns.~\ref{eqn:fint} and \ref{eqn:eqfor} that the bond
forces $(f_l)_{1\le l\le N}$ satisfy equations of the form
\be
(1\le\mu\le N_f)\ \ \ \sum_{l=1}^N H_{\mu l}f_l = F^{ext}_\mu,
\label{eqn:defH} \ee
defining a linear operator, $H: \R ^N\rightarrow \fff$. 
Bond forces $(f_l)_{1\le l\le N}$ are then said to be \emph{statically
admissible} with the load $(F^{ext}_\mu)_{1\le \mu\le N_f}$. Bond forces
that are statically admissible with a load equal to zero (in equilibrium
without any external action) are the elements of a subspace $S_0$ of $\R ^N$,
the null space of operator $H$. Its dimension, that we
denote as $h$, is the number of linearly independent such self-balanced sets
of internal forces, or, in other words, the \emph{degree of indeterminacy of
bond forces} in the system (also called the \emph{degree of hyperstaticity}).
If not empty, the set of statically admissible bond forces is an affine space
of dimension $h$.

The relative normal velocity of the grains $i$ and $j$ joined by a bond is
\be
\delta V _{ij} =
{\bf n}_{ij}\cdot\left({\bf V}_i-{\bf V}_j+{\bf \Omega}_i\wedge
{\bf R}_{ij}-{\bf \Omega}_j\wedge{\bf R}_{ji}\right),
\label{eqn:deltaV}
\ee
with the convention that it is positive when the particles are approaching
each other. Eqn.~\ref{eqn:deltaV} defines a linear operator, $G$, acting on
$\vvv$ into $\R ^N$. The \emph{range} of $G$ is the subspace $\ccc$ of
\emph{compatible} relative normal velocities, \emph{i.e.}, those N-vectors for
which one can effectively find values for the velocities,
relations~\ref{eqn:deltaV} being satisfied. The \emph{null space} of $G$ is
the vector space $M$ of `mechanisms', also  called `floppy modes',
\emph{i.e.,} motions that do not alter the lengths $h_l$ of the bonds. Its
dimension, denoted as $k$ in the sequel, is the number of independent such
motions, or, in other words, regarding the bonds as rigid, the \emph{degree of
indeterminacy of velocities}, also called \emph{degree of hypostaticity}.
Imposing the condition $\delta V_{ij}=0$ in all bonds of the structure
restricts the possible values of velocities $(v_\mu)_{1\le \mu\le N_f}$ to a
vector space of dimension $k$. Depending on the type of load and boundary
conditions,  the whole set of grains might keep some overall rigid body
kinematic degrees of freedom. System B, for instance, has 3 independent such
motions, as any solid body in 2D. If $k_0\le d(d+1)/2$ denotes the number of
such particular motions allowed by the boundary conditions, the system is said
to be \emph{rigid} when it does not have other mechanisms, \emph{i.e.,} when
$k=k_0$. 

An important and useful result, the classical \emph{theorem of virtual power}
states the following. Let $(\delta V_l)_{1\le l\le N}$ be any element of
$\ccc$, corresponding to the velocity vector $(v_\mu)_{1\le \mu\le N_f}$, and
let $(f_l)_{1\le l\le N}$ be a set of bond forces statically admissible with
the load $(F^{ext}_\mu)_{1\le \mu\le N}$. One then has: \be
\sum_{l=1}^{N}f_l\delta V_l=\sum_{\mu=1}^{N_f}F^{ext}_\mu v_\mu.
\label{eqn:thvirpow}
\ee
Equality~\ref{eqn:thvirpow}, for arbitrary (`virtual') equilibrium set of
internal forces and velocities, stresses the \emph{geometric} meaning of
forces and the \emph{mechanical} meaning of velocities. It is easily
established in two steps: first use the force balance equations in the
right-hand side; then transform the sum over degrees of freedom into a sum
over bonds. 

As a direct consequence of the theorem, one deduces that operator $H$ is in
fact (as one might check directly, reading the matrix elements in
eqns.~\ref{eqn:deltaV} and \ref{eqn:defH}) the transpose of $G$ :
$H=G^T$. This follows from the sequence of equalities
$$\left( f\vert \delta V\right)=\left( f\vert Gv\right)=
\left( Hf\vert v\right) =\left( G^T f\vert v\right) ,$$
valid for arbitrary $v$ (such that $Gv=\delta v$) and $f$ (such that
$Hf=F^{ext}$), in which a bracket notation is used
for scalar products. Consequently, $S_0$, the null space of $G^T$, is the
orthogonal complementary to $\ccc$, the range of $G$, in $\R ^N$:
\be
S_0=\ccc ^\perp \label{eqn:ortho}.
\ee
Thus to check that some values $\delta V_l$ that one might try to assign to
the relative normal velocities are compatible, it is sufficient to ensure the
orthogonality of $N$-vector $(\delta V_l)_{1\le l\le N}$ to all $N$-vectors of
self-balanced bond forces (or a spanning subset thereof):
\be
(\delta V_l)_{1\le l\le N} \perp \sss _0. \label{eqn:compatible}
\ee
One thus uses \emph{forces} (elements of $S_0$) as cofactors in a set of
\emph{geometric} compatibility conditions. 

Recalling $k$ (the number of
mechanisms) is the dimension of the null space $M$ of $G$, one has
$$N_f=k+\dim(\ccc).$$ As $h=\dim(S_0)$, from~\ref{eqn:ortho}, one also has:
$$N=h+\dim(\ccc).$$
Elimination of the dimension of $\ccc$ from those two equalities yields the
following relationship between the degree of hypostaticity, $k$, the degree
of hyperstaticity, $h$, the number of bonds, $N$, and the number of degrees of
freedom, $N_f$:
\be
N+k=N_f+h. \label{eqn:relkh}
\ee
As we will check on examples below, relation~\ref{eqn:relkh} holds whatever
the choice of the list of bonds between objects, although it is of course
desirable in practice to define bonds according to the interaction law.
One may, for example, declare a bond to join two grains whenever their
surfaces are separated by a minimum distance smaller than some threshold
$h_0>0$. The choice of a larger $h_0$, thereby increasing $N$, will decrease
$k$ and/or increase the degree of hyperstaticity $h$.

Let us remark that the properties we have just dealt with in the case of bonds
that carry normal forces, are very easily generalized to the case of
arbitrary  contact forces, at the cost of minor modifications. Relative normal
velocities and normal contact forces are replaced by d-vectors, $\R ^{dN}$
replaces $\R ^N$, equalities~\ref{eqn:thvirpow} (with, now, a scalar product
within the sum in the left-hand side) and \ref{eqn:ortho} are still satisfied.
Instead of \ref{eqn:relkh}, one ends up with $dN+k=N_f+h$. Adding friction
increases $h$ and/or decreases $k$.

Returning to frictionless systems, the case of spheres or discs deserves a
special treatment: no normal force is able to exert any torque, and all
rotational degrees of freedom are therefore mechanisms. It is convenient to
ignore them altogether. Their number $n\frac{d(d-1)}{2}$ ($n$
is the number of particles) is then subtracted both from $N_f$ and from $k$,
and eqn.\ref{eqn:relkh} still holds. Such granular systems are then analogous
to `central-force networks':
networks of freely articulated bars, or systems of threads tied
together, in which only the translational degrees of freedom of the nodes
matter. One should be aware, however, that the presence of friction
reinstates rotations into the problem.

We now illustrate the notions and properties introduced in this section with
examples of structures defined in systems A, B and C, ignoring, as explained
just above, disc rotations.

First consider system B. Three different structures are apparent on
figure~\ref{fig:sysB}. The first one, that we denote as SB1, is the set
of bonds that are drawn as thick lines; the second, SB2, contains all bonds of
SB1, plus those that are drawn with thin continuous lines on the figure; and,
finally, the third structure, SB3, comprises all possible bonds between
nearest neighbours in the system, \emph{i.e.,} all those of SB2 plus the
dotted lines. Ignoring rotations, one has $N_f=2n=38$.

\begin{figure}[htb]
\centering
\includegraphics[width=8cm]{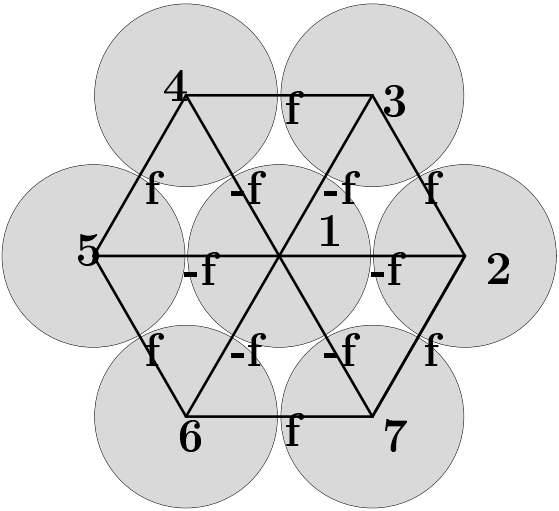}
\caption{A set of self-balanced normal forces. The 6 bonds of the
regular hexagon perimeter carry some normal force $f$, while the 6 ones
involving the central disc labelled 1 carry the opposite force.
\label{fig:hyper}}
\end{figure}

Structure SB3 is a set of rigid triangles sharing common edges with their
neighbours. It is devoid of mechanisms, except the 3 overall rigid body
degrees of freedom of the system. Thus $k=3$. $N=42$ bonds are present. In
view of eqn.~\ref{eqn:relkh}, one has $h=7$. One can exhibit 7 linearly
independent systems of self-balanced normal forces, as follows. The small
structure, with 12 bonds, involving 7 discs, depicted on fig.~\ref{fig:hyper},
allows to define one such set of forces. Noting that 7 such patterns
are present on SB3 (centered on discs 5, 6, 9, 10, 11, 14 and 15), the right
count is reached.

Structure SB2 is made of $N=35$ bonds. It can be shown (on studying the
properties of the corresponding matrix $G$) to be devoid of self-balanced sets
of forces, $h=0$, and of mechanisms other than rigid body motions, $k=3$.
Thus $N+k=N_f+h$.

\begin{figure}[htb]
\centering
\includegraphics[width=8cm]{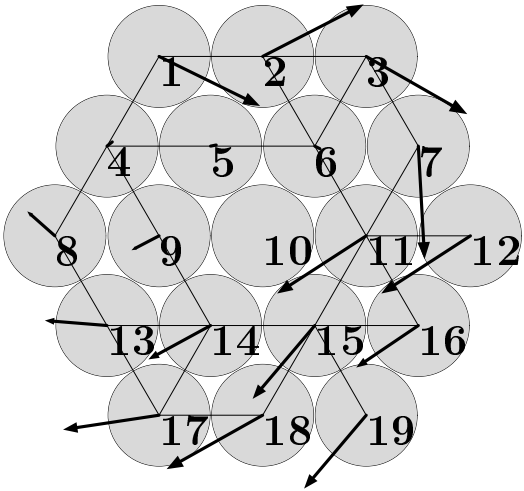}
\caption{A collective mechanism on structure SB1. Arrows represent disc
velocities. \label{fig:meca}}
\end{figure}

Structure SB1, comprising $N=25$ bonds only, still has $h=0$. According to
eqn.~\ref{eqn:relkh}, it should possess 10 additional independent mechanisms.
2 of them are due to disc 10, which is now completely free. 4 others involve
discs 5, 9, 12, and 19, which are still free to move in one direction. In the
case of a divalent disc like 5, this is due to the exact alignment, on the
regular lattice, of bonds 4-5 and 5-6. Four less trivial mechanisms are
more collective. One of them is shown on figure~\ref{fig:meca}.
Two structures, SA1 and SA2, are defined, on fig.~\ref{fig:sysA}, in system A.
SA1 is made of all bonds drawn with continuous lines, and SA2 contains, in
addition, the two bonds drawn with dotted lines (19-24 and 32-34). Depending
on the boundary condition, discs 1 to 8 either possess collectively one degree
of freedom (for BC2) and then $N_f=57$, or none (for BC1) and $N_f=56$.

SA2 has 57 bonds. It is devoid of mechanism ($k=0$) for whatever BC. For BC2, 
one also has $h=0$ and eqn.\ref{eqn:relkh} holds as an equality between the
number of bonds and the number of degrees of freedom. For BC1, one has $h=1$.
Indeed, one may recognize, in the bottom left corner of the pile, with discs
1, 2, 3, 9 and 10, part of the hyperstatic pattern of fig.~\ref{fig:hyper}.
With BC1, one needs not care about equilibrium of discs 1, 2 and 3 that are
perfectly fixed. A system of self-balanced bond forces is thus found on
attributing a common value to the normal forces in bonds 1-9, 9-10, 10-3, and
the opposite value to the normal forces in bonds 2-9 and 2-10. In the case of
BC2, those forces do not balance, since the equilibrium equation for the
collective degree of freedom of the bottom row (a combination of
eqns.~\ref{eqn:defQ1BC2} and \ref{eqn:fint}) is not satisfied.
As to SA1, it has the same properties as SA2, with 2 additional mechanisms
(collective ones like that of fig.~\ref{fig:meca}).

Consider now structure SC that is shown, in system C, on
figure~\ref{fig:sysC}, with the lines connecting disc centers, or joining
discs to the wall, that define $N=70$ bonds. Taking into account the degree
of freedom of the wall, one has $N_f=2n+1=75$. One may show $h=0$. Thus one
has $k=5$. Two discs (10 and 14) are entirely free, hence 4 mechanisms.
The missing one is a global rotation, as a solid body, of the set of all
particles around the center of the circular container, the wall remaining
immobile. Such a motion would not be possible if the same boundary condition
was used with another container shape.

\subsection{The problem: the structure and the load.}

Once a list of bonds is chosen, thus defining the structure,
we shall refer to the situation of the structure submitted to a given
load as `the problem'. 

Solving the problem would mean finding the motion or equilibrium state of the
system (determining, \emph{e.g.,}
new equilibrium positions and intergranular forces), once the load, from
an initial state of rest with no external force,
has been applied. We are not, of course, able to do
that at this stage, since no contact law relating the forces to the relative
motion of neighbouring particles has been introduced. The only information
available is that the internal forces are required to belong to some vector
space that is known once the structure is defined, and to be exerted on given
points on the grain surfaces.

It is said that \emph{the load is supported} by the structure if its
application leads to an equilibrium state in which internal forces, carried
by the bonds of the structure, balance the external ones. 

We can state a
necessary condition for the load to be supported: it must be possible to find 
statically admissible intergranular forces. Necessarily, the $N_f$-vector
of external forces must lie in the range of operator $G^T$, \emph{i.e.}, it
must be orthogonal to the null space $M$ of $G$:
\be
(F_\mu)_{(1\le\mu\le N_f)} \perp M. \label{eqn:suppor}
\ee
This simply means that if the load is to be supported, it must
not set the mechanisms into motion. Such a load is said to be
\emph{supportable}. All supportable loads are not always supported.

By definition, the \emph{backbone} of a structure is the
set of bonds $l_0$ such that a list of statically admissible internal forces
$(f_l)_{1\le l\le N}$ exists with $f_{l_0}\ne 0$. In the following we shall
also refer, as `the backbone', to the set of grains reached by such bonds.

In general, a full mechanical characterization of the equilibrium properties
of the system requires some constitutive law in the contacts. 
However, there are interesting situations in which 
\begin{itemize}
\item
condition~\ref{eqn:suppor} being fulfilled, the load is supportable;
\item
if it is supported, then all intergranular forces are uniquely determined by
the equations of equilibrium. \end{itemize}
These two conditions define an \emph{isostatic problem}.

Further restrictions on internal forces are often enforced in the form of
inequalities. The definition of a supportable load is then modified
accordingly, imposing additional conditions, to be satisfied simultaneously
with~\ref{eqn:suppor}. Their consequences will be discussed in sections III
and VII.

\subsection{Isostaticity: various definitions.}

In section VI we shall see that equilibrium configurations of assemblies of
rigid frictionless grains interacting via contact forces only are generally
such that the problem is isostatic. 

Here, we first insist on the difference between an \emph{isostatic problem},
as defined just above, and an \emph{isostatic structure}, to be defined
below. Once condition~\ref{eqn:suppor} is satisfied, the set of possible bond
forces is an affine space of dimension $h$. One has an isostatic problem if
both conditions ~\ref{eqn:suppor} and $h=0$ are fulfilled. Some mechanisms
might still exist in the structure ($k\ne 0$), provided they are orthogonal to
the load direction.

Structure SA1 (figure~\ref{fig:sysA}), with discs exactly centered on the
sites of a regular triangular lattice, is such that the problem, denoted as
PA1 in the following,  defined with BC2 and the following load: \footnote{
The load, in that case, is supportable if, and only if, $42p\le Q_1\sqrt{3}
\le 74p$ }
 \be
\left\{\begin{array}{rc}(9\le i\le 36)\ \ {\bf F}_i^{ext}&= -p{\bf e}_y \\
Q_1&={294\over 5\sqrt{3}}p,\\
\end{array}
\right.
\label{eqn:Aload}
\ee
where $p$ is the weight of one disc and ${\bf e}_y$ is the vertical upwards
unit vector, is isostatic, although 2 mechanisms are present. 

Analogously, structure SB1, along with the load shown on
figure~\ref{fig:sysB}, defines an isostatic problem PB1 in spite of the
$k=10$ mechanisms. In particular, the load direction (provided discs sit right
on the regular lattice sites) is exactly orthogonal to the velocity vector
represented on figure~\ref{fig:meca}. Structure SC, submitted to the following
load:

\be \left\{\begin{array}{rc}(1\le i\le 37)\ \ {\bf F}_i^{ext}&=0 \\
Q_1&=Q_1^0,\\ \end{array}
\right.
\label{eqn:Cload}
\ee

where a prescribed value $Q_1^0$ is imposed to generalized force $Q_1$ defined
in eqn.~\ref{eqn:Qcerc}, yields an isostatic problem.

Isostatic \emph{structures}, on the other hand, are such that all problems are
isostatic, whatever the choice of the load. More precisely, one requires all
loads orthogonal to the overall rigid-body degrees of freedom to be
supportable with a unique determination of internal forces. Equivalently, both
conditions $h=0$ and $k=k_0$ are to be satisfied. Both the degree of
hyperstaticity and the degree of hypostaticity (excluding rigid-body motions)
should be equal to zero. This entails the well-known condition \be N=N_f -
k_0,\label{eqn:maxwell} \ee
stating that the number of equilibrium equations ($N_f-k_0$) is equal to
the number of unknowns ($N$).

Equality~\ref{eqn:maxwell} is a necessary condition for the structure to be
isostatic, not a sufficient one. For example, in the structure defined by the
addition of the bond joining discs 19 and 24 to SA1 with the first boundary
condition (BC1), one has $k_0=0$, $N=N_f=56$, while $h=k=1$.

Structure SA2, with BC2, is isostatic. SB2, with $N_f=38$ and $k_0=3$, is
isostatic. As to SC,  it would be isostatic upon removal of
grains 10 and 14, only if the global rotation of the set of grains with
respect to the wall were ignored.  Of course, all those structures, as we are
dealing with discs, are only isostatic if rotations are ignored. Only problems
with no external torque exerted on the grains are isostatic. This should be
remembered on comparing $h$ and $k$ with and without friction in such systems.

As we shall see, isostatic problems, rather than isostatic structures,
naturally occur in some model granular systems. The distinction is relevant,
for it accounts for disconnected or `dangling' parts in disordered structures
like SC, and for the peculiarities of lattice models. Moreover, some systems
can also spontaneously, as we shall see,  select a non-rigid ($k>k_0$)
equilibrium configuration. 
\subsection{Generic versus geometric properties.}
The distinction between isostatic problems and isostatic structures should not
be confused with another one: that between \emph{geometric} and \emph{generic}
isostaticity.  We have used a \emph{geometric} definition of a structure, as
associated to one particular position of the system in configuration space,
and accordingly the definition we gave is that of geometric isostaticity.  A
\emph{topological} one can be introduced which, irrespective of particle
positions, is only sensitive to the connectivity of the network of bonds.  In
the case of spheres or discs, when rotations can be ignored, this amounts to
regarding the structure as a graph: a set of edges (bonds) joining at vertices
(grains).  Operator $G$, spaces $S_0$, $M$ and their dimensions $h$ and $k$
smoothly depend on the coordinates of the grains, \emph{via} vectors ${\bf
n}_{ij}$ and ${\bf R}_{ij}$. However, the rank of a parameter-dependent matrix
stays at its maximum  except for special values of the parameters.
Equivalently, the dimension of the null space is generically equal to its
minimum value. Applying this to both $G$ and $G^T$, one may define the generic
degree of indeterminacy of velocities (with due account to the $k_0$
rigid-body degrees of freedom) $k$ and the generic degree of indeterminacy of
forces $h$ as the respective generic (minimum) dimensions of their null
spaces. This allows to define a suitable isostaticity notion for topological
structures: a \emph{generically isostatic structure} is one for which both
numbers $h$ and $k-k_0$ are equal to zero.

It follows from the definitions that a geometrically isostatic structure is
always, once regarded as a topological structure, a generically isostatic
one, but that the reciprocal property is not true. Ref.~\cite{GRHBTC90} gives
a counterexample for a system  of discs (like systems A and B, equivalent to a
network of articulated bars) on the regular triangular lattice. In specific
configurations (like that of a regular triangular lattice), one might
exceptionnally have $h=k-k_0>0$ on generically isostatic structures. 

In two dimensions, there exists some powerful algorithms~\cite{JT95,MD95} to
evaluate the generic degrees of force and velocity indeterminacy in
central-force networks (or systems of frictionless discs). Such computational
methods only deal with connectivity properties, they do not manipulate
floating-point numbers and are therefore devoid of numerical round-off errors.
They were successfully applied to systems of up to $10^6$ nodes. However they
are of course unable to compute position-dependent quantities like force
values.   
\section{Contact law and potential minimization.} 
So far, the only
restriction on intergranular forces was that they should be normal to the
grain surfaces.\footnote{In fact, all the properties of Sections
II, IV, VI hold true provided the \emph{direction} of each intergranular force
is imposed.}
In this section we consider some more specific cases of frictionless grains,
in which some ``contact law'', relating normal forces to relative positions,
is known. This provides some limited additional information, that is not
sufficient in general to predict the grain trajectories once they are
submitted to external forces, for all dynamical aspects are still unknown and
the characterization of equilibrium might even be incomplete. Our aim is to
deduce as much as possible on the global properties of the granular assembly
from as little information as possible on the detailed mechanical laws of the
contacts, in order to stress the importance of geometrical aspects. Thus we
first present the simplest case of rigid, frictionless and cohesionless
grains, in which contacts simply behave as struts. Then we introduce and
briefly discuss other possible laws in which unilaterality or rigidity
constraints are modified or relaxed. Most of those frictionless systems
possess a potential energy that is stationary at equilibrium states and
reaches then a minimum if they are stable. Throughout this section, it is
assumed that a one-parameter loading mode has been defined for varying
particle positions and orientations, with constant external forces, and that
the potential energy of external forces, $W$, can be written in the forms of
eqns.~\ref{eqn:defW} and~\ref{eqn:defWoplm}. 
\subsection{Rigid frictionless grains, no cohesion.} 
In this case, the contact law takes the form of the
so-called Signorini condition:
\be 
\left\{\begin{array}{ll}f_{ij}=0&\mbox{if
$h_{ij}>0$} \\ f_{ij}\ge 0&\mbox{if $h_{ij}=0$}\\ \end{array} \right.
\label{eqn:loicont}
\ee
It should be noted that this law does not express a functional dependence of
$f_l$ on $h_l$. Let us study the variations of $W$ near equilibrium states.
First, consider such a state, in which some non-negative contact forces
$f_l^*$, in closed contacts ($h_l=0$)  balance the external load $Q$. Let us
apply the theorem of virtual power with statically admissible force set
$(f_l^*)_{1\le l\le N}$, and arbitrary particle velocities, corresponding to
relative normal velocities $\delta V_l = -{dh_l\over dt}$ and a value $\lambda
= {d\Lambda \over dt}$ for the kinematic parameter conjugate to $Q$. For any
$l$ such that $f_l^* >0$, the Signorini condition requires that $h_l=0$ and
one must have $\delta V_l \le 0$ to comply with the impenetrability
constraints. Then, from 
\[
{dW\over dt} = -Q{d\Lambda\over dt} = -Q\lambda =
-\sum_l f_l^* \delta V_l
\]
it follows that any motion that does not lead to grain interpenetration can
only, to first order in $t$ (any parameter on the trajectory in configuration
space)  \emph{increase} the potential energy. This non-negative first-order
variation might be equal to zero if $ \delta V_l = 0$ for any active contact
$l$, \emph{i.e.,} if a mechanism exists on the backbone of the contact
structure. Whether the equilibrium state corresponds to a minimum of $W$
depends then on the sign of second or higher order variations. If the backbone
of the contact structure is rigid, then $W$ is necessarily minimized at
equilibrium.

Conversely, let us assume that a configuration of the grains has been reached,
that locally minimizes $W$ under the constraints $h_l \ge 0$. There must then
exist some non-negative \emph{Lagrange multipliers} $f_l$, such that, for any
coordinate $x_\alpha$, \be -{\partial \over \partial x_\alpha} (Q\Lambda)= -
\sum_l f_l {\partial h_l\over \partial x_\alpha}. \label{eqn:mineq}
\ee
Only for such indices $l$ that $h_l=0$ do the $f_l$ take non-vanishing values.
The partial derivative in the right-hand side of~\ref{eqn:mineq} is the
opposite of matrix element $G_{l,\alpha}$, while, from~\ref{eqn:defW}, that of
the left-hand side is the external force conjugate to $x_\alpha$.  Thus, we
have just written that parameters $f_l$ are in fact equilibrium contact forces
satisfying~\ref{eqn:loicont}, and reaction forces stem from geometrical
constraints.  

We now introduce a few other related contact laws and mechanical models.
\subsection{Systems with tensile or bilateral forces.}
Networks of rigid strings or cables are analogous to frictionless spheres
(ignoring their rotations) if the sign of forces is reversed and if the
distance constraint $h_l\ge 0$ is replaced by $h_l\le 0$. The Signorini
condition~\ref{eqn:loicont} becomes
\be
\left\{\begin{array}{ll}f_{ij}=0
&\mbox{if $h_{ij}<0$} \\ f_{ij}\le 0&\mbox{if
$h_{ij}=0$},\\ \end{array}
\right.
\label{eqn:loitens}
\ee
and the whole treament of the preceding subsection straightforwardly applies.

In the case of non-spherical grains, an analogous system supporting tensile
forces is an idealized chain, in which `grain' -chain links-  perimeters are
free to cross. Pairs of neighbouring links (interpenetrating `grains') exert a
force on one another, opposing their separation, when their intersection
reduces to a contact point.

A \emph{bilateral} contact law:
\be
\left\{\begin{array}{ll}f_{ij}=0&\mbox{if $h_{ij}\ne 0$} \\
f_{ij}\ \mbox{unknown}&\mbox{if $h_{ij}=0$},\\
\end{array}
\right.
\label{eqn:loibil}
\ee
might model rigid cohesive grains, that `stick' to one another. The sticking
force might be limited by an unequality:
\be
\left\{\begin{array}{ll}f_{ij}=0&\mbox{if $h_{ij}\ne 0$} \\
f_{ij} \ge -f_0&\mbox{if $h_{ij}=0$},\\
\end{array}
\right.
\label{eqn:coh}
\ee
When one simply uses the form~\ref{eqn:loibil}, assuming the pairs that are
stuck in contact will not come apart, the conclusions of subsection III.A
still hold, if unilateral conditions on relative velocities and displacements
are replaced by bilateral ones, and if all sign constraints on contact forces
are removed. Equilibrium configurations are characterized by stationarity of
potential energy $W$. Minimization of $W$  ensures stability. A sufficient,
but not necessary condition for minimization of $W$ is the rigidity of the
backbone of the contact structure.

Reciprocally, statically admissible normal contact forces naturally appear
as Lagrange multipliers associated with bilateral constraints $h_l=0$ at a
potential energy minimum. 

However, contact law~\ref{eqn:coh} does not lend itself to a potential energy
formulation.

\emph{Tensegrities}é\cite{Vassart} (with rigid elements) are by definition
mixed networks of struts (satisfying condition~\ref{eqn:loicont}) or bars
(bilateral) on the one hand, and cables (satisfying~\ref{eqn:loitens}), on the
other hand. Their potential energy has the same properties as stated above.
\subsection{Systems with a smooth interaction potential.} 
The model of
perfectly rigid grains is physically reasonable when contact deformations
($h_l<0$) are negligible in comparison with any other relevant length in the
problem. When this is no longer the case, or when one wishes to model sound
propagation, it is appropriate to deal with contact laws that involve elastic
deformations, \emph{e.g.,}
\be
 \left\{\begin{array}{ll}f_{ij}=0&\mbox{if
$h_{ij}> 0$} \\ f_{ij}=K_{ij} \vert h_{ij}\vert^m&\mbox{if $h_{ij}\le 0$},\\
\end{array}
\right.
\label{eqn:hertz}
\ee
in which $K_{ij}$ is a stiffness constant that depends on material properties
and on the geometry of contact $i,j$. The exponent is $m=3/2$ (Hertz law) for
smooth surfaces in 3D, and other values might model roughness and the presence
of conical asperities~\cite{GO90,JO85}.

Such contact forces derive from an elastic potential energy:
\be
W^{el}=\sum_{l=1}^N w(h_l),
\mbox{with}\  w(h_l)={K_l\over m+1}\vert h_l\vert ^{m+1}.
\label{eqn:defpotel}
\ee
Likewise, rigid cables as introduced in subsection III.B could be replaced by
elastic ones. That stable equilibrium states correspond to minima, in the
absence of frictions, of the total potential energy \be
W^{tot} = W^{el}+W,
\label{eqn:wtot}
\ee
sum of the elastic potential~\ref{eqn:defpotel} and the potential energy of
external forces~\ref{eqn:defW} or~\ref{eqn:defWoplm}, is an extremely
familiar property. The Signorini condition might physically be regarded as the
limit of the interaction law expressed by equation~\ref{eqn:hertz} when the
stiffness constants become very large, or, equivalently, when the level of
intergranular forces approaches zero. Alternatively, it is mathematically
possible to introduce a regularized contact law of the form~\ref{eqn:hertz} as
an approximation,  when contacts are stiff enough, of the ideal
impenetrability constraint. Such a point of view is adopted in optimization
theory: the procedure known as penalization of the constraints amounts,
instead of minimizing $W$ subject to impenetrability constraints, to searching
for unconstrained minima of $W+W^{el}$. 

\begin{figure}[ht!]
\includegraphics[width=8cm]{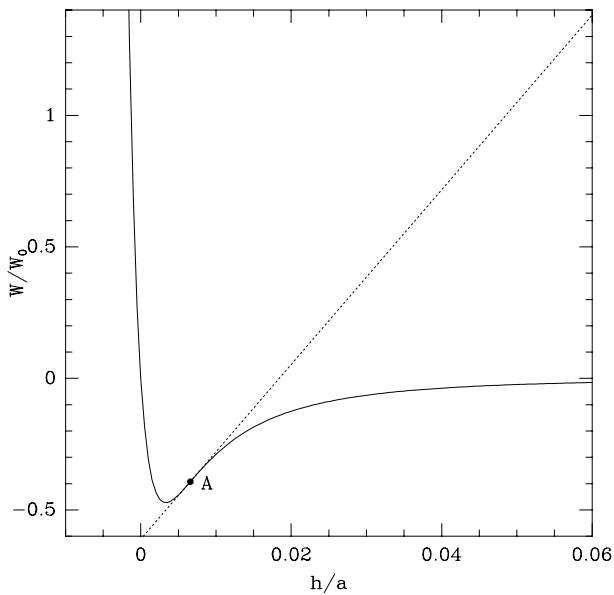}
\caption{Interaction potential, with an attractive tail, as a function of interstitial thickness $h$.
The curve has an inflexion point $A$, corresponding
to the maximum attraction force (equal to the slope of the dotted line).
\label{fig:attraction}} 
\end{figure}

Tensile contact forces of limited intensity, as in contact law~\ref{eqn:coh},
might result from some attractive interaction of finite, but small, range, as
depicted on fig.~\ref{fig:attraction}.
It is interesting to note that the addition of an attractive tail has turned
the potential $w(h)$ into a non-convex function of interstitial thickness
$h$. At the inflexion point, A,  the attractive force reaches its maximum
$f_A$. If one pulls, with a growing force, on two grains in contact in order to
separate them, an instability, in which the contact suddenly breaks open, is
reached as the pulling force reaches the value $f_A$. When the corresponding
intergranular distance, $h_A$, is so small that it is negligible in comparison
to all other relevant lengths in the problem, one might then replace the
smooth attractive potential by contact law~\ref{eqn:coh}, with $f_0=f_A$. On
doing so, one loses  however the possibility to exploit minimization
properties.

We shall see that the potential minimization properties have important
consequences in terms of the possible uniqueness of the equilibrium state
under a prescribed load, and, eventually, as to the possible origins of
macroscopic plastic dissipation. But, first, we have to extend the properties
we have stated for velocities (or infinitesimal displacements) to small
displacements, around a given reference configuration. 

\section{The approximation of small displacements.}
\subsection{Definition.}
We wish to use the concepts we have introduced in the preceding sections
while allowing some motion of the grains, of small but finite extent, which
might alter the list of closed intergranular contacts. Consequently, we
introduce the assumption that displacements, from a reference configuration,
are small enough as to be regarded as infinitesimal quantities. This
\emph{approximation of small displacements} (ASD) is a crucial step that is
very often taken in solid state mechanics. Indeed, it is indispensable if one
wishes to deal with linear problems: adding up two displacement fields, for
instance, in continuum mechanics, is otherwise a meaningless operation. In the
case of granular systems, it will also lead to a linearization of the
problems, for the curvature of configuration spaces will be ignored.  Its
range of validity has to be assessed a-posteriori, but is of course presumably
larger in dense systems, where contacts might open and close with only tiny
changes of the relative positions of neighbouring grains. 

Specifically, we assume the coordinates of the grains to stay close to
reference values. Quantities pertaining to the reference configuration will
be labelled with a superscript `$0$'. It is often convenient, then, to work
with a fixed structure --the list of contacts that might close, and transmit a
force, is a-priori known. 

Interstitial thicknesses $h_l$ are written as $h_l = h_l^0-\delta u_l$, with
a relative normal displacement $\delta u_l$ that is \emph{linear} in the grain
displacements (and rotations), regarded as small quantities. Vectors ${\bf
n}_{ij}$, ${\bf R}_{ij}$, ${\bf R}_{ji}$ are regarded as constant, equal to
${\bf n}_{ij}^0$, ${\bf R}_{ij}^0$, ${\bf R}_{ji}^0$.  As they appear as
cofactors of the displacements, taking their variations into account would
introduce second order terms. All changes of the structure geometry are
ignored. Spaces $\ccc$, $\sss$, operators $G$, $G^T$, are assumed to be the
same in the actual as in the reference configurations. Displacements are now
endowed with the same linear algebraic structures as velocities. $G$ operates
on displacements, yielding relative normal displacements $\delta u_l$, the
compatibility condition for relative normal displacements is the orthogonality
to the space  of self-balanced internal forces $\sss_0$, a theorem of virtual
work can be stated instead of the theorem of virtual power, etc\dots 

Within the framework of the ASD, the specificity of mechanical problems
disappears: as the effect of the displacements of the grains (variations of
the coordinates) on the positions (coordinates) themselves are ignored, one
can find analogies with various other local properties of a list of fixed
points, nodes or lattice sites. Forces now appear as unknown vectors carried
by fixed directions, and the sum of incoming forces on a node has to vanish.
Part V introduces the analogy with scalar transport on a fixed network.
\subsection{Lattice models.} 
Regular packings of monodisperse spheres in 3D
(or discs in 2D) on FCC or hexagonal compact (respectively, triangular in 2D)
lattices are simple systems that are often studied theoretically,
experimentally~\cite{DM57,Rennes2} and
numerically~\cite{SHR87,RH89,OR95,OR97a,OR97b,LU97,JNR97a,JNR97b,HHR97,MO98a}.
Because truly monodisperse systems do not exist, and because of possible
elastic deformations of the grains, one cannot expect such lattices to remain
perfectly regular and undisturbed. However, as lattice perturbations will be
small, it is a common practice~\cite{SHR87,RH89,OR97a,OR97b,JNR97a,JNR97b} to
resort to the ASD, with a perfect lattice as the reference configuration from
which displacements and strains are evaluated.

Consider \emph{e.g.,} the case of slightly polydisperse discs on a triangular
lattice, as in systems A and B. A perfect lattice can be chosen as the
reference state, in which the spacing between neighbouring sites is the lowest
upper bound $a$ of the diameter distribution. Diameters are assumed to be
distributed between $a(1-\alpha)$ and $a$, with a small parameter $\alpha \ll
1$. The diameter of disc $i$ is thus  \be a_i=a(1-\delta_i\alpha),
\label{eqn:randdiam} \ee
$\delta_i$ being a random number, drawn independently for each $i$ between $0$
and $1$. When a certain number of intergranular contacts is created, as it is
often necessary (cf. section III) in order to sustain some external forces,
the lattice will be slightly distorted, with displacements of order $\alpha$.
The ASD amounts to deal with all relevant quantities to leading order in
$\alpha$. In all possible contacts, the normal unit vector is kept parallel to
one of the three directions of dense lines in the triangular lattice. It is
convenient to work with a fixed structure $S_0$ that comprises all bonds
between nearest neighbours on the lattice. If grains are required to touch to
exert a force on one another, forces, in a state of equilibrium under a
supported load, will be carried by some contact structure, the bonds of which
form a subset of $S_0$. 

One might then regard problem PA1, in system A, as defined on $S_0$. Once the
random radii were fixed, we found, within the ASD, an equilibrium
configuration for problem PA1, satisfying the Signorini
condition~\ref{eqn:loicont}, in which the contact structure was SA1.
Similarly, once the values of the radii were known in system B, SB1 was found,
within the ASD, as the contact structure corresponding to a solution of
problem PB1, posed on SB3=$S_0$. Within the ASD, all displacements and
deformations are proportional to $\alpha$, and the problem is, apart from a
scale factor $\alpha$ for displacements, only sensitive to parameters
$(\delta_i)_{1\le i\le n}$.

Such is not the case, of course, without the ASD, if one takes into account
the rotations of unit vectors ${\bf n}_l$ of the bonds due to the deformation
of the lattice.  
\section{Analogy with scalar problems.} 
We briefly recall the
analogy between the mechanical problems we have been discussing, within the
approximation of small displacements, and that of current transport on a
resistor network. Such an analogy was presented \emph{e.g.,} in
ref.~\cite{GRHBTC90}. It is useful because some properties are more
immediately intuitive in scalar models, and because statistical models
(percolation, directed percolation, minimum paths\dots ) have been more
extensively studied and are more familiar in the scalar case. The term
`scalar' refers to the transport of a scalar quantity (current) as opposed to
a vectorial one (force) in mechanical problems. Currents entering one node by
the conducting bonds of the network should balance the external current fed
into that node, just like bond forces balance external efforts. The analog of
the displacement vector (which, in the general case, also involves angular
displacements) is the (scalar) potential of a node, and the duality between
forces and displacements translates into the duality between currents and
potentials. All the developments of section II, adapted within the ASD to
displacements instead of velocities, are valid for resistor networks. $\delta
u_l$ is the potential drop in bond $l$.  One may define spaces $\fff$, $\ccc$,
$\vvv$, $\sss_0$, $M$ operators $G$ and $G^T$, state the theorem of virtual
power, etc\dots The analog of a system of self-balanced bond forces is a set
of currents satisfying the conservation law without any external source,
\emph{i.e.}, a combination of current loops. One may define as many linearly
independent elements of $M$ as there are disconnected parts in the network.
The number of degrees of freedom $N_f$ is now equal to the number of nodes. It
is related  to the number of bonds $N$, the number of independent loops $h$
and the number of disconnected parts ($1$ for a connex network) $k$ by the
scalar version of eqn.~\ref{eqn:relkh}: \[ N+k=N_f+h, \] a simple topological
identity valid for an arbitrary graph.

\section{The isostaticity property.}
\subsection{Statement and context.}
We consider an assembly of rigid, frictionless grains that only exert normal
contact forces on one another. Those forces might however, be attractive or
repulsive. We assume that the system, submitted to a prescribed load, has
evolved to an equilibrium configuration in which the contact structure
supports the load. We also regard the geometric definition of particles as
incompletely known, thereby introducing randomness: such parameters as grain
diameters or radii of curvature are to be regarded as distributed over small
intervals.

Then one can state the following remarkable property: {\em  with probability
one, the problem, posed on the contact structure, is isostatic}.

Such an isostaticity property was (more or less explicitly) reported in
ref.~\cite{GRHBTC90} and articles cited therein, in the case of triangular
lattice systems, within the ASD, with grains satisfying the Signorini
condition~\ref{eqn:loicont}. Isostaticity was also stated in
refs.~\cite{OR95,OR97a,OR97b,JNR97b}, that deal with the same model.
Moukarzel~\cite{MO98a,MO98b} then argued that systems of frictionless grains
interacting by repulsive elastic contact forces should become isostatic in the
limit of large contact stiffnesses. And ultimately, Tkachenko and
Witten~\cite{TW99} derived an isostaticity property for disordered systems of
rigid frictionless spheres in arbitrary dimension, each grain being submitted
to an external force (\emph{e.g.,} to its weight), whatever the sign of
contact forces.

Here, we will establish the isostaticity of the \emph{problem} ($h=0$), rather
than the isostaticity of the structure ($h=0$ and $k=k_0$), in quite general
situations. As we shall see in section VIII, full rigidity ($k=k_0$) in
addition to absence of hyperstaticity ($h=0$), is a less general property, of
\emph{geometric}, as opposed to \emph{topological}, origin. 
\subsection{General arguments.} 
The arguments we give below to establish the
isostaticity property emphasize the peculiarity of equilibrium states, in
which sufficiently many intergranular contacts should be created in order to
resist the externally imposed forces. Thus such states belong to  a subset of
configuration space of vanishing measure. Grains have been brought to rest by
some unspecified dynamic dissipative process. Our  derivation admittedly
retains a heuristic flavor, for a definitive proof would require much more
specific mathematical assumptions. Readers that demand more mathematical
rigour will have realized that arguments presented by other
authors~\cite{MO98a,MO98b,TW99} are not without reproach either, and may refer
to the next paragraph. There, within the ASD (and thus at the expense of
additional assumptions about the magnitude of displacements from a reference
configuration), isostaticity is rigourously deduced.  

To ease the presentation of our arguments, let us introduce a few compact
notations. We denote as $(q_i)_{1\le i\le N_f}$ a set of coordinates in
configuration space $\eee$. The geometry of the grains depends on some random
parameters (sizes, shapes\dots ), collectively denoted as $\zeta$. $\zeta$
might be regarded as a vector with a large number, say $p$,  of components:
$\zeta \in \R^p$. The evolution of the granular system can be modelled as a
function $\Phi$ that maps an initial configuration $(q_i)^{(0)}_{1\le i\le
N_f}$ to the actual equilibrium configuration $(q_i)_{1\le i\le N_f}$. The
motion of the grains from $(q_i)^{(0)}_{1\le i\le N_f}$ to $(q_i)_{1\le i\le
N_f}$ might \emph{e.g.,} be described by a differential equation. $\Phi$ then
expresses the dependence on initial conditions. $\Phi$ also depends on
$\zeta$, which has the role of a set of parameters. To proceed, on has to
assume that this dependence is sufficiently regular: $\Phi : \eee \times \R^p
\rightarrow \eee $ is generally a  smooth function. Although the evolution of
a pack of grains is expected to exhibit a high sensitivity to parameters and
initial conditions, it is dissipative and will bring the system very close to
equilibrium in a finite time. Chaotic trajectories deviate fast from one
another, but the evolution in a finite time is expected to be expressed by a
smooth mapping, that also depends continuously on parameters $\zeta$, except
perhaps for peculiar values that correspond to bifurcations between different
sets of final states or `attraction basins'. If, for instance, one reproduces
the same dynamical evolution from the initial to the final configurations and
gradually change the size of one particle, one expects, physically, the final
state to change only gradually, until for some value of the geometrical change
some rearrangement of finite extent will suddenly take place. We assume such
bifurcations only occur for isolated values of the parameters, such that
around the actual  $\zeta \in \R^p$, there exists generically a neighbourhood
$\Omega$ within which the parameter set might vary without creating any
discontinuity or closing any additional contact in final configuration
$(q_i)_{1\le i\le N_f} \in \eee$.

Consider now the set $L$ of intergranular contacts corresponding to this
configuration (the contact structure, as defined in section II). As  $\zeta$
changes within $\Omega$, maintained contacts form some non-empty subset of
$L$, which is sufficient to carry the load.

If $\zeta \in \Omega $ varies along a curve parametrized by $u$, so does, via
the mapping $\Phi$, $(q_i)_{1\le i\le N_f}$ in $\eee$.  If a contact $(i,j)\in
L$  is to be maintained in this motion, one must have:
\be
{dh_{ij}\over du} = 0.\label{eqn:keepcontact}
\ee
This means that the coordinates of grains $i$ and $j$ have to adjust to the
change in grain geometry  $\zeta$.  If parameter $u$ is formally regarded as
\emph{time}, relative normal \emph{velocities} ${\displaystyle \delta
V_{ij}=-{dh_{ij}\over du}}$, in all contacts that are maintained, are required
to balance the effect of the change of $\zeta$, to ensure that equality
\ref{eqn:keepcontact} is still satisfied. Increasing, if needed, the number
$p$ of $\zeta$ components, it is natural to assume that such conditions on
relative velocities are independent from contact to contact, for the required
value of $\delta V_{ij}$ only depends on those geometric parameters that
govern the shape of grains $i$ and $j$ in the immediate vicinity of their
contact point.  Therefore, for a list $L$ of $N$ contacts to be maintained for
arbitrary $\zeta \in \Omega$, any $N$-vector  $(\delta V_l)_{1\le l\le N} \in
\R^N$ of possible relative normal velocities in the contacts of $L$ must be
compatible. In view of condition~\ref{eqn:compatible}, only such contact
structures $L$ that are devoid of self-balanced sets of internal forces
(\emph{i.e.,} such that $h=0$ or $\sss _0 = \{0\}$) can be maintained. If,
exceptionnally, the equilibrium configuration $(q_i)_{1\le i\le N_f}$ admits
one non-vanishing element $(\gamma_l)_{1\le l\le N}$ of $\sss _0$, then, as
the condition  \[ \sum_{1\le l\le N} \gamma _l \delta V_l = 0 \] cannot be
ensured for arbitrary $(\delta V_l)_{1\le l\le N} \in \R^N$, and grains cannot
interpenetrate,  one at least of the contacts $l$ such that $\gamma _l \ne 0$
will open ($\delta V_l < 0$) upon slightly tampering  with geometric
parameters $\zeta$. 

We have thus shown that, with probability one, the contact structure in the
equilibrium configuration cannot be hyperstatic, the degree of indeterminacy
of forces $h$ is equal to zero.

The above derivation relies on rather specific assumptions about mapping
$\Phi$. One should be aware, however,  that we are free to choose any initial
configuration that does not violate impenetrability conditions. The
assumptions we have relied upon are quite natural when the initial and final
equilibrium configurations are close to each other. Basically, one has then to
accept the idea that the fine geometrical details of grain surfaces, in the
vicinity of their contact points at equilibrium, do not significantly
influence their trajectories except in the very final stage. Thus they can be
regarded as randomly chosen during this ultimate stage of the approach to
equilibrium, as though the system `realized' then what their actual values
are. In the next subsection it is assumed that the `initial' and final state
are so close that the motion between them might correctly be described within
the ASD. Other derivations might resort to fictitious construction processes
of the granular assembly, in which $\Phi$ is replaced by a simpler function.
One might consider, \emph{e.g.,} sequentially bringing the grains, one by one,
to their equilibrium position, thus gradually enlarging the list of contacts.
If, at any stage in the process, $h$ is strictly positive, some of the
contacts cannot be maintained on slightly altering some of the geometrical
details of grain surfaces near the most recently created contacts. 

The equilibrium state, as we have just concluded, is devoid of hyperstaticity
($h=0$). What about its possible mechanisms ? We have assumed that it can
support the load. It is tempting to conclude that mechanisms do not exist in
the generic case, since the orthogonality condition~\ref{eqn:suppor} would
have to be maintained as the shape of the grains is altered. However, one has
to keep in mind that equilibrium configurations are very peculiar ones, and we
shall see that the existence of mechanisms in the equilibrium state depends in
general on the sign of intergranular forces, and on the shape of the grains.   
\subsection{Alternative derivation within the ASD.
\\The special case of lattice models.}

A slightly different point of view may be adopted in the framework of the ASD:
within the approximation, the problem being replaced by a simplified one, the
isostaticity property can be established in a rigourous way. Also, the analogy
with the scalar problem might make the result more immediately intuitive. Let
us assume the ASD to be valid with a reference configuration in which all
contacts are slightly open: a list of bonds is defined, with strictly positive
values of interstitial thicknesses $h_l ^0$. $h_{ij}^0$, the distance
separating the surfaces of grains $i$ and $j$ is to be regarded as a random
number that depends on fine details of their geometry.  $h_{ij}^0$ values for
the different bonds are independent and continuously distributed. Once the
system has been brought to an equilibrium configuration, forces are carried by
contacts, \emph{i.e.} bonds $l$ for which $h_l=0$. If $(\gamma_l)_{1\le l \le
N}$ is a set of self-balanced forces carried by those contacts, the theorem of
virtual work, applied with such bond forces on the one hand, and with the
displacements from the reference to the equilibrium  configurations on the
other hand, yields :
\be
\sum_{l=1}^N \gamma _l (h^0_l - h_l)=\sum_{l=1}^N \gamma _l h^0_l = 0.
\label{eqn:hyperasd}
\ee
Thus a certain linear combination of the random distances
$h_l ^0$ has to be equal to zero. Coefficients
$(\gamma_l)_{1\le l \le N}$ are fixed once the reference configuration is
known. Moreover, \emph{via} an iterative dilution process, they can be chosen
among a finite set, as we now show: assume a set of self-balanced forces
$(\gamma_l)_{1\le l \le N}$ to exist, and define the set $B_0$ of bonds $l$
for which $\gamma _l \ne 0$. Then, as long as it is possible, proceed to
successive `dilutions' of this set, defining $B_1$, $B_2$, etc\dots requesting
that there is one bond less in $B_{k+1}$ than in $B_k$, but that it is still
possible to find self-balanced forces localized on the bonds of the reduced
set. The final $B_{k_0}$, that can non longer be diluted, will be such that
the values of $\gamma _l$ will be uniquely determined for each $l\in B_{k_0}$,
up to a common factor, which is fixed if one imposes the condition that the
largest $\gamma _l$ is equal to one. In this way, one thus defines
\emph{irreducible sets of self-balanced forces}, that are put in one-to-one
correspondence with certain substructures of the whole contact structure. In a
finite system, one thus has a finite number of such irreducible sets of bond
forces. If a system of self-balanced forces can be carried by the contacts
that are closed, then equation~\ref{eqn:hyperasd} has to be satisfied with one
of the irreducible systems of self-balanced forces, an occurrence of
probability zero.

\begin{figure}[ht] 
\centering
\includegraphics[width=6cm]{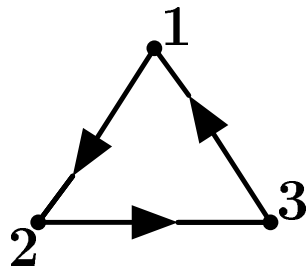}
\caption{Three bonds forming a loop in the resistor network.
A current might circulate as indicated by the arrows.
\label{fig:loop}}
\end{figure}

The scalar analog of this derivation is especially straightforward. To the
requirement that only particles in contact exert a force on one another
corresponds the condition that a bond between sites $a$ and $b$ on the
resistor network can only carry a current when the potential difference
$v_a-v_b$ is equal to a prescribed value, $v_{ab}^0$. Parameters $v_{ab}^0$
are to be regarded as random, chosen according to a continuous probability
distribution and independent from bond to bond. Then, the appearance, once
some current is injected at one node of the resistor network and extracted at
another, of a loop of current-carrying bonds is to be discarded as an
occurrence of zero probability. (One may of course define irreducible loops,
as the ones that carry a unit current and do not contain stricly smaller
subloops). Assume three bonds, making a loop between three sites, say
$1\rightarrow 2\rightarrow 3\rightarrow 1$, to carry a non-vanishing current
(figure~\ref{fig:loop}).
This implies an exact relation of the form $\pm v_{1,2}^0 \pm v_{2,3}^0 \pm
v_{3,1}^0=0$, which has no chance to be satisfied.

Let us consider now, as an example, returning to granular systems, the small
hyperstatic structure of fig.~\ref{fig:hyper}, and assume the 7 grains have
been brought, from the reference configuration of the triangular lattice model
defined in section IVB, in which all interstices are open ($h^0_{ij}>0$), to
an equilibrium configuration in which the 12 bonds are closed contacts, with
$h_{ij}=0$. Labelling the grains as on the figure, equation~\ref{eqn:hyperasd}
reads:
\[
\sum_{i=2}^7h_{1,i}^0 - \sum_{i=2}^6h_{i,i+1}^0,
\]
which is true with probability zero for continuously distributed independent
random numbers $h^0_{ij}$. Within the lattice model with random diameters, as
introduced in section IVB, one has 
\be
h^0_{ij}={a\over 2}(\delta _i +\delta
_j)\alpha \label{eqn:hepsilon},
\ee
one obtains a relationship between $\delta_i$'s:
\[
\delta _1 = {1\over 6} (\delta_2 + \delta_3 + \delta_4 +
\delta_5 + \delta_6 + \delta_7),
\]
which, once again, is satisfied with probability zero.

It is less obvious, however, that the disorder on the radii of discs that
remain exactly circular (or of perfect spheres in 3D) is sufficient, because
of the induced disorder on $h_{ij}$'s, as in eqn.~\ref{eqn:hepsilon}, to
forbid the existence of \emph{any} set of self-balanced contact forces. The
problem is that, because of ~\ref{eqn:hepsilon}, interstitial thicknesses are
no longer independent. On transforming \ref{eqn:hyperasd} into a relation
between $\delta_i$'s, one gets  \[ \sum_i  \left( \sum_{j\ne i}\gamma_{ij}
\right) \delta_i = 0, \]
which might well be satisfied if $ \sum_{j\ne i}\gamma_{ij}=0$ for each $i$.
This latter condition  has no chance to be obeyed in a disordered system, but
may be achieved on a regular lattice. This does not occur, however, with
nearly monodisperse discs on the regular triangular lattice in 2D, because 3
independent conditions per disc are to be satisfied, and the number of
contacts, at most three times the number of discs on this 6-coordinated
lattice, has to be strictly smaller, because hyperstatic configurations like
that of fig.~\ref{fig:hyper} cannot exist. 

The situation is different for the analogous 3D model, defined with slightly
polydisperse spheres on the sites of an FCC lattice. Each sphere has 12
nearest neighbours, and one may find hyperstatic structures in which contacts
will be maintained with polydisperse spheres. A simple example of such a
structure can be found,  with 24 spheres and 64 contacts\footnote{The
interested reader might obtain the list of sphere positions from the author.}.
Although a small amount of polydispersity eliminates hyperstaticity in 2D
triangular lattices of discs, it does not do so in FCC lattices of spheres,
provided the grains, in spite of the distribution of radii, remain perfectly
spherical. If the shape of the grains is also affected by  the slight
geometric disorder, then (with the notations of fig.~\ref{fig:defbond}), one
has $\Vert {\bf R}_{ij}\Vert \ne \Vert {\bf R}_{ik}\Vert$ for $j\ne k$,
interstitial thicknesses $h_{ij}$ become independent in all bonds of the
lattice, and hyperstaticity is forbidden. (Within the ASD, it is consistent to
ignore the rotation of unit vectors ${\bf n}_{ij}$ due to small departures
from sphericity).

\subsection{Consequences. Remarks.}
Once the list of active contacts in an equilibrium state is known,
isostaticity of the problem enables a purely geometric determination of the
forces, independently of material properties. As an example, system C was
brought in equilibrium under the load defined by eqn.~\ref{eqn:Cload}, with
conditions~\ref{eqn:loicont}. As soon as the list of contacts (structure SC)
is known, the set of normal contact forces is entirely determined.

This gives a meaning to the limit of rigid particles: in generic situations,
when the sizes and shapes of the grains are affected by some amount of
randomness, there is no problem of force indeterminacy once an equilibrium
configuration has been reached. The actual value of contact forces will not
depend on the detail of the contact law, provided it might be regarded as
rigid, but it will be sensitive to fine geometrical details. As an example,
consider frictionless elastic contacts obeying eqn.~\ref{eqn:hertz}. Let us
assume a stable equilibrium state of the grain assembly, regarding the grains
as perfectly rigid (condition~\ref{eqn:loicont}), has been reached. One thus
has a local minimum of $W$ (defined in eqns.~\ref{eqn:defW}
or~\ref{eqn:defWoplm}). Then, let us take into account the finite, but small,
deformability of the contacts. The same list of contacts will carry forces
that, to first order in the small displacements, do not change. Evaluation,
within the ASD, of relative normal displacements $h_l<0$ in force carrying
contacts yields $h_l = -({f_l\over K_l})^{1/m}$, such relative displacements
are compatible because of the isostaticity property, and the resulting elastic
energy,
$$
W^{el}={1\over m+1} \sum K_l \vert h_l \vert ^{m+1} = {1\over m+1}
\sum K_l^{-1/m}f_l^{(m+1)/m},
$$
tends to zero as stiffness constants $K_l$
tend to infinity. Thus the actual values of constants $K_l$ and exponent $m$
(these data might vary from contact to contact) are irrelevant.

Once an equilibrium state has been reached,
force values do not depend on the details of the
contact law: this is an important step on the way to the {\em reduction of
the mechanics of granular systems to geometry}--the basic goal of the present
paper. This contributes to ease
the derivation of generic mechanical properties of granular systems. 

The simplification that results from the isostaticity property should however
be balanced with the two following difficulties.

Firstly, configurations of granular systems, due to the same isostaticity property,
are necessarily quite sensitive to fine geometric details: tiny variations of
grain dimensions or positions might lead to opening of some contacts. As all
contacts are indispensable to support the load, the system has to rearrange
somehow to create other contacts that compensate for one that were lost. This
is the origin of a property known as \emph{fragility}, to be more accurately
defined, and discussed, in part IX. 

Secondly, one should be aware that the choice of an equilibrium configuration
among several possible ones might depend on other physical parameters than
the geometry of the grains. The reduction to geometry is thus not complete. In
section VII below, the consequences of the ASD are studied, and it is shown
that mechanical problems are entirely geometric within the approximation.

As a consequence of the absence of hyperstaticity ($h=0$), one readily
obtains, from~\ref{eqn:relkh}, a bound on the number of contacts $N$ that
carry a force, involving the number $N_f$ of degrees of freedom of the
particles belonging to the backbone of the force-carrying structure: $N\le
N_f-k_0$. Neglecting the effect of boundary conditions on the count of $N_f$
in large granular systems, one gets an upper bound on the coordination number
$c = {2N\over n}$:
\be
\left\{\begin{array}{ll}c\le 2d&\mbox{for spheres} \\
c\le d(d+1)&\mbox{in the general case},\\
\end{array}
\right.
\label{eqn:ubc}
\ee
Particles in 3D that possess an axis of revolution, like spheroids, also have
one trivial rotational free motion (in the absence of friction). Thus one
should subtract one degree of freedom for each, hence the bound $c\le 10$,
instead of the general 3D value $12$. 

Interestingly, an estimate $c\simeq 11$ for the coordination number of long
rods or fibers was given by Philipse~\cite{AP96}, on the basis of some
statistical assumptions about the random packings of such particles.

What we have established is in fact the {\em absence of hyperstaticity of a
generically disordered assembly of rigid grains, regarded as frictionless}.
Forces, in the derivation, only appear as convenient auxiliary quantities
(`virtual' forces) to deal with a purely geometric problem. The conclusions
thus holds in the presence of solid friction. Assemblies of rigid grains with
friction therefore abide by inequality~\ref{eqn:ubc}. (It is of course well
known, from numerical simulations in particular~\cite{BR90,ZDG95,OSCS98}, that
the contact coordination number is a decreasing function of the friction
coefficient). 

It is also worth pointing out that~\ref{eqn:ubc} does not depend on the
polydispersity of the grains. Grains that are much larger than their
neighbours will often touch a large number of them. However, this effect
should be compensated in the average coordination number by an opposite one,
affecting  small grains.  When they touch a large one, this latter effectively
occupies half of the surrounding space, thereby reducing the possibility for
other contacts.

On the ground that force-carrying structures should be rigid (devoid of
mechanisms, $k=k_0$) the \emph{opposite} inequality, $N\ge N_f$, whence the
\emph{lower} bound $d(d+1)$ ($2d$ for spheres or discs) for the coordination
number, is sometimes quoted in the literature~\cite{SA98,TW99}. We regard it
as wrong in general (although true for systems of non-cohesive rigid
frictionless spheres, as we shall see). As pointed out by
Alexander~\cite{SA98}, the physically relevant concept is not rigidity, but
stability (under a given external load). This is discussed in section VIII
below. First, section VII is devoted to the exploitation of potential
minimization properties within the ASD. 
\section{Equilibrium and potential minimization within the ASD.} 
The approximation of small displacements
introduced in section IV has several important consequences. Finding an
equilibrium state amounts, in some cases, to solving a convex minimization
problem, for which optimization theory provides useful properties and tools.
The relationship with percolation or minimum path models are also to be
discussed within the ASD. 
\subsection{Convexity.} 
When the potential energy is
a convex function of displacements or positions, and when the rigid
constraints define a convex set in configuration space, then the search for a
stable equilibrium state is a convex optimization problem, and the following 
important properties can be exploited~\cite{Tucker}.
\begin{enumerate}
\item
The equilibrium conditions, which express the \emph{stationarity} of the
potential, are not only \emph{necessary} conditions for potential minimization
(\emph{i.e., }stability), they are also \emph{sufficient}.
\item
A local minimum of potential $W$ is a global minimum. $W$ is flat, equal to
its minimum value, over a convex set of possible equilibrium configurations. 
\item
A structure being given, a supportable load will be supported.
\item
Equilibrium forces are the solution to another optimization problem
(the so-called \emph{dual problem}).
\item
Rigid laws and elastic ones can be dealt with in the same way.
\end{enumerate}

Let us, among the contact laws presented in section III, distinguish the ones
that lead to convex problems. It should be remarked first that standard
convexity is defined in vector spaces, not on manifolds. In order to exploit
the classical results of convex optimization theory to grains of arbitrary
shape, it is necessary to place ourselves within the frame of the ASD, which
replaces the curved configuration space by its flat tangent space $\R^{N_f}$.

As intergranular distances $h_l$ are, within the ASD, affine functions of
displacements, it follows that both rigid constraints $h_l \ge 0$ or $h_l \le
0$ define a convex set (and so does $h_l = 0$): the accessible part of
configuration space is a simplex, a convex set whose boundaries are a
collection of flat sections (parts of affine spaces). Since the potential
energy of external forces, $W$, is linear
in the displacements, its minimization belongs to the class of \emph{linear
optimization problems}, that are the subject of a large literature in applied
mathematics and operational research. This important case --granular systems
within the ASD with contact laws of type~\ref{eqn:loicont}, or systems abiding
by~\ref{eqn:loitens} or~\ref{eqn:loibil} , or tensegrities--is dealt with in
detail in section VIIB.

Still within the ASD, contact laws involving smooth interaction potentials
will lead to convex problems if the potential function $w$ is convex. This is
the case for unilateral elasticity, as defined in~\ref{eqn:hertz}
and~\ref{eqn:defpotel}, but not for intergranular potentials that possess an
attractive tail like on figure~\ref{fig:attraction}.

\emph{Outside} the ASD, convexity can be discussed in the case of
spheres or discs, since, ignoring rotations, their configuration
space is flat. One immediately checks, then, that impenetrability constraints
$h_l\ge 0$, once $h_l$ is no longer approximated as an affine function of
displacements, define a non-convex set of admissible configurations. The
opposite inequality $h_l\le 0$, on the contrary, does lead to convex problems.
As we shall see, frictionless spheres on the one hand, and systems of strings
tied together on the other hand behave exactly in the same way, upon reversing
the sign of forces and deformations, \emph{within} the ASD, but strongly
differ \emph{without} the ASD.

\subsection{Rigid, unilateral contact law.}
\subsubsection{Context. Notations}
The properties of convex problems enumerated above are valid, in particular,
in the case of linear optimization problems, for which they are sometimes
presented in particular forms~\cite{Tucker,JE86}. Here, in order to stress
their physical meaning, we shall directly rederive them. We consider an
assembly of rigid frictionless grains, satisfying the Signorini
conditions~\ref{eqn:loicont}, dealt with within the ASD. We assume a structure
has been defined, and if the load is supported, some of its $N$ bonds will, at
equilibrium, close ($h_l=0$) and transmit a force ($f_l>0$). The following
also applies if condition~\ref{eqn:loicont} is replaced by~\ref{eqn:loitens}
or~\ref{eqn:loibil}. 

Keeping the same notations as in sections II and IV, we know that the
impenetrability constraints are expressed with matrix $G$
\be
\mbox{For\ } 1\le l\le N,\ \ \sum_{\mu = 1}^{N_f} G_{l\mu} u _\mu \le h^0_l,
\label{eqn:impen}
\ee
the transpose of which appears in the equilibrium equations
\be
\mbox{For\ } 1\le \mu\le N_f,\ \ \sum_{l=1}^N G_{l\mu}f_l= F^{ext}_\mu.
\label{eqn:equil}
\ee
Throughout this section, compact notations will be used for vectors of
external forces (${\bf F}^{ext}$ for $(F^{ext}_\mu)_{1\le \mu\le N_f}$)
contact forces (${\bf f}$ for $(f_l)_{1\le l\le N}$),  interstices (${\bf h}$
for $(h_l)_{1\le l\le N}$), and displacements (${\bf u}$ for $(u_\mu)_{1\le
\mu\le N_f}$), the bracket notation (\emph{e.g.,} $({\bf f} \vert {\bf h})$)
is used for scalar products, while operator notations and abbreviation for
inequalities reduce~\ref{eqn:impen} to $G {\bf u} \le {\bf h}^0$.
\subsubsection{Minimization in displacement space.} 
We now show that finding
equilibrium displacements is \emph{equivalent} to solving the following linear
optimization problem:
$$
\ppp _1\left\{\begin{array}{l}
\mbox{Minimize }W=-Q\lambda=-\sum F^{ext}_\mu u_\mu\\
\mbox{with constraints: (\ref{eqn:impen})}\\
\end{array}
\right.
$$
We know from section III that a solution to problem $\ppp _1$ provides a set
of Lagrange parameters $(f_l)_{1\le l\le N}$ that satisfy both
conditions~\ref{eqn:loicont} and~\ref{eqn:equil} (or~\ref{eqn:mineq}), and are
therefore equilibrium forces.

Conversely, in the case of a linear optimization problem such as $\ppp _1$,
the stationarity condition is sufficient to ensure that $W$ is minimized. 

This can be checked as follows: let ${\bf u}^* \in \vvv$ represent one
solution for displacements, and, likewise, let us denote equilibrium contact
forces as ${\bf f}^* \in \R^N$. To ${\bf u}^*$ corresponds the set of values
${\bf h^*}$ for interstitial distances, and the Signorini condition might be
expressed as
$$
({\bf f}^* \vert {\bf h}^*) =0,
$$
while any displacement vector ${\bf u} \in \vvv$, corresponding to ${\bf h}$,
satisfies
$$
({\bf f}^* \vert
{\bf h}) \ge 0.
$$
From the theorem of virtual work, one then has
$$
W({\bf u})-W({\bf u}^*)= -({\bf f}^* \vert {\bf h}^*)+({\bf f}^* \vert {\bf
h}) \ge  {\bf 0}
$$ 
and displacement ${\bf u}^*$ minimizes the potential energy.

\begin{figure}[hbt]
\centering
\includegraphics[width=8cm]{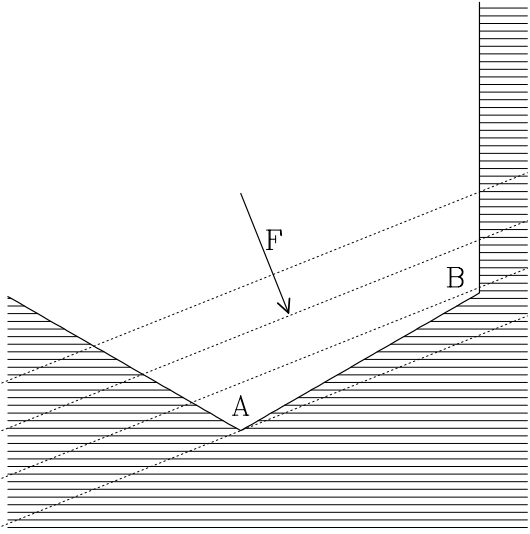}
\caption{
Aspect of simplex of variables satisfying affine constraints
like~\ref{eqn:impen}, cut by the plane of the figure. $W$ is constant on
parallel hyperplanes (sketched as dotted lines, orthogonal to $F$, projection
of load direction onto the plane). $W$ reaches its minimum at one extreme
point at least (like A and B) or on `faces' or` edges', included in an affine
space of dimension $k$, that are part of the simplex boundary, (like segment
AB). The hatched region is forbidden by impenetrability constraints.
\label{fig:simplex1}} 
\end{figure}

Figure \ref{fig:simplex1} is a schematic representation of problem $\ppp _1$.
A simplex, defined by a set of affine constraints like~\ref{eqn:impen}, is
limited by flat faces, where some of the constraints are active. Its extreme
points (the `corners') are where a maximum list of constraints are
simultaneously active. The criterion to be minimized is itself an
affine function, it is constant on hyperplanes
that are orthogonal to the load.
Equilibrium is achieved on the simplex boundary, at least in one extreme
point, in general on a simplex $A$ in a space that is orthogonal to the load
direction. Let $k$ (smaller than $N_f$) denote the dimension of this space.
Within the set of solutions, $W$ is constant, and a certain number $N^*$ of
contacts are maintained closed. Let us denote this structure as $S^*$: it is
the list of contacts that are closed for all equilibrium configurations. For
those equilibrium states that are on the boundary of $A$, some additional
contacts are created. It follows from its definition that $k$ is the degree of
velocity (here, within the ASD, of displacement) indeterminacy of $S^*$.
Since, from part VI, its degree of hyperstaticity is zero,
one has $k=N_f-N^*$. 
\subsubsection{Supportable loads will be supported.}
In general, displacements are thus determined up to some motion within convex
set $A$.

Let us now show that $A$ is not empty if the load is supportable. We assume
some statically admissible forces $(f^0_l)_{1\le l\le N}$ to be defined on the
bonds of the complete structure that was defined a-priori. Then a finite lower
bound for $W$ on the whole simplex of admissible displacements can be obtained
upon writing the variation of $W$ from the reference configuration as
$$
\Delta W= -\sum_{1\le l\le N} f^0_l \delta u_l \ge -\sum_{1\le l\le N} f^0_l
h_l^0.
$$ 
$W$, thus, cannot decrease to $-\infty$ within the simplex, and has
to reach a finite minimum somewhere on the boundary. Moreover, one can show
that $A$ is also bounded, except for \emph{marginally} supportable loads. We
say the load is \emph{not} marginally supportable if there exists a small
neighbourhood of $(F^{ext}_\mu)_{1\le \mu \le N_f}$ in force space $\fff$
within which all loads are supportable. Let us now consider a situation in
which $A$ is not bounded. One can then find one direction along which
displacements go to infinity within $A$. Now let us assume
the load is not marginally supportable. One can apply a
small load increment $(\delta F^{ext}_\mu)_{1\le \mu \le N_f}$,
such that $(F^{ext}_\mu +\delta F^{ext}_\mu)_{1\le \mu \le N_f}$
is still supportable, with $(\delta F^{ext}_\mu)_{1\le \mu \le N_f}$
in the direction for which $A$ is not bounded, which leads to a
contradiction. Therefore the load has to be marginally supportable if $A$ is
not bounded. 
\subsubsection{Dual problem in bond force space} 
We now turn to
the dual optimization problem, to which equilibrium contact forces are the
solution, \emph{viz.} 
\be
\ppp _2\left\{\begin{array}{l}
\mbox{Maximize }Z({\bf f})=-({\bf f} \vert {\bf h}^0)=-\sum_l h_l^0 f_l\\
\mbox{with constraints: (\ref{eqn:equil}) and ${\bf f} \ge {\bf 0}$}\\
\end{array}
\right.
\label{eqn:probp2}
\ee
We know that equilibrium displacements ($\bf{u^*}$) and contact forces
($\bf{f^*}\ge 0$) respectively satisfy~\ref{eqn:impen} and~\ref{eqn:equil},
and are such that 
\be
\left({\bf f}^* \vert (G.{\bf u}^*-{\bf h}^0)\right)=0.
\label{eqn:saddle}
\ee
Thus, any possible set of non-negative bond forces $\bf{f}$ balancing the load
is such that
$$\left({\bf f}\vert (G.{\bf u}^* -{\bf h}^0)\right)\le 0=
\left({\bf f}^* \vert (G.{\bf u}^*-{\bf h}^0)\right)
$$
on the one hand, and
$$
({\bf f}\vert G.{\bf u}^*)=(G^T{\bf f}\vert {\bf u}^*)=({\bf F}^{ext}\vert
{\bf u}^*)
$$
on the other, which entails $Z({\bf f})\le Z({\bf f}^*)$: ${\bf f}^*$ is a
solution to problem $\ppp _2$.

Conversely, if one starts from problem $\ppp _2$, and consider a solution
$\bf{f^*}$, then it is possible to define an $N_f$-vector $\bf{u^*}$ of
Lagrange parameters corresponding to constraints~\ref{eqn:equil},  and an
$N$-vector $\bf{h}$ of non-negative Lagrange parameters corresponding to
constraints $\bf{f}\ge 0$, such that \be -{\bf h}^0+G{\bf u}^* + {\bf h}=0.
\label{eqn:euler2} \ee
Moreover, $h_l$ vanishes whenever $f_l> 0$. This means that $\bf{u^*}$ is
actually a displacement vector abiding by~\ref{eqn:impen}, and
equation~\ref{eqn:euler2}  entails that the Signorini condition, in the
form~\ref{eqn:saddle}, is also satisfied. We know then that $\bf{u^*}$ is a
solution to $\ppp _1$.

Equilibrium displacements and contact forces thus coincide with the respective
solutions to $\ppp _1$ and $\ppp _2$, a pair of \emph{linear optimization
problems in duality}. We have shown that:
\begin{itemize}
\item
If $\bf{u^*}$ is a solution to $\ppp _1$, then it is possible to find a
solution $\bf{f^*}$ to $\ppp _2$, \ref{eqn:saddle} being satisfied.
\item
If ${\bf f}^*$ is a solution to $\ppp _2$, then it is possible to find a
solution ${\bf u}^*$ to $\ppp _1$, \ref{eqn:saddle} being satisfied.
\item
If ${\bf u}^*$ and ${\bf f}^*$ respectively abide by the constraints of
optimization problems $\ppp _1$ and $\ppp _2$, and if, in addition,
\ref{eqn:saddle} (equivalent to the Signorini condition~\ref{eqn:loicont}) is
satisfied, then ${\bf u}^*$ and ${\bf f}^*$ are respectively solutions to
$\ppp _1$ and $\ppp _2$.
\item
The optimum value of the criteria are equal in both problems:
condition~\ref{eqn:saddle} ensures $W({\bf u}^*)=Z({\bf f}^*)$.
\end{itemize}
\subsubsection{The uniqueness property.}
Within the affine space of bond forces satisfying~\ref{eqn:equil}, constraints
$f_l\ge 0$ define a simplex, and, just like for $\ppp _1$, the set of
solutions to $\ppp _2$ is a convex part $B$ of its boundary. Let $h$ denote
the dimension of the affine space spanned by $B$. Since $B$ is the set of
possible equilibrium forces, $h$ is in fact the degree of force
indeterminacy of the problem. Generically, one has, from part VI, $h=0$,
and the only solution to problem $\ppp _2$ is an
extreme point of the simplex of admissible forces. We have thus shown that
{\em in terms of forces, the solution is uniquely
determined}. This is a stronger conclusion that the sole isostaticity of
the problem established in part VI: in general, contact
forces are uniquely determined {\em once the list of contacts is known}.
In the case of a system of rigid grains, with
contact law~\ref{eqn:loicont}, dealt with within the ASD, {\em the list of
force-carrying contacts itself} (the list of bonds, among those that are
defined a-priori in the reference configuration, for which neighbouring grains
will actually touch and exert a force on each other) {\em is uniquely
determined}. Forces are carried by contact structure $S^*$, which was defined
in connection with the discussion of the solutions to problem $\ppp _1$, and,
if some mechanisms exist ($k>0$), the other contacts that might be created
will not carry any force.

If the contact law is~\ref{eqn:loicont}, if geometrical changes from a
reference configuration are small enough for the ASD to be valid, if the load
is supportable (but not marginally so), then the system will reach an
equilibrium state, which apart from bounded displacements within convex set
$A$ (that do not change $W$) is totally independent of all dynamical
properties of the system, and entirely determined by the sole geometry.

\subsubsection{Examples.}
Systems A and B introduced in part II, were treated within the lattice model
defined in section IV.B, with the ASD, and condition~\ref{eqn:loicont}.
Structure SA1, once the random numbers $\delta_i$ were known, was obtained as
the uniquely determined list of force-carrying contacts at equilibrium under
the load defined by~\ref{eqn:Aload}. Within the ASD, it is possible to close 2
other contacts, \emph{e.g.,} those that belong to SA2. However, they will not
transmit any force. Likewise, for specific values of the $\delta_i$'s, SB1
was obtained as the list of force-carrying contacts in system B submitted to
the load that is represented on figure~\ref{fig:sysB}. It is possible to
close some other contacts (such as those that belong to SB2), but they cannot
carry (within the ASD) any force. Uniquely determined force-carrying
structures, depending on the load, will possess a varying degree of
displacement indeterminacy $k$. Once system B, in addition to the forces on
the perimeter, was submitted to small (randomly oriented) external forces
exerted on each grain, then isostatic structure SB2 was obtained.

In ref.~\cite{OR97b}, the triangular lattice model, as in section IV.B, was
studied for isotropic loads. As an application of the \emph{global}
minimization property, it was shown, within the ASD (to first order in
$\alpha$) that \emph{the} maximum packing fraction of polydisperse discs is,
in the limit of large systems, equal to
\be
 \Phi _{max}={\pi \over 2 \sqrt{3}}
(1 - k \alpha ), \label{eqn:phimax}
\ee
with $k=0.314\pm 0.003$ in the case of a uniform distribution of radii.
\subsubsection{Minimal structures. Analogies with other problems.}
As equilibrium contact forces are the coordinates of an extreme point of the
simplex of problem $\ppp _2$, a maximum set of inequality contraints $f_l \ge
0$ are simultaneously satisfied as equalities, $f_l=0$. This means that
force-carrying structure $S^*$ is minimal with respect to the equilibrium
requirement~\ref{eqn:equil}. In section VI.C, we invoked an iterative dilution
process to define irreducible sets of self-balanced forces. Likewise, one can
define minimal structures, such as $S^*$, as irreducible by further dilution,
since it is impossible to require more bond forces to vanish if the load is to
be balanced. Any such irreducible structure $S$ might carry a unique set of
bond forces balancing the load, it geometrically determines one solution to
equations~\ref{eqn:equil}.

Recalling we have defined a loading parameter $Q$, to which all external
forces are proportional, there exists for each minimal force-carrying
structure $S$ a set of coefficients $(\beta_l^L)_{1\le l\le N}$, such that the
forces carried by $S$ that balance the load are \be f_l=\beta_l^S Q .
\label{eqn:coeffpart} \ee
By definition, one has
\[
\left\{\begin{array}{ll}\beta_l^S\ne 0&\mbox{if $l\in S$} \\
\beta_l^S= 0&\mbox{if $l\notin S$}\\
\end{array}
\right.
\]
Among all minimal structures $S$ with non-negative coefficients
$\beta_l^S$, $S^*$ minimizes
$$
\sum_{l\in S} \beta_l^S h^0_l.
$$

Let us now recall the analogy with a problem of current transport on a
resistor network, as introduced in part V, with the following constitutive
law. To the requirement that contact forces are repulsive corresponds an
\emph{orientation} of the bonds, which behave as diodes rather than resistors.
Bond $a\rightarrow b$ between nodes $a$ and $b$ carries some current
$i_{ab}\ge 0$  that is related to the potential difference $v_a-v_b$ by the
analog of the Signorini condition:
\be
\left\{\begin{array}{ll}i_{ab}>0&\mbox{if $v_a-v_b=v^0_{ab}$} \\ i_{ab}=
0&\mbox{if $v_a-v_b<v^0_{ab}$}\\ \end{array} \right.
\ee
The bond becomes a supraconductor (the analog of a rigid contact) when the
threshold potential difference $v^0_{ab}$ is reached, and it is an insulator
if $v_a-v_b$ is smaller. 

It is customary to define a scalar analog of the mechanical load by injecting
some external current $I$ in one node, that we denote as $i$, and extracting
it from another one, that we denote as $o$. $I$ is then the analog of the
mechanical parameter $Q$. A minimal structure (\emph{i.e.,} one that cannot be
further diluted), to carry the current, is a \emph{path} from $i$ to $o$. If
its coefficients $\beta$ cannot be negative, it is a \emph{directed path}, on
which the current flow respects the a-priori orientation of the bonds. On such
a path $S$, all bonds $l\in S$ carry the total current $I$, hence $\beta_l^S
=1 \ \mbox{for all } l\in S$. In the analogous scalar problem,  the current is
carried by the directed path $S^*$ that minimizes, among all directed paths
$S$ from $i$ to $o$, the criterion $$\sum_{l\in S} \beta_l^S v^0_l=\sum_{l\in
S} v^0_l.$$ In the scalar problem, the criterion reduces to a sum of `costs'
associated with the bonds of the network.

The analogous problem to $\ppp _2$ in the scalar case is thus the well-known
\emph{minimum directed path} (or \emph{directed polymer}) problem on a
network~\cite{HHZ95}. This analogy was introduced in~\cite{GRHBTC90}, for
problem $\ppp _1$, upon transforming the minimum path problem into the dual
problem, which consists in maximizing the potential drop $v_i-v_o$, knowing
that in each bond $l$ $v_l$ cannot exceed the threshold value $v^0_l$. The
dual point of view adopted here--the analogy for problem $\ppp _2$--  stresses
the geometric origin of equilibrium forces, as coefficients characterizing the
maximum localisation of efforts onto structure $S^*$. Contact forces in
granular packings have often been studied in the recent
literature~\cite{CLMNW96,EC97,SO98}. It is interesting to be able to define
them as the solution to a well-defined optimization problem  of random
geometry~\cite{JNR97b}. 

Some statistical properties of structures $S^*$ were studied in
refs.~\cite{OR97b,OR97a}, in the case of the 2D triangular lattice model, as
defined in section IV.B, with a uniform distribution of $\delta_i$'s. It was
shown, in particular, for isotropic loads in the limit of large systems, that
the density of force-carrying bonds tends to a non-vanishing limit, and the
distribution of contact force values was evaluated.

The statistical properties of the solution to the `directed polymer' problem
are related to those of  directed percolation~\cite{HHZ95}. Likewise, one can
expect, in the case, in particular, of a very wide distribution of values of
$h_0$ in the mechanical problem, minimization problem $\ppp _2$ to be related
to some unilateral percolation problem. Such a percolation model was never
studied to our knowledge. It is a \emph{geometric} problem, unlike generic
central force percolation~\cite{MD95}, for which (in 2D at least) only the
topology of a diluted structure matters.  
\subsubsection{Some macroscopic results for the triangular lattice model.}

To see what macroscopic mechanical behaviour might result from the properties
stated in this section, we briefly recall here some results obtained by
numerical simulation of the triangular lattice model~\cite{JNR97b}, as
presented in section IV.B, with a uniform distribution of parameters
$\delta_i$ (eqn.~\ref{eqn:randdiam}).

Samples of up to 12600 discs were submitted to varying states of stress. The
following inequalities, in which coordinate label 1 corresponds to one of the
three directions of dense rows in the triangular lattice, and compressive
stresses are conventionnaly positive, define the domain of supported loads, as
macroscopically expressed in terms of stresses.
\be
\left\{\begin{array}{lll}
&\phantom{\le} \sigma_{22}&\le 3\sigma_{11}\\ 
{\displaystyle -{\sigma_{22}\strut \over \sqrt{3} \strut}}&\le \sigma_{12}
&{\displaystyle \le {\sigma_{22}\strut \over \sqrt{3} \strut}}\\
\end{array}
\right.
\label{eqn:suppsigma}
\ee

All intensive quantities, like, \emph{e.g.,} distributions of force values,
density of the contact structure, distribution of contact orientations,
etc\dots were found to possess well-defined thermodynamic limits,  
independently of the details of the boundary conditions, provided a uniform
state of stress is imposed, and the stress tensor $\t2{\sigma}$ satisfies
conditions~\ref{eqn:suppsigma} \emph{as strict inequalities}. Correlation
lengths or, in other words, sizes of representative volume elements, or of
independent subsystems, are finite, but appear to diverge as marginally
supported loads (for which one of conditions~\ref{eqn:suppsigma} holds as an
equality) are approached.

Taking, as in section IV.3, the undisturbed lattice, in which the spacing
between sites is equal to $a$, the maximum disc diameter, as the reference
state, a strain tensor $\t2{\epsilon}$ can be identified. It is related to
displacement field ${\bf u}$ by
\be
\epsilon_{\alpha \beta} = -{1\over 2}
\left( {\partial u_\alpha \over\partial x_\beta} +{\partial u_\beta
\over\partial x_\alpha}\right), \label{eqn:epsu}
\ee
and the potential energy per unit surface area is
(summation over repeated indices implied)
\be
W = -\sigma_{\alpha \beta} \epsilon_{\beta \alpha}=-\t2{\sigma}:\t2{\epsilon}.
\label{eqn:wmacro}
\ee

Coordinates of tensor $\t2{\epsilon}$ are found to be expressible as linear
combination of the average of bond elongations $\delta u_l$ for the three
bond orientations of the triangular lattice. In $\t2{\epsilon}$ space
(3-dimensional for a 2D system), impenetrability conditions define, in the
thermodynamic limit, a strictly convex accessible domain $\ddd$, limited by a
smooth surface $\Sigma$, the equation of which we denote as
\be
f(\t2{\epsilon})=0, \label{eqn:defSigma}
\ee
while the interior of accessible region $\ddd$ corresponds to
the strict inequality:
$$
f(\t2{\epsilon})<0.
$$
As a macroscopic consequence of the variational properties stated in part VII,
the relationship between tensors $\t2{\sigma}$ and $\t2{\epsilon}$ is the
following:
\begin{eqnarray}
\sigma_{ij}=\lambda {\partial f \over \partial
\epsilon _{ij}},& \mbox{with $\lambda \ge 0$ , if }&f(\t2{\epsilon})=0
\label{eqn:loimacro}\\
\sigma_{ij}=0& \mbox{if } &f(\t2{\epsilon})<0
\nonumber
\end{eqnarray}
Wherever the granular system transmits stress, the value of $\t2{\epsilon}$
is as far as possible in the direction of $\t2{\sigma}$ within $\ddd$,
\emph{i.e.}, where the tangent plane to its boundary $\Sigma$ is orthogonal to
$\t2{\sigma}$, thus minimizing potential energy~\ref{eqn:wmacro}.

$\ddd$ is unbounded in the direction of non-supported loads.
Strains go to infinity on surface $\Sigma$ when the stress tensor approaches
one of the marginally supported directions. $\Sigma$ has three asymptotic
planes, respectively orthogonal to those three marginally supported load
directions.

The one-to-one correspondence between supported stress \emph{directions} on
the one hand, and strain tensors such that $f(\t2{\epsilon})=0$ on the other
hand, is a macroscopic translation of the uniqueness property stated in
paragraph VIIB5. The potential energy density has a finite thermodynamic limit
(a result that generalizes to non-isotropic states of stress the one of
equation~\ref{eqn:phimax}), and possible variations of $\t2{\epsilon}$ within
convex set $A$, discussed in VII.B.3, shrink to a vanishing range
($\t2{\epsilon}$  becomes uniquely determined) as the system size grows.

Constitutive law~\ref{eqn:loimacro} can be used to solve for stress and
displacement fields whenever a sample of the model material is submitted to
some external forces that do not lead to unbounded displacements and overall
failure. The field  of $\lambda$ values should be obtained on solving the full
boundary value problem. 
\subsection{Systems with bounded tensile forces.} 
\begin{figure}[ht]
\centering
\includegraphics[width=8cm]{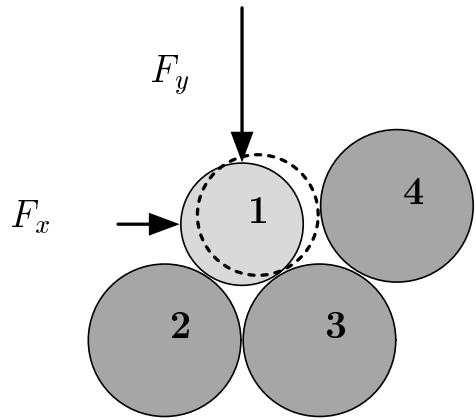}
\caption{A small sample of the triangular lattice model, in which the only
mobile disc, marked 1, is slightly smaller than discs 2, 3 and 4, and is
submitted to an external force. Disc 1 is shown in one possible equilibrium
position, in contact with 2 and 3. The other one is sketched with a dotted
line.
\label{fig:excoh}}
\end{figure}

If the unilateral contact law~\ref{eqn:loicont} is replaced by~\ref{eqn:coh}, the
remarkable properties stated above in VII.B are lost. Let us illustrate this
on a simple example. Consider the system depicted on figure~\ref{fig:excoh},
to be dealt with, within the ASD, as a triangular lattice model in the sense
of IV.B, the contact law being~\ref{eqn:coh}. Only one disc is mobile (number
1), and we first consider the case of a vertical force of intensity $F_y$
oriented downwards like on the figure, keeping $F_x=0$. (Later in part IX we
come back to this simple example and discuss its behaviour when $F_x$ is
altered).
Two equilibrium positions are possible: disc 1 might either be in contact with
discs 2 and 3, or with 3 and 4. As grains are rigid and only exert normal
forces on one another when they exactly touch, the problem is isostatic in
both equilibrium configurations, in agreement with the general property of
section VI. The load, defined with $F_y>0$, is always supportable on structure
$S_1$, consisting in bonds $1-2,\ 1-3$, and it is also supportable on
structure $S_2$, consisting in bonds $1-3,\ 1-4$ as long as $F_y<f_0\sqrt{3}$.

Thus, for $0<F_y<f_0\sqrt{3}$, even within the ASD, {\em the equilibrium state
and the list of force-carrying contacts are not uniquely determined}. Whether
$S_1$ or $S_2$ will be chosen depends on the trajectory of disc $1$ from its
initial (reference) position.

Likewise, {\em supportable loads are not necessarily supported}. To check
this, let us remove disc $2$. In its motion, disc $1$ might come into contact
with both $3$ and $4$, and, provided $0<F_y<f_0\sqrt{3}$, reach an equilibrium
position, maintaining those two contacts. However, it might as well never meet
disc $4$, and find a trajectory, past disc $3$, on which its potential energy
will keep decreasing forever.

\subsection{Smooth, convex interaction potentials.}
In the case of the elastic contact law~\ref{eqn:hertz}, within the ASD, all
properties of convex problems enumerated in section VII.A are valid. Let us
state the `elastic' versions of the `rigid' optimization problems of VII.B.
$\ppp _1\ $ is simply replaced by
$$
\ppp _1 ^{el}:\ \mbox{Minimize $W^{tot}$ defined in~\ref{eqn:wtot}},
$$
while contact forces are the solution to
\be
\ppp _2^{el}\left\{\begin{array}{l}
\mbox{Maximize }-{\displaystyle \sum _l
\left(h^0_lf_l- {m\over m+1} K^{-1/m}f_l^{(m+1)/m}\right)}\\
\mbox{with constraints: (~\ref{eqn:equil}) and $\bf{f}\ge 0$}\\
\end{array}
\right.
\label{eqn:probp2el}
\ee

The function of contact force $f$ that appears within the sum is the opposite
of the Legendre transform of the elastic energy $w$, regarded as a function
of relative displacement $\delta u$, \emph{i.e.}, $f\delta u - w(\delta u)$,
taken with $f={\displaystyle dw\over d(\delta u)}$. Thus, solving $\ppp
_2^{el}$ amounts to `minimizing the complementary energy', a common procedure
to find the forces in an elastic problem.

In fact, one could have defined a potential energy, in the rigid case, equal
to $+\infty$ if grains interpenetrate, and treat rigid problems exactly like
elastic ones, constraint~\ref{eqn:impen} being taken care of by the definition
of the potential. If the region, in phase space, that is forbidden by the
constraints is convex, then such a potential can still be regarded as a convex
function. Both the condition~\ref{eqn:loicont} and elastic law~\ref{eqn:hertz}
are then expressed by \[ f\in \partial w(\delta u), \]
in which $ \partial w(\delta u)$ denotes the \emph{subdifferential}
of $w$ at $\delta u$, \emph{i.e.}, the set of all $f$
such that
$w(\delta u') \ge w(\delta u) + f(\delta u'-\delta u)$ for any $\delta u'$.
This mathematical possibility to unify rigid and elastic laws is specific to
convex problems. This is the precise meaning of property 5 cited in section
VII.A. Here, we preferred to resort to a separate presentation of the rigid
case in section VII.B, to stress the physical consequences of the variational
properties. The reader may refer to~\cite{JJM74} for a more systematic
approach.

Comparing $\ppp _2$ and $\ppp _2 ^{el}$, as defined by~\ref{eqn:probp2} and
~\ref{eqn:probp2el}, one may expect the following behaviour for the
distribution of contact forces, as a set of grains with elastic contacts is
submitted to a constant load, but the stiffness constant $K$ is gradually
reduced. (Similarly, one could also increase $Q$, keeping $K$ constant). When
$K$ is very large, the elastic term is negligible in comparison with
$Z(\bf{f})$, and the values of the forces should coincide with the (unique)
rigid contact solution of $\ppp _2$. Thus the contact structure should barely
suffice to carry the load (isostatic problem), the forces should exhibit the
characteristic disorder of granular systems, with large fluctuations, force
chains, etc\dots On the other hand, let us assume that the list of possible
contacts (structure $S_0$) is well-coordinated, that there are many more
contacts that are easy to close upon increasing the confining forces or
decreasing the contact stiffness parameters. Then, in the limit of small $K$,
$Z(\bf{f})$ will, in turn, become small in comparison with the elastic energy.
The elastic term tends to share equally the forces between contacts. Thus, a
narrow distribution of force values is expected in this limit, and spatial
heterogeneities should be strongly reduced. Knowing that the minimum structure
$S^*$ and the complete list of possible contacts $S_0$ are of comparable
densities, the order of magnitude of the average force $f_0$ does not change
as grains are made softer. The two extreme regimes of stiff and soft contacts
should thus be respectively defined by the conditions $K\gg {\displaystyle f_0
\strut \over h_0^m \strut}$ and $K\ll {\displaystyle f_0 \strut \over h_0^m
\strut}$,  involving a typical interstitial distance $h_0$. 

Those two limits, and the transition regime, in which the contact density
increases, were observed~\cite{JNR97a} on the 2D triangular lattice model, as
defined in section IV.B, with contact law~\ref{eqn:hertz}.

\subsection{Remarks. The `elasticity' of rigid grains.}
As announced beforehand, we have exhibited, in this section, model granular
systems for which, at the expense of several assumptions, including the
validity of the ASD, mechanical properties are entirely determined by geometry.

We have seen that the distinction between systems made of rigid or deformable
grains is not  necessarily as important as one might have expected: similar
potential energy minimization properties might be stated, the limit of large
contact stiffnesses might safely be taken without any singularity (subsection
D), and macroscopic stress-strain relationships might be written for some
systems of rigid grains, as recalled in paragraph B.8. The difference between
the systems such that the search for an equilibrium state is a convex
minimization problem (in which case the properties listed in subsection A are
satisfied) and the others, such as the example of subsection C, is finally
more relevant.

Constitutive law~\ref{eqn:loimacro} expresses a one-to-one correspondence
between the direction of stress tensor $\t2{\sigma}$ and strain tensor
$\t2{\epsilon}$, which is restricted to belong to surface $\Sigma$. It is
quite similar to a macroscopic elastic law, even though it applies to systems
of rigid discs. The response to a supported stress increment will be
reversible. If this increment, $\t2{\delta \sigma}$ is in the direction of the
preexisting stress tensor $\t2{\sigma}$, then no additional displacement or
stress will result for rigid grains. For deformable grains, if contact
law~\ref{eqn:loicont} is replaced by~\ref{eqn:hertz}, a small deformation,
inversely proportional to constant $K$, will follow. If, on the other hand,
$\t2{\delta \sigma}$ is orthogonal to the initial stress tensor, its
application will entail a small strain increment $\t2{\delta \epsilon}$, such
that the new strain tensor will be exactly the point of $\Sigma$ where the
orthogonal direction is that of the new stress tensor. In this second case,
the apparent elastic modulus is thus inversely proportional to the curvature
of surface $\Sigma$.

In spite of the analogy, presented in paragraph B.7, between the backbone of
the force-carrying structure and cost-minimizing directed paths for scalar
transport, the statistical properties of those two systems are quite
different. In agreement with various results on disordered systems of
grains~\cite{RJMR96,BG91} the triangular lattice system was
found~\cite{OR97a,OR97b,JNR97b} to possess a standard thermodynamic limit:
intensive quantities like the density of the backbone, the strains, the
distribution of contact force values have limits in the limit of large system
size (except for marginally supported loads). On the other hand, unlike the
force-carrying structure in the mechanical problem we have been studying, the
optimal directed path in the corresponding scalar problem is a critical
object. 

The validity of the ASD --that might at first sight appear as a mere
technical aspect-- is finally a crucial ingredient of the model granular
systems that we are studying here. The next section examines some stability
properties that are important as soon as one does not resort to the
approximation.

\section{Outside the ASD: questions of stability.}
We now enforce, on physically acceptable equilibrium states, another
requirement: that they should be \emph{stable}. We limit ourselves to the
cases when stability can be discussed in terms of a potential energy. If the
equilibrium state is a local minimum of the potential energy, then there
exists a region of finite extent in displacement space, around equilibrium
positions, within which the system is spontaneously attracted to the
equilibrium configuration.

Within the ASD, one can only discuss potential variations that are of first
order in displacements. When floppy modes exist ($k>k_0$), they appear as
marginally unstable and one cannot tell whether, to higher orders, they
actually destabilize the equilibrium configuration.  The mechanical response
to small perturbations or load increments is strongly dependent on these
stability questions.

In general, we will show, with examples (section A), that the answer might
depend on quite specific geometrical features of the granular system, and on
the contact law. We are only able to give general answers for spheres or
discs, as shown in section B. Section C discusses some consequences on the
geometry and coordination of granular packings at equilibrium, and on the
macroscopic mechanical behaviour.
\subsection{Simple examples.}
We consider rigid frictionless particles of various shapes, and discuss the
stability of simple configurations, that depends on the ability of contacts
to withstand tension, and on the shape of the grains. 
\subsubsection{Bond alignments.} 
\begin{figure}[ht]
\centering
\includegraphics[width=7.5cm]{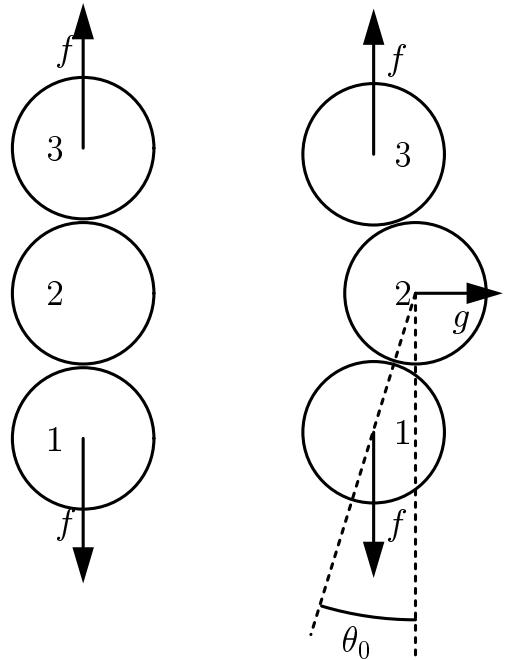}
\caption{An alignment of 3 spheres (left), the middle one touching the other
2. Spheres 1 and 3 are submitted to equal and opposite forces along the line
of centers. The new equilibrium configuration, upon exerting a lateral force
$g$ on the midle sphere, is shown on the right.
\label{fig:buckle}}
\end{figure}

Assume three spheres, or three discs in 2D,
to have their centers aligned as on fig.~\ref{fig:buckle}, the two extreme
ones being submitted to opposite forces in the direction of the line of
centers. Let us discuss the problem in 2D.
The determination of contact forces is an isostatic problem, and there is,
apart from rigid body motions, a trivial mechanism corresponding to free
lateral motion of the middle disc 2. This is of course well known to lead to
the familiar buckling instability if one pushes the extreme discs towards each
other, and to be stable if one pulls on them, provided the contacts can resist
tensile forces. In the latter case, assuming one controls the forces parallel
to line 1-2 exerted on particles 1 and 3, while their position in the other
direction is fixed, the system will respond \emph{elastically} to a small
additional force exerted on disc 2, even though the contact law is rigid.
After the system reaches its new equilibrium state, the orientation of
contacts is such that the new load is orthogonal to the floppy mode.
Specifically, if $g$ is the lateral force pulling disc 2 away from the line
1-3, and if $f$ denotes the external force exerted on 1 and 3, the new
position of the center of disc 2 is such that, assuming equality of the 3
radii, the angle $\theta _0$ between 1-3 and 1-2 (fig.~\ref{fig:buckle}) is
given by 
$$
\theta _0=\tan^{-1}({g \over 2f}),
$$
while contact forces (tensile, and therefore negative) are
$$
f_{12}=f_{23}=-f\cos(\theta_0).
$$
The potential energy, as a function of $\theta$ ($\theta$ parametrizes the
free motion that maintains the two contacts), reads
$$
W=-2af\cos(\theta)-ag\sin(\theta)=-a\sqrt{4f^2+g^2}\cos(\theta-\theta _0),
$$
and has its minimum for $\theta=\theta _0$.

This elastic behaviour is similar to that of a rigid string under tension,
which will deform in response to lateral sollicitations.

On carrying out the same calculations in the case of compressive forces, with
 $f<0$, one will notice that $g$ and $\theta_0$, corresponding to the
equilibrium position of disc 2, are now of opposite signs. One then has
$$
W=a\sqrt{4f^2+g^2}\cos(\theta-\theta _0),
$$
which is \emph{maximized} in the unstable equilibrium position
$\theta=\theta_0$. 

In section VIIIB, we show that the conclusions reached on this simple example
are general: any floppy mode in a system of discs or spheres that admits only
compressive contact forces leads to an instability. If, on the contrary, all
contact forces are in fact tensile, the system being thus analogous to a
network of tight strings, any floppy mode is stable, and an elastic response
to small load increments can be observed.

\begin{figure}[t]
\centering
\includegraphics[width=7.5cm]{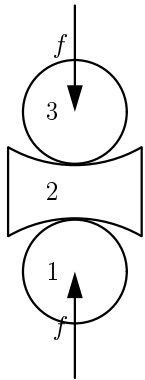}
\caption{An alignment like that of fig.~\ref{fig:buckle}, the middle sphere
being replaced by an object turning concave parts of its surface towards
spheres 1 and 3. \label{fig:concave}} 
\end{figure}

Let us now replace disc 2 by a particle presenting concave surfaces toward
discs 1 and 3, as shown on fig.~\ref{fig:concave}. The system is similar to
that of fig.~\ref{fig:buckle}, the free lateral motion of the middle particle,
maintaining the contacts, is a mechanism. 
It is not difficult to show, however, that the configuration of
fig.~\ref{fig:concave} has, compared to the alignment of discs,
\emph{opposite} stability properties: the mechanism is stable for compressive
forces, unstable for tensile ones. Thus stability properties are quite
sensitive to particle shape.

\subsubsection{Arches.}
Systems submitted to gravity provide other familiar examples of non-rigid
equilibrium states. A string of circular, or spherical, particles, each of
them tied to two neighbours by a frictionless contact condition that supports
tension, behaves as a chain, and will eventually adopt a stable equilibrium
configuration if one fixes  its two extremities and let it dangle under its
weight. The number of mechanisms in this system is equal to the number of free
particles. 

\begin{figure}[ht] 
\centering
\includegraphics[width=8cm]{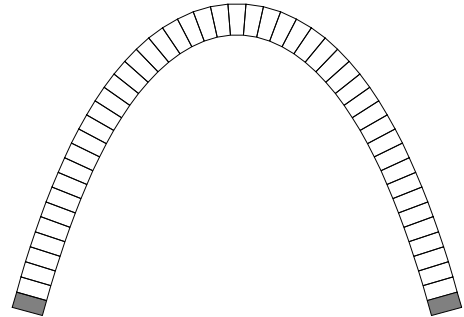}
\caption{An arch built with stones sharing flat lateral surfaces. 
\label{fig:arch}}
\end{figure}

The analogous system to the chain, in which contacts transmit compressive
forces, is the arch, fig.~\ref{fig:arch}. The general result for spheres
entails that all arches made of spheres are unstable. However, one usually 
builds  arches with appropriately shaped stones,  \emph{e.g.,} carving them to
share common flat lateral surfaces with their neighbours, as on
fig.~\ref{fig:arch}. 
Such an arch is a system that possesses one floppy mode per stone (still
assuming no friction), but its geometry might be adequately chosen to support
the load. In such a case, any free motion of the stones, that slide on their
flat common surfaces, all contacts being maintained, does not change the
potential energy. One thus has an example of \emph{marginal stability}. Such
an arch is only able to carry the one particular load for which it was
specifically designed. (Any amount of friction, however, stabilizes the
system).

\subsubsection{A stable mechanism with strictly convex cohesionless grains.}
\begin{figure}[ht]
\centering
\includegraphics[width=8cm]{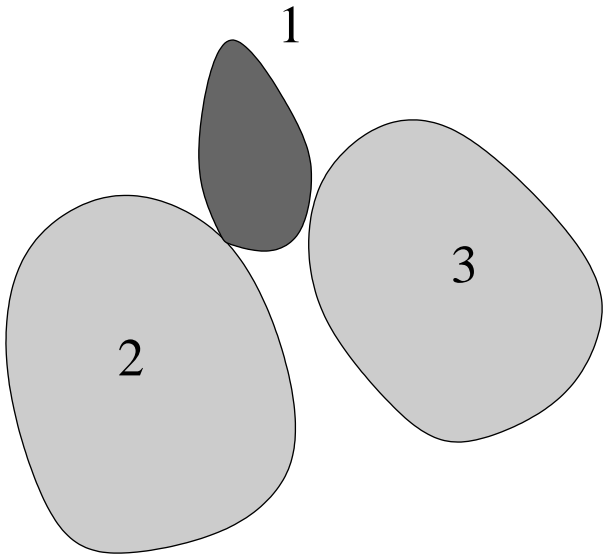}
\caption{The upper grain 1 relies on two of its neighbours and is submitted to
its weight, oriented downwards. Its rotation is a stable floppy mode.
\label{fig:strict}}
\end{figure}

In view of the previous examples, one might be tempted to infer that
mechanisms, when contacts only support compression, can be stable with concave
grains (fig.~\ref{fig:concave}), are sometimes marginally stable with flat
surfaces (fig.~\ref{fig:arch}), but are always unstable  with stricly convex
grains (fig.~\ref{fig:buckle}). This is however not true, as shown by
the simple example of fig.~\ref{fig:strict}.
We are not aware of other general answers to this question of stability than
the ones that are given for spheres below.
 \subsection{General results for spheres and discs.}

\subsubsection{Tensile contact forces (systems of cables).}
In the case when all contacts, at equilibrium, carry a tensile force, then
stability is immediately proved once it is realized, as remarked in section
VIIA, that minimizing the potential energy is a convex optimization problem
(see property 1 stated in section VIIA). 

Just like for the simple example of figure~\ref{fig:buckle}, floppy modes can
exist in stable equilibrium configurations. Then, the system will respond
elastically to small load increments that provoke small motions of those
floppy modes. Applying such load increments amounts to slightly deform the
potential energy landscape on the manifold of configurations that maintain the
initially existing contacts.
A new minimum is found, close to the previous one.

Systems of rigid cables, \emph{whatever the level of deformation}, should
therefore possess exactly the same kind of elasticity, due to preexisting
stresses, as assemblies of rigid frictionless particles without cohesion
\emph{within the ASD} (whose mechanical response to load increments was
discussed in section VII.E).

Those properties were in fact discussed by Alexander~\cite{SA98}, in his
monograph on the elasticity of various kinds of networks and amorphous
systems, in the case when the contact law is \emph{elastic}. Alexander pointed
out that stable configurations are not necessarily rigid. He stressed that
force-carrying bonds or contacts always have a stabilizing effect when they
transmit a traction, and a destabilizing one when they transmit a compression.

Our present study, in this subsection, might be regarded as complementary to
his, since we deal with  \emph{rigid} contacts.

\subsubsection{Cohesionless grains.}
Let us now show that, in the absence of tensile force in the contacts,
an equilibrium configuration of
rigid, frictionless discs or spheres is necessarily unstable if the backbone
is not a rigid structure.

We shall do so by yet another application of the theorem of virtual power,
as follows.

We assume a packing of spheres to be in equilibrium under a prescribed load.
Spheres are rigid, and the problem is therefore isostatic, $h=0$. Flat walls
can also exist, \emph{e.g.,} as a device to enforce some kind of boundary
condition on the packing, but we assume that they cannot rotate. We assume
there is at least one mechanism: $k\ge 1$. Consequently, it is possible to
move the grains (and the walls) while maintaining the whole list of contacts.
(The possibility that a mechanism could exist for the considered equilibrium
configuration alone, and disappear as soon as the grains are displaced is to
be discarded as non-generic. This would, in particular, due
to~\ref{eqn:relkh}, entail $h\ge 1$). We now study the variation of the
potential energy in one such motion, with a `time' $t$ parametrizing the
trajectories, and show that it decreases.

\begin{figure}[ht]
\centering
\includegraphics[width=7cm]{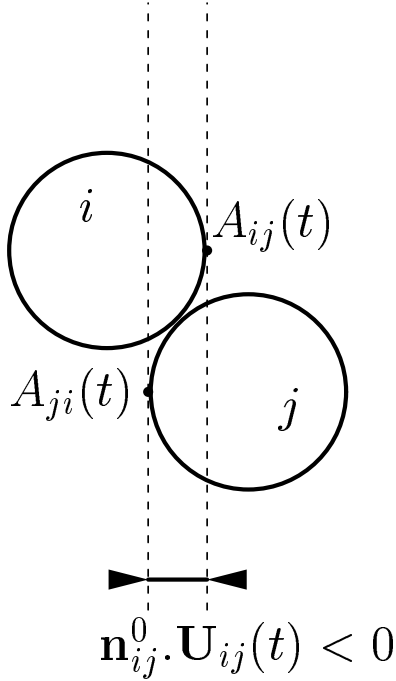}
\caption{Sketch of the position, at time $t$ of two spheres in contact.
\label{fig:aijaji}}
\end{figure}

Objects do not rotate in this motion (this is an assumption for walls, and
rotations of frictionless spheres are ignored anyway). Particle $i$ has a
time-dependent velocity ${\bf V}_i(t)$, and initially, in the equilibrium
configuration from which the motion starts at $t=0$, touches its neighbour $j$
in a point $A_{ij}^0$, where the normal unit vector to its surface, pointing
to the center of $j$, is ${\bf n}_{ij}^0$, the equilibrium contact force being
$f_{ij}$.  Let $A_{ij}(t)$ denote the material point of the surface of grain
$i$ that was at  $A_{ij}^0$ initially. Similarly, following the material
motion of $j$, one defines $A_{ji}(t)$, which does not coincide in general
with $A_{ij}(t)$. It is possible, at each time $t$, to apply the theorem of
virtual power, thus evaluating $W^{\prime}(t)$, the time derivative of
potential energy $W$ at time $t$, as follows. The definition of a structure,
in part II, was in fact completely arbitrary. Here, let us use this one: at
time $t$, although objects $i$ and $j$ that are in contact effectively touch
each other by a different point, define a bond to exist between $A_{ij}(t)$
and $A_{ji}(t)$, oriented by ${\bf n}_{ij}^0$, which, because objects do not
rotate, is still carried by the common normal direction to the surfaces of $i$
and $j$ at these two points. This structure might be used to define virtual,
fictitious bond forces, that we choose equal to the initial equilibrium
contact forces, \emph{i.e.,}  $f_{ij}$, carried by ${\bf n}_{ij}^0$ in the
bond between $A_{ij}(t)$ and $A_{ji}(t)$. These forces are now used in the
theorem of virtual work, with the real velocities. This is perfectly valid,
because for each $t$ \begin{itemize} \item the virtual internal forces balance
the constant load \item in the bond between $i$ and $j$, the force exerted on
$i$ is still equal to the opposite of the force exerted on $j$. \end{itemize}
One obtains:
\[
W^{\prime}(t)=
\sum_{i<j}f_{ij}{\bf n}_{ij}^0.\left({\bf V}_j(t)-{\bf V}_i(t)\right),
\]
the sum running over all bonds. As $f_{ij}{\bf n}_{ij}^0$ does not depend on
$t$, this is easily integrated. Denoting as ${\bf U}_{ij}(t)$ the vector of
origin $A_{ij}(t)$ and extremity $A_{ji}(t)$, the net variation of potential
energy at time $t$, from the beginning of the motion is
\be
W(t)-W(0)=\sum_{i<j}f_{ij}{\bf n}_{ij}^0.{\bf U}_{ij}(t).
\label{eqn:wtmw0}
\ee
In the motion, $A_{ij}(t)$ and $A_{ji}(t)$ are still extreme points of solids
$i$ and $j$ in the respective directions ${\bf n}_{ij}^0$ and
$-{\bf n}_{ij}^0$.  As spheres $i$ and $j$ have stayed in contact, it follows
that, as shown on figure~\ref{fig:aijaji}, the contribution of bond $i-j$
to~\ref{eqn:wtmw0} is strictly negative, unless $A_{ij}(t)=A_{ji}(t)$, in
which case it is zero.
The same conclusion holds true for a contact between a sphere and a flat wall
that does not rotate. Consequently, one must have:
\[
W(t)-W(0) <0,
\]
unless all intergranular contacts that carry non-vanishing equilibrium forces
are maintained, in the motion, via the same material points. This latter
condition means that the backbone of the contact structure in the equilibrium
configuration moves as a rigid body. 

Mechanisms that only affects grains that do not carry any force, without
altering the geometry of the backbone will not, of course, change the value
of $W$ and lead to instabilities.

Otherwise, the instability is always present. We have shown that {\em the
backbone of the contact structure}, in a stable equilibrium configuration of
a packing of rigid, frictionless spheres that do not support tensile forces in
the contacts, {\em is devoid of mechanisms other than rigid-body motions:
$k=k_0$}.  As we already knew, from part VI, that it cannot possess
self-balanced contact forces ($h=0$), one reaches the conclusion that {\em it
is an isostatic structure}.

\subsection{Consequences. Discussion.}
\subsubsection{Coordination of packings.}
The isostaticity of the force-carrying structure in packings of rigid
frictionless \emph{spheres} with contact law~\ref{eqn:loicont} thus results
from a stability analysis. The opposite inequality to the ones established
in section VI.D, can, in this case, be stated: one has $N\ge N_f$, and
consequently, $N=N_f$, on the backbone of the contact structure. For large
systems, the absence of floppy mode  implies a \emph{lower bound} on the
coordination number:  $$c\ge 2d\ \ \mbox{ on the backbone}.$$ This is equal to
upper bound~\ref{eqn:ubc}, hence the equality: $c=2d$.

However, for frictionless grains with different shapes, or for spheres with
cohesion, one cannot expect in general inequality~\ref{eqn:ubc} to hold as an
equality, even on the sole backbone.

Returning to cohesionless packings of spheres, when each one is submitted to
an external force, it has to belong to the force-carrying backbone, and the
whole system satisfies $N=N_f$ (or, asymptotically for large sizes, $c=2d$).
This happens in system A, treated without resorting to the ASD. The
force-carrying structure that was obtained, SA2, is isostatic and spans the
whole system. When external forces are transmitted from the boundary, as in
system C, floppy modes can exist, typically as isolated spheres, like discs
$10$ and $14$ on fig.~\ref{fig:sysC}, or small sets of spheres, that are
not or insufficiently connected to the backbone. If not too widely
polydisperse systems of spheres, regions that are totally shielded from force
transmission are usually quite small. According to our experience in numerical
simulations, if the radio of the largest to the smallest radius is $2$ in a
polydisperse assembly of discs, then one very rarely sees more than 3 discs
together in such regions. In 2D, ring-like arrangements surrounding discs that
carry no force, such as $29-30-31-15-6-5-13-28$ and $11-3-9-22-23-24$ on
fig.\ref{fig:sysC}, cannot easily be made very large: the curvature of the
`ring' would then decrease, increasing the risk of inward buckling.

\subsubsection{Lattice models with and without the ASD.}
The triangular lattice model, as defined in section IV.B, of which systems A
and B are particular samples, provides vivid examples of the difference
between tensile contacts (systems of strings, satisfying~\ref{eqn:loitens})
and compressive ones (rigid grains obeying~\ref{eqn:loicont}), once dealt with
outside the ASD. \emph{Within} the ASD, both types of systems share the same
properties, and an equilibrium state of one of them can be mapped onto an
equilibrium state of the other, as follows. In the reference state, rigid
discs do not touch, since $h^0_{ij}=(a/2)(\delta_i+\delta_j)\alpha>0$. This
can be mapped onto a string network system, in which the `contact law'
is~\ref{eqn:loitens}, on replacing each $\delta_i$ by $-\delta_i$ and
attributing the length $a(1+\alpha (\delta_i+\delta_j)/2)$ to the string
joining $i$ and $j$. On reversing the sign of external forces, an exact
correspondence is achieved between equilibrium states.

\begin{figure}[ht] 
\centering
\includegraphics[width=8cm]{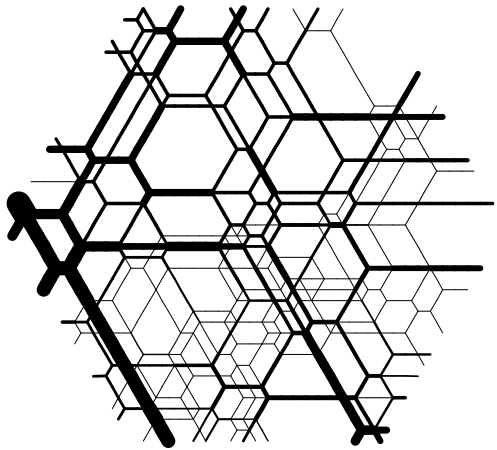}
\caption{Triangular lattice model, within the ASD: force-carrying structure
$S^*$ in  a hexagonal sample submitted to an isotropically compacting load.
Line widths are proportional to force intensities. The very same structure is
observed in a corresponding system of strings undergoing  isotropic tension. 
\label{fig:hex20asd}}
\end{figure}

Fig.~\ref{fig:hex20asd} shows the force-carrying structure, as obtained within
the ASD, in a hexagonal sample (for one random choice of $\delta _i$ values,
drawn according to a uniform distribution) of 1141 discs. This system is
submitted to  an isotropic pressure, via a imposed homogeneous shrinking of
the perimeter. 

As established in section VIIB, such a structure is, within the ASD, only
dependent on the random parameters $\delta_i$. The dynamics ruling the motion
of the particles from the reference to the equilibrium positions, and the
actual value of $\alpha$ are both irrelevant. In the corresponding system of
strings submitted to isotropic tension, exactly the same force pattern is
obtained at equilibrium. We denote as $S^*$ the backbone of the contact
structure, as displayed on fig.~\ref{fig:hex20asd}.  Just like in structure
SB1, which carries the force in a similar sample of smaller size, many discs
do not belong to $S^*$, which only contains 619 of them, thus possessing 1239
degrees of freedom (counting the one of the `wall'). Many floppy modes are
present, 381 of them are associated with bond alignments (discs having two
contacts in opposite positions), and the remaining 5 are more collective (like
the one of fig.\ref{fig:meca}). Some statistical properties of $S$ structures
in the large system limit were studied in~\cite{OR97a}.

\begin{figure}[ht]
\centering
\includegraphics[width=8cm]{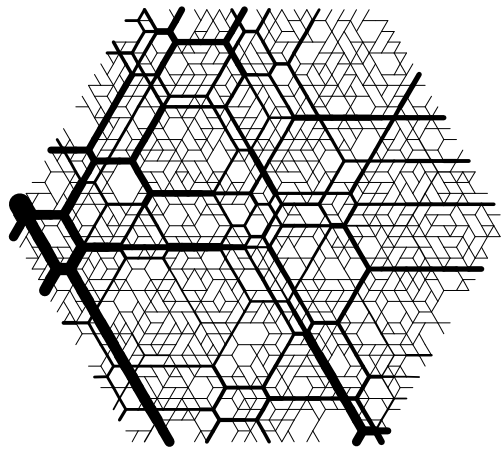}
\caption{Structure $SC$ that replaces $S^*$ ouside
the ASD in the case of contacts resisting compression.
\label{fig:hex20comp}}
\end{figure}

\begin{figure}[ht]
\centering
\includegraphics[width=8cm]{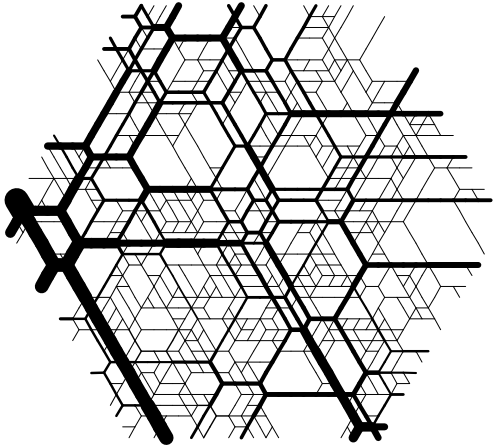}
\caption{Structure $ST$ that replaces $SC$ for an analogous system of
cables resisting tension.
\label{fig:hex20tens}}
\end{figure}

We numerically determined force-carrying structures in the rigid disc system
under compression, and in the corresponding system of strings under tension,
without the ASD. Those structures, that were obtained with $\alpha=1/48$ (this
value is now relevant), are respectively denoted as $SC$ and $ST$, and shown
on figures~\ref{fig:hex20comp} and~\ref{fig:hex20tens}.
Slight distortions of the regular triangular lattice, although not apparent on
the figures, were taken into account in the calculations. From part VII, we
know that $ST$ is still determined by the sole system geometry: since forces
are the solution to a convex optimization problem, the uniqueness property
still holds. This is not the case for $SC$, and the result now depends on the
actual dynamics (the rule that was adopted to move the discs to their final
equilibrium positions). The calculation was carried out with the `lubricated
granular dynamics' method of refs.~\cite{OR97a,OR97b}. 

As expected, $SC$ is devoid of mechanisms: it is an isostatic structure, with
1052 discs, 2105 degrees of freedom, and exactly 2105 contacts. Only 89 grains
out of the total number 1141 do not belong to $SC$. Most of them are isolated
grains, or pairs of neighbours (slightly larger regions shielded from the
forces appear near the perimeter, due to a boundary effect).

On the other hand, $ST$ stays more tenuous, with 840 discs only, and 1401
contacts. Thus 280 floppy modes still live on $ST$, 232 of which are simple
bond alignments and 48 are collective.

In spite of those differences between the density of $S^*$, $SC$ and $ST$, it
does appear on the figures that the spatial distribution of the forces is
very similar, the strongest `force chains' remaining unaltered.  The
distributions of force values in $S^*$ and $SC$, in the limit of large systems
were evaluated in ref.~\cite{OR97a}, and shown to coincide, within statistical
uncertainties, except for the small forces that appear on $SC$ in the
additional contacts created by the buckling instabilities in $S^*$.

Thus, resorting to the ASD is quite a legitimate procedure, provided $\alpha$
is small enough as to allow to regard the differences between $SC$ or $ST$ on
the one hand, and $S^*$ on the other, as refinements that can be neglected.

In the limit $\alpha \to 0$, any contact force on $SC$ is expected to tend to
its value in $S^*$, although the density of force-carrying contacts is
discontinuous. 

In the system of strings under tension, on the other hand,
mechanisms do not lead to instabilities, and the density of the backbone
itself should continuously approach that of $S^*$ as $\alpha \to 0$.
\subsubsection{Role of grain shape: are spheres special ?}
We have seen that it is necessary to examine, beyond the ASD, questions of
stability, to find qualitative differences between intergranular contacts
that resist compression and cables that resist tension, and between spheres
and other shapes.

Of course, one expects macroscopic properties of granular assemblies to
smoothly depend on grain shape: packings of nearly spherical grains will
resemble packings of spheres. Experimentally, it has sometimes been observed
that systems of spheres, in a quasi-static experiment, yield particularly
noisy responses. It is also empirically known in civil engineering that
granulates made of smooth and rotund particles, like river-bed gravel, are
especially unstable and prone to large plastic deformations. 

Unfortunately, detailed data at the microscopic level on non-spherical
grains close to equilibrium are scarce.

Although detailed analyses of such features are lacking, and our study of
granulate stability should be extended to the case of spheres with friction,
one might speculate that such particular behaviours of rotund objects could be
related to the specific property we have established here: whenever some
motion is smoothly initiated (\emph{i.e.,} with a very small initial
acceleration), while existing force-carrying contacts are maintained, then it
will entail some loss of potential energy, and thus accelerate further. Hence
probably the jerky aspect of system trajectories in configuration space.

Section IX discusses, precisely, when and how a system jumps from one
equilibrium state to another.
\section{Mechanical response to load increments: towards macroscopic
behaviour.}

So far, we have mainly dwelt on mechanical properties of model granular
systems. Those can be proved directly. We wish now to discuss possible
macroscopic consequences in terms of the constitutive laws that are relied
upon in a continuum mechanics description. We thus have to infer some of the
properties of granular packings in the limit of large systems. To be
quantitative, some statistical knowledge of the geometry of large granular
systems is needed, which requires experiments or numerical simulations. Here,
as we do not present new experimental or statistical studies,  we shall focus
on qualitative properties, extrapolating on the characteristics of finite
systems we have been presenting so far, and exploiting some recent numerical
results, especially those of ref.~\cite{JNR97b}, recalled in paragraph VIIB.8. 

Some macroscopic aspects of granular mechanics are recalled in part A.
Possible origins of plasticity are discussed in part B, in relation to
grain-level  characteristics. Part C examines some consequences of the strong
isostaticity property of systems of frictionless spheres without cohesion, in
which case some response functions to load increments are related to the
operator $G$, defined in section II in relation to equation~\ref{eqn:deltaV},
corresponding to the isostatic structure. Part D exploits the results of
ref.~\cite{JNR97b}, deriving the form of the macroscopic equations to be
solved when a small load increment is applied. Finally, these results are
compared, in part E, to some other approaches and theories, that were put
forward by several authors in the recent literature, both at the
microscopic~\cite{MO98a,MO98b,TW99} and the
continuum~\cite{BCC95,WCC97,CWBC98,Claudin} level.

\subsection{Macroscopic granular mechanics: known features, conflicting models.}

\begin{figure}[ht]
\centering
\includegraphics[width=8cm]{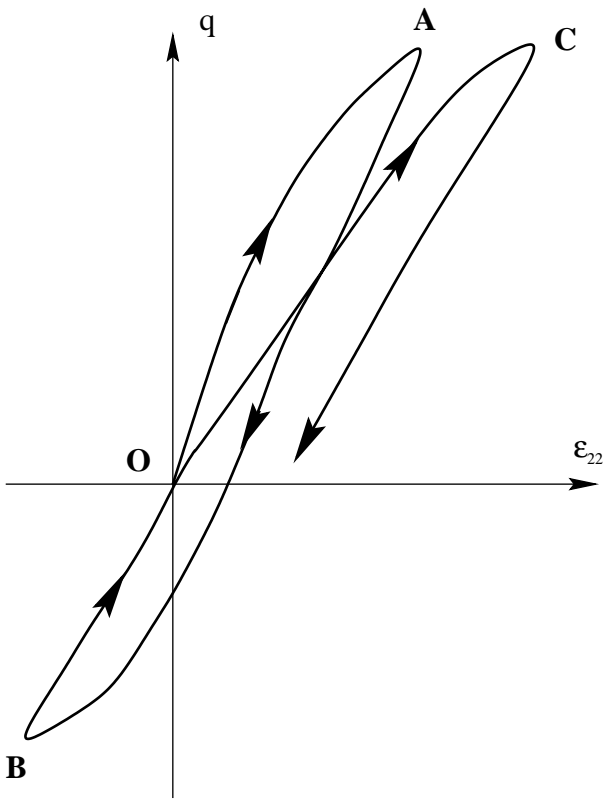}
\caption{Schematic aspect of response to cyclic variations in $q$
in the $\epsilon _{22}$ - $q$ plane.
\label{fig:hyster}}
\end{figure}

A classical way (see, \emph{e.g.,}, in~\cite{Muirwood}) to study the
macroscopic mechanics of granulates is to submit a sample to a triaxial test.
Such a device is designed to impose a uniform state of stress throughout the
sample. It does not matter, for our discussion, whether this macroscopic
stress is imposed via a fluid pressing on a flexible membrane (as in a
laboratory apparatus, for the lateral confinement) or via a control of the
position of a rigid wall (as in some numerical simulations). We just need to
remember that a varying load is imposed, and depends on two parameters $p$ and
$q$, the axial stress ($\sigma _{yy}$ on the figure) being equal to $p+q$ and
the lateral one  ($\sigma _{xx}$), to $p$. A typical experiment consists in
gradually increasing $q$ at constant $p$. One may then observe the resulting
strains.
\begin{figure}[ht]
\centering
\includegraphics[width=8cm]{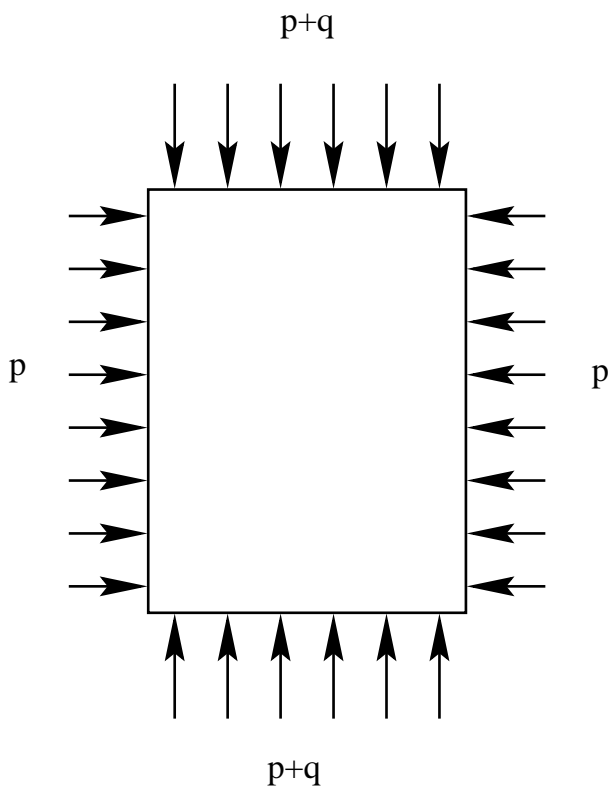}
\caption{The triaxial experiment.
\label{fig:triax}}
\end{figure}
The classical elasto-plastic constitutive laws that are applied to granular
materials are incremental, which means that they do not relate stresses and
strains directly, but predict the increment of strain resulting from an
increment of stress, given the current state of the system (the definition of
which might require other, `internal' variables). Cycling sollicitations of
small amplitude usually yield, in the stress-strain plane, loops with some
amount of hysteresis. The surface area of such a loop as OABO on
fig.~\ref{fig:hyster} is the plastically dissipated energy associated to
deviatoric stresses (to which the work due to volume changes has to be added
to get the total plastic work).

In marked contrast with classical soil mechanics approaches, some authors
recently proposed a new type of macroscopic mechanical description for the
statics of granular packings~\cite{BCC95,WCC97,CWBC98,Claudin}. According to
them, resorting to strain variables should be avoided and one should look for
direct relationships between the components of the stress tensor, so that it
is possible to determine the whole stress field in a granular sample by
solving hyperbolic second-order partial differential equations. Those, like
wave equations, possess characteristics, preferred directions along which they
reduce to simpler, first order forms. To solve the problem, one may integrate
along the characteristics that emerge from every point where some external
force is applied. Consequently, in a packing in which the forces
exerted on the top boundary (wall or set of particles) are
known, a perturbation (external force increment), 
will \emph{propagate} downwards, but will not be felt above the point where it
is applied. The exact relation between stresses to be used should then depend
on the actual process by which the sample was made. If the current stress
level is changed, by, say, a manipulation of the boundary conditions, like in
the triaxial test, then the granular system rearranges until the new
constitutive relation, corresponding to its new state, agrees with the new
externally imposed stress values. Those theories, in their current state of
development, do not predict the extent to which the system has to rearrange,
or, in other words, the magnitude of the ensuing strain increment.  
It has been recently proposed~\cite{TW99} that isostaticity could justify
such theories for frictionless assemblies of
grains. These suggestions are discussed in section IX.E below.

We now turn to a discussion of some possible microscopic origins of plastic
dissipation.

\subsection{Origins of plastic dissipation.}
When a given supported external load places the system in a uniquely
determined equilibrium state, one has to expect a mechanical behaviour devoid
of plastic dissipation. Hysteresis loops like those of fig.~\ref{fig:hyster}
cannot occur. Plasticity is related to the lack of uniqueness of equilibrium
states. At the level of continuum mechanics, it is sometimes termed `internal
friction', since the material behaves as if different layers of matter slided,
with friction, on one another within the bulk of the sample. We have thus
identified two microscopic origins of \emph{internal friction in systems of
frictionless grains}.
\begin{enumerate}
\item
bounded tensile forces in the contacts (as in section  VI.C)
\item
rearrangements of finite extent (\emph{i.e.,} the ASD is no longer valid)
between equilibrium position of assemblies of spherical grains.
\end{enumerate}

Let us illustrate these different behaviours on the simple example of
fig.~\ref{fig:excoh} (section VI.C).

\begin{figure}[ht]
\centering
\includegraphics[width=8cm]{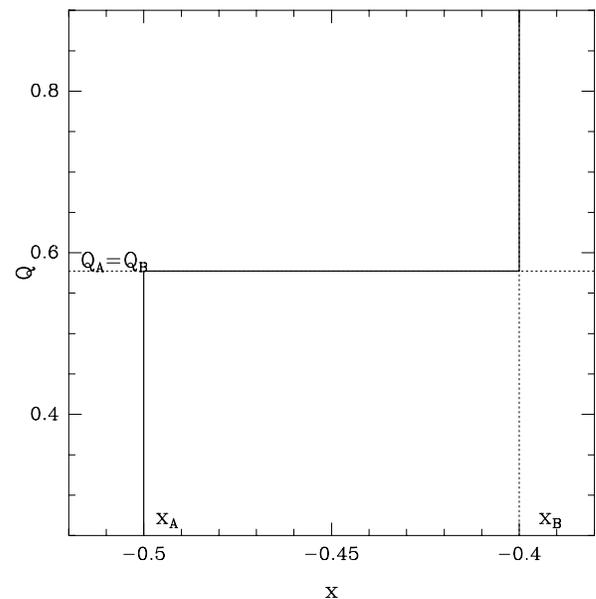}
\caption{Loading parameter
$Q={F_x\over F_y}$ versus coordinate $x$ of the mobile disc of the system of
fig.\ref{fig:excoh}.
\label{fig:courbeasd}}
\end{figure}

Starting from an equilibrium configuration in which the external force on
disc $1$, in contact with $2$ and $3$, is vertical, let us gradually increase
its horizontal component $F_x$. We first discuss the problem  within the ASD.
It is then a particular example of $\ppp _1 $ discussed in section
VII, a linear optimization problem with two unknowns (the coordinates of disc
1). In fact, the simplex within which potential energy $W$ has to be
minimized is exactly the one that was shown on fig.~\ref{fig:simplex1}.
Points A and B on that figure are respectively the equilibrium positions of
the center of disc $1$ when it is in contact with $2$ and $3$, and with $3$
and $4$.  Changes from one position to the other happen when the direction of
${\bf F}$ is orthogonal to that of segment AB. One may monitor the abscissa of
the mobile disc, $x$, which, as presented on fig.~\ref{fig:courbeasd}, is
related to loading parameter $Q={F_x\over F_y}$ \emph{via} a step-like
function.

\begin{figure}[ht]
\centering
\includegraphics[width=8cm]{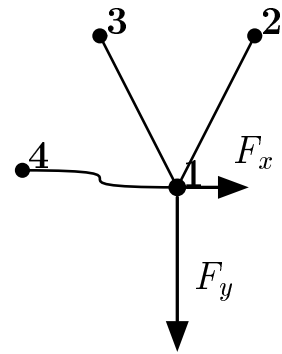}
\caption{System of cables, equivalent, within the ASD, to the system of discs
of fig.~\ref{fig:excoh}, with the same values of external forces. Here, the
cables joining $1$ to $2$ and $3$ are taut, while the one joining $1$ to $4$
is not.
\label{fig:petitt}}
\end{figure}

In analogy with this problem of rigid grains, one may build a system of rigid
cables (resisting tension, but not compression), which, if treated within the
ASD, yields exactly the same simplex of accessible configurations, the same
optimization problem ($\ppp _1$) as that of fig.~\ref{fig:simplex1}. This
system of cables is shown on fig.~\ref{fig:petitt}. Node $1$ is now tied to
$2$, $3$, and $4$, by cables that are slightly longer than the common distance
between $2$ and $3$, and between $3$ and $4$.

\begin{figure}[ht]
\centering
\includegraphics[width=8cm]{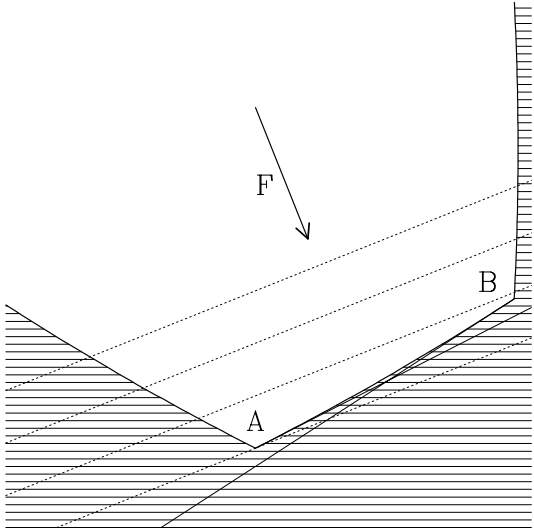}
\caption{Minimization problem in the plane of coordinates of node $1$,
for the system of figure~\ref{fig:petitt} without the ASD. The accessible
part of configuration space (outside the hatched zone) is convex. Its
boundary has sharp corners ($A$ and $B$), but, unlike on
fig.~\ref{fig:simplex1}, corresponding to the same problem within the ASD,
displays curvature in between. Tangents to that curve at $A$ and $B$ are
drawn. \label{fig:trajt}} 
\end{figure}

\begin{figure}[ht]
\centering
\includegraphics[width=8cm]{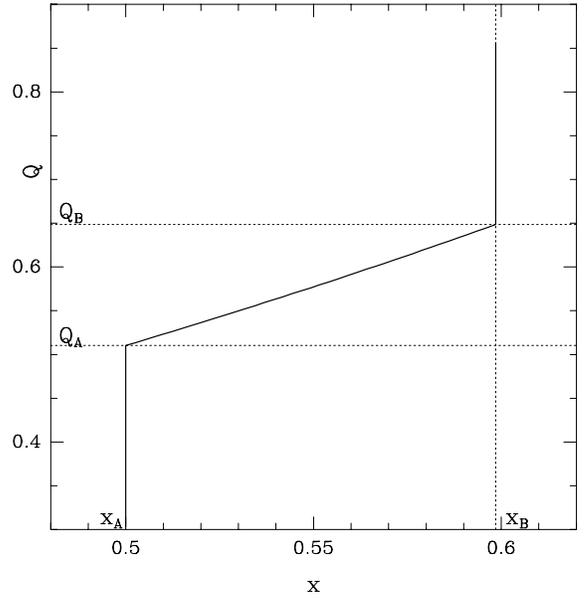}
\caption{Force ratio $Q$ versus coordinate $x$ of node $1$ for the system of
fig.~\ref{fig:petitt}, without the ASD.
\label{fig:courbet}}
\end{figure}

Outside the ASD, the potential minimization problem for the system of cables
is no longer a linear optimization problem, but, according to the general
properties discussed in section VIII, is still a convex problem. In the plane
of the coordinates of node $1$, the simplex of fig.~\ref{fig:simplex1} changes
into a domain limited by curved faces, as shown on fig.~\ref{fig:trajt}. The
curvature of the faces being oriented inwards, this domain of accessible
configuration is convex.
When the orientation of force ${\bf F}$ is such that, on fig.~\ref{fig:trajt},
the direction of constant potential energy lines lies between those of
tangents to the accessible domain in $A$ and $B$, the equilibrium position is
a point on arc $AB$, and only one cable is taut, the one joining $1$ to $3$.
In this case, the motion along arc $AB$ is a mechanism, but stability is
maintained, just like in the example of fig.~\ref{fig:buckle}.  There is still
a one-to-one correspondence between $Q={F_x\over F_y}$ and $x$, as shown on
fig.~\ref{fig:courbet}. As the difference between cable lengths and distances
2-3 and 3-4 decreases, displacements get smaller and smaller.
The difference $Q_B-Q_A$ tends to zero, the curvature of the accessible region
boundary on fig.~\ref{fig:trajt} vanishes, and the curve of
fig.~\ref{fig:courbet} approaches the ASD case, fig.~\ref{fig:courbeasd}.
Over a finite interval between $Q_A$ and $Q_B$, the force-displacement
relationship is  a smooth function, unlike the stepwise dependency shown on
fig.~\ref{fig:courbeasd} (corresponding to the limit of very small motions).

\begin{figure}[ht]
\begin{center}
\includegraphics[width=8cm]{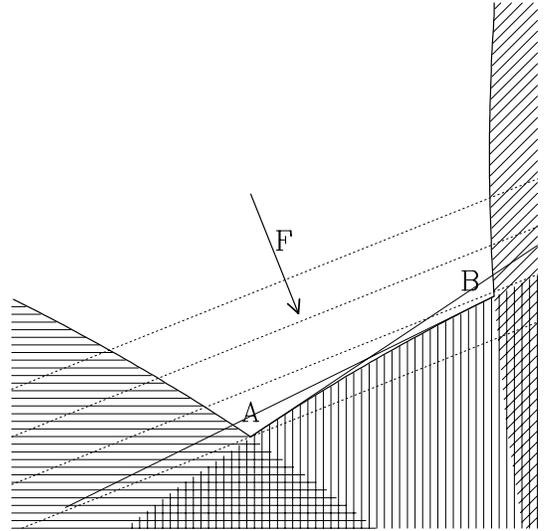}
\caption{Same as figs.~\ref{fig:simplex1} and~\ref{fig:trajt}, in the case of
the system of fig.~\ref{fig:excoh}, without the ASD.  The straight lines are
the tangents to the boundary curve at points $A$ and $B$. 
\label{fig:trajc}}
\end{center}
\end{figure}

\begin{figure}[ht]
\begin{center}
\includegraphics[width=8cm]{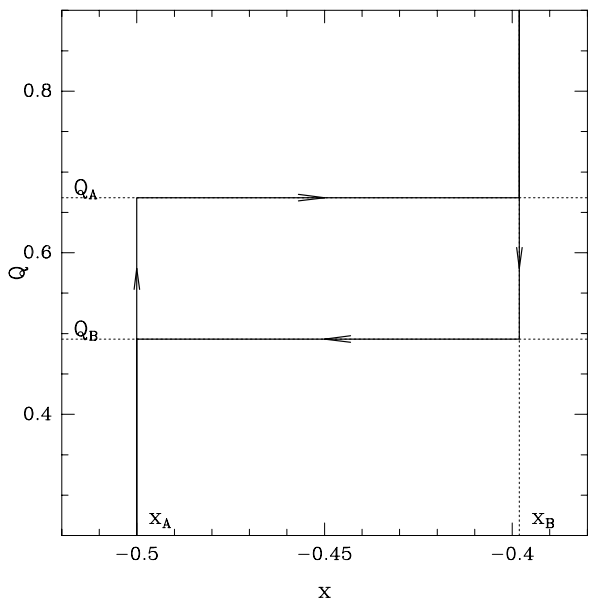}
\caption{Same as figs.~\ref{fig:courbeasd} and~\ref{fig:courbet},
in the case of the system of fig.~\ref{fig:excoh} outside the ASD. The
force-displacement relation is now history-dependent, as shown by the arrows.
\label{fig:courbec}}
\end{center}
\end{figure}

Let us now deal with the system of fig.~\ref{fig:excoh} (with rigid,
impenetrable discs and frictionless contacts that do not resist tension)
\emph{outside the ASD}. The accessible domain in the coordinate plane is, as
opposed to the previous cases, no longer convex, as shown on
fig.~\ref{fig:trajc}. 
The upper limit $Q_A$ of the $Q$ interval for which position $A$ is stable is
now larger than the lower limit $Q_B$ of the $Q$ interval for which position
$B$ is stable. Because of this \emph{bistability} for $Q_B\le Q\le Q_A$, the
$Q$ versus $x$ relation now exhibits hysteresis, as shown on
fig.~\ref{fig:courbec}. 

As shown in section VI.C, contact law~\ref{eqn:coh}, that allows for some
bounded tensile forces in the contacts, is such that both equilibrium
positions $A$ and $B$ will be simultaneously possible for some values of $Q$,
in the system of fig.~\ref{fig:excoh}. $Q$ then varies with $x$ exactly as
shown on fig.~\ref{fig:courbec}, with $Q_A={\displaystyle 1\strut \over
\sqrt{3}\strut}+{\displaystyle f_0 \strut \over F_y \strut}$  and 
$Q_B={\displaystyle 1\strut \over \sqrt{3}\strut}-{\displaystyle f_0 \strut
\over F_y \strut}$.

One may note, however, that the plasticity due to cohesion of finite strength
differs from the one due to geometric rearrangements in the two following
respects.
\begin{itemize}
\item
With contact law~\ref{eqn:coh}, plasticity does not disappear in the limit of
small motions (when the ASD becomes valid).
\item
It is sensitive to the \emph{magnitude} of external forces, not only on their
\emph{direction}. The figure analogous to~\ref{fig:courbec}, in the
($Q={F_x\over F_y}$, $x$) plane, now depends on the value of $F_y$. When $F_y$
is very much larger than $f_0$,  the cohesive strength of contacts might be
neglected, and vanishes as a source of plastic dissipation.
\end{itemize}

It might be expected, on going, from the elementary example dealt with in
this section, to larger and larger systems, that curves like
fig.~\ref{fig:courbec}, forces (like $F$) averaging to stresses and
displacements (like $x$) to strains, will gradually look like
fig.~\ref{fig:hyster}. In larger systems,  the curve of
fig.~\ref{fig:courbec} will look like a staircase. Presumably, as the system
size increases, the number of the steps, and their amplitude, if expressed in
terms of intensive quantities, will tend to zero. Then the smoothness of the
curves sketched on  fig.~\ref{fig:hyster} \emph{might} be recovered in the
thermodynamic limit. Whether it actually \emph{will} is of course not obvious
a-priori, a careful statistical analysis~\cite{CRprep} is required. In the
case of systems treated within the ASD, each step of the resulting staircase
will be retraced back and forth, without any irreversibility. Such models can
be expected to share the properties of the lattice system of
ref.~\cite{JNR97b} and paragraph VII.B.8, in which the staircase does indeed
approach a smooth stress-strain curve in the thermodynamic limit. (But this
curve is unique, one cannot obtain fig.~\ref{fig:hyster} in such a case).

The difference between plasticity of cohesive and non-cohesive grains
that was pointed out above is reminiscent of the difference in the behaviour,
under growing hydrostatic pressure, of sands and clays~\cite{Muirwood}. As the
magnitude of the load increases (but its direction is fixed), the level of
plastic deformation in the cohesive material (clay) is much higher than in the
non-cohesive one (sand). 

It is also interesting to  note that some theories of friction between solid
surfaces~\cite{CV97} are, just like the mechanisms for \emph{internal}
friction that we invoke here, based on the history-dependent selection of one
among several possible stable equilibrium configurations.
\subsection{Consequences of isostaticity.}

We focus here on systems of frictionless, cohesionless and rigid spheres (the
contact law being~\ref{eqn:loicont}) in equilibrium under a given load, for
which it was shown, in two steps (sections VI and VIII), that the
force-carrying backbone is an isostatic structure.  We discuss some specific
consequences of this property. In the simple example treated in subsection B
just above, both equilibrium configurations A and B correspond to isostatic
contact structures, and it is easy to predict for which value of the loading
parameters the system will change from one to the other. Exploiting the
isostaticity property, we will show here that such a prediction can, to some
extent, be done in an arbitrary system.

In this subsection, we only consider the backbone, ignoring the rest of the
system. We suppose that grains have been renumbered, so that index $\mu$,
with $1\le \mu \le N_f$ only label the degrees of freedom of objects that
belong to the backbone. We shall also adopt the convention that the whole
backbone does not move as a rigid body (thus excluding the $k_0$ corresponding
degrees of freedom from the list). Likewise, $1\le l\le N$ here only labels
the force-carrying contacts ($N=N_f$).  
\subsubsection{Response to perturbations, without rearrangement.} 
Isostaticity of the whole structure
means that matrix $G$, and its tranpose $G^T$ are square and have an inverse.
Not only are equilibrium forces, given the load, uniquely determined, but it
is also possible to predict how small external force increments (on the
backbone) will be distributed in the existing contacts. Changing the load from
$(F_{\mu}^{ext})_{1\le \mu\le N_f}$ to $(F_{\mu}^{ext}+\delta F_{\mu}^{ext}
)_{1\le \mu\le N_f}$ will result, in contact $l$, in force increment $\delta
f_l$, given by (summation over repeated indices implied)
\be
\delta f_l =
(G^T)^{-1} _{l\mu} \delta F^{ext}_\mu = G^{-1} _{\mu l} \delta F^{ext}_\mu.
\label{eqn:resp1}
\ee

The backbone being rigid, this change in forces does not entail any
displacement: $u_\mu = 0$ for each $\mu$. This correctly describes the
mechanical response of the granular assemblage as long as all contacts forces
remain positive.  This should be the case, in a finite system, for
sufficiently small perturbations of the initial load.

\subsubsection{Dual response of velocities to bond length variations.}
Parallel to the one-to-one correspondence between contact forces and external
loads expressed by eqn.~\ref{eqn:resp1}, is the inversible linear mapping
between velocities and relative normal velocities in the contacts. There is no
compatibility condition in the absence of hyperstaticity, and one may impose
arbitrary values to relative normal velocities $(\delta V_l)_{1\le l\le N}$
for the whole list of contacts. The resulting velocities of the spheres are
then (summation over $l$ implied) :
\be
V_\mu = G^{-1} _{\mu l} \delta V_l.
\label{eqn:resp2}
\ee
On comparing to \ref{eqn:resp1}, it appears that the same matrix element
$G^{-1} _{\mu l}$ is both equal to the force increment in  contact $l$
created when a unit external force is exerted on the coordinate $\mu$ on the
one hand, and to the velocity coordinate $\mu$ when $\delta v$ is equal to one
in contact $l$ and to zero in all other contacts,  on the other hand. Such a
symmetry in response functions was remarked by Moukarzel~\cite{MO98b}, who
derived it by different means.
\subsubsection{Response to perturbations: structural rearrangements.}

The particular form of mechanical response expressed by eqn.~\ref{eqn:resp1},
in which no motion occurs and the load increment is supported by the
initially existing contacts, ceases to be relevant as soon as negative contact
forces appear. In the case of a two-parameter loading mode, such as the
biaxial experiment at constant $p$, in which $q$ is gradually increased from
its initial value $q=0$, one may write in each contact $l$ $$f_l = \beta _l p
+ \gamma _l q,$$ where $\beta _l$ and $\gamma _l$ are, due to isostaticity,
geometrically defined coefficients. In general one finds that some of the
$\gamma _l$ are negative. Let us denote as $L^-$ the set of such contacts. The
load will no longer be supported as soon as $q$ reaches the value
\be
q _{max} = \min _ {l\in L^-} {-\beta _l\over \gamma _l} p.
\label{eqn:qmax}
\ee

For larger $q$'s, the theorem of virtual power shows that it is possible to
decrease the potential energy upon opening the contact $l_0$ for which the
minimum in the right-hand-side of~\ref{eqn:qmax} is reached, all other
contacts remaining closed. The system will then rearrange, until a new set of
contacts is created, such that the new load $(p,q)$ is supported with positive
contact forces. If one uses the ASD to describe this motion, then, within this
approximation, the new list of contacts, as shown in section VI, is entirely
determined by the sole system geometry, as the solution to a simplex problem.
Outside the ASD, the new equilibrium state, after the system rearranges, 
might depend on specific dynamical laws. In general, the range of validity of
the ASD and the influence of the dynamics are to be tested, in experiments or,
perhaps more easily, in numerical simulations. However, we have just shown, in
fact, that \emph{the direction of velocities at the beginning of the
rearrangement} is determined by purely geometrical conditions, at least if
$l_0$ is unique: to find those directions, just impose $\delta V_{l_0} =-1$
(thus opening contact $l_0$), and  $\delta V_l =0$ for any $l\ne l_0$, from
which all velocity components are deduced as $v_\mu = -G^{-1}_{\mu l_0}$, from
equation~\ref{eqn:resp2}. 

Simulations of disordered systems of discs~\cite{CR99} suggest that $l_0$ is
generically unique, except in situations when the opening contacts involve a
cluster of d+1-coordinated spheres in d dimensions. Examples of such clusters
are sets of discs 8, 19 and 2, or 6 and 15, or 12 alone on
figure~\ref{fig:sysC}. It is easily realized that once one contact force
involving \emph{e.g.,} disc 8 is known, then all contact forces involving
discs 8, 19, or 2 are also known, and proportional to the first one. Thus,
they all vanish simultaneously. This means that all matrix columns
$(G^{-1}_{\mu l})_{1\le \mu \le N_f}$ are proportional to one another for all
indices $l$ that label contacts of d-spheres belonging to the same
d+1-coordinated cluster. Returning to the determination of the motion when the
load ceases to be supported by the initial list of contacts, it follows that
even though, in such a case, several contacts, involving the same cluster
of d+1-coordinated spheres, may simultaneously open, the uniqueness of the
initial velocities, up to a common amplitude factor, is preserved for all
spheres that do not belong to the said cluster.

\subsubsection{Fragility.}
When a rearrangement occurs after a load increment, the mechanical response of
the granular assembly, unlike the one expressed by equation~\ref{eqn:resp1},
involves both force changes \emph{and} displacements. It depends on the
possibility of closing contacts that are not present in the initial
equilibrium configuration. This geometric information is not contained in
matrix $G$, which only depends on the network of initially existing contacts.
One could thus study a second type of response to perturbations, that involves
displacements. To see which of the two kinds of response is more relevant for
the macroscopic mechanical behaviour, one has to impose perturbations that
possess some macroscopic  meaning, such as changes of $q$ in a biaxial
experiment. Then, assuming, to fix notations, $q$ is increased from zero, two
cases need be considered. Either the thermodynamic limit of $q_{max}$, as
defined in~\ref{eqn:qmax}, is positive, or it is equal to zero. In the first
case, there exists a finite interval of stress for which no motion occurs in
the continuum limit, and the mechanical response discussed in the preceding
paragraphs in terms of the sole matrix $G$ is macroscopically relevant. In the
second case the granular material might be appropriately termed
\emph{fragile}, since, in the thermodynamic limit, arbitrarily small
macroscopic perturbations provoke rearrangements of the contact structure.
Then, any macroscopic mechanical experiment involves displacements, the sole
knowledge of one network of contacts that corresponds to a given value of the
loading parameters is not sufficient. The response expressed by the sole
matrix $G$ is not the macroscopically relevant one.

Our simulations of frictionless rigid discs~\cite{JNR97b,CR99,CRprep} show
that such systems are indeed fragile in this sense.\footnote{The fragility
property is in fact contained in the results stated in paragraph VII.B.8, as
any stress increment, however small, that is not parallel to the preexisting
stress, entails some additional strain in the thermodynamic limit}.

\subsubsection{An algorithm to compute a sequence of
equilibrium configurations.}

This suggests the following procedure to determine the sequence of equilibrium
states reached by an assembly of rigid, frictionless, cohesionless spheres
under varying load (p,q), without resorting to \emph{any} dynamical parameter
(without introducing any inertia, or mechanism of dissipation).

\begin{itemize}
\item
1) Starting from an equilibrium configuration, increase loading parameter q
until contact force $f_{l_0}$ vanishes.
\item
2) Move grains in the direction determined by the opening of contact $l_0$,
the others remaining closed. Keep the same prescription for the grain
trajectories as for the initial velocities, taking into account the rotation
of vectors ${\bf n}_{ij}$, until some new contact $l_1$ is created, such that
the new contact list, replacing $l_0$ (now open) by $l_1$, defines an
isostatic structure.
\item
3) If, in the new contact structure, the contact
forces that balance the load are all positive, a new equilibrium state,
corresponding to the new load, has been reached: one may go back to step 1)
and further increase q. Otherwise, some contact forces are negative. Pick up
the one with the highest tensile force, call it $l_0$ and go back to step 2),
with the new contact list.
\end{itemize}

This algorithm has been implemented by G. Combe and the present
author~\cite{CR99}. We propose to name it the `geometric quasi-static method'
(GQSM).  It does involve arbitrary ingredients: there is no reason to forbid
other openings of contacts once interstice $h_{l_0}$ has reached a finite
positive value. Its  great advantage is the possibility to compute
trajectories from the sole knowledge of the system geometry.

The system evolution, under a varying load, appears as a sequence of
equilibrium states that are separated by `jumps' or rearrangements, in which
the list of active contacts is altered. In a phase of equilibrium, the forces
are carried by a minimum list of contacts. In a phase of motion, normal
relative velocities, among the whole bond list, are localized on \emph{one}
bond (several if a structure -- a list of bonds-- larger than the contact
structure, is considered). Both \emph{maximum localization phenomena} are
related to geometric constraints.

The predictions of the GQSM algorithm were compared with those of other
methods that resort to dynamical models (and, as argued in the introduction,
also involve arbitrary, non-physical features). The results will be presented
elsewhere. As mentioned above, mechanical properties, at the level of
individual trajectories in configuration space, cannot be expected, outside
the ASD, to be uniquely determined.  However, in view of the important role of
the geometry, which determines exactly the value of the loading parameters for
which system should rearrange and the direction of the initial velocity
vector, it can be hoped that the statistical properties of such trajectories
that are relevant for the macroscopic laws will present little dependence on
dynamical features of the system (such as masses or dissipative shock laws).

\subsubsection{Rearrangements within the ASD.}

Within the approximation, as the equilibrium state corresponding to a given
load is unique, there is no need to resort to an incremental approach. If one
however does so, then the whole rearrangement event is geometrically
determined. It can be computed with the GQSM as presented above. Then, it will
be observed, on performing step 3) of the algorithm, that the new contact
structure, as soon as a new contact is created, supports the load with only
positive contact forces. Thus, unlike in the general case~\cite{CR99}, no
cascade of successive rearrangements occurs in step 3). Rearrangements are
simpler events in which one element of the contact structure is replaced by
another. 

Let us prove this statement.

Let $S_0$ denote the old list of contacts, and $S_1$ the new one. Both
structures are isostatic, and for any given load one can find unique values
of both sets of bond forces $(f_l)_{l\in S_0}$ and $(f_l)_{l\in S_1}$ that
ensure equilibrium. In the following members of these two sets, in order to
distinguish them, are written down with a superscript: $f_l^{(0)}$ and
$f_l^{(1)}$ respectively denote the force carried by bond $l$, as computed
with structure $S_0$ and with $S_1$. 

Recalling also the notations of the preceding paragraph, $S_1$ is equal to
$S_0$, deprived of contact $l_0$, to which contact $l_1$ is added. When the
value $q_{max}$ of the loading parameter is reached, $f^0_{l_0}$ has decreased
to zero. This means that, exceptionally, the smaller structure $S_0\setminus
\{l_0\}=S_1\setminus \{l_1\}$ can support the  load, and one has
$f^{(1)}_{l_1}$, while $f_l^{(1)}=f_l^{(0)}$ for each $l\in S_0\setminus
\{l_0\}$. As we assume, for simplicity, that contact forces reach zero
separately, there exists a finite range of positive increments $\delta q$ such
that one has $f_{l_0}^{(0)}<0$, while $f_l^{(0)}>0$ for $l\in S_0\setminus
\{l_0\}$, for $q =q_{max}+\delta q$. Likewise,  reducing the $\delta q$
interval if needed, we require the condition $f_{l}^{(1)}>0$ for $l\in
S_1\setminus \{l_1\}$.

We now pick up one such value of $q$, and evaluate the variation $\delta W$ 
of the potential energy (that corresponds to this value of $q$) in the
rearrangement. 

On the one hand, one may obtain $\delta W$ on applying the theorem of virtual
work to structure $S_0$. As contact $l_0$ has opened, the corresponding
relative normal displacement is negative: $\delta u_{l_0}<0$, while $\delta
u_{l}=0$ for each $l\ne l_0$. Therefore, because $f_{l_0}^{(0)}<0$, one has
$$
\delta W = -f_{l_0}^{(0)} \delta u_{l_0} <0.
$$

On the other hand, one may obtain $\delta W$ on applying the theorem of
virtual work to structure $S_1$. As contact $l_1$ has closed, the
corresponding relative normal displacement is positive: $\delta u_{l_1}<0$,
while $\delta u_{l}=0$ for each $l\ne l_1$. Therefore, because $\delta W =
-f_{l_1}^{(1)} \delta u_{l_1} <0$, one has
$$
f_{l_1}^{(1)} > 0.
$$

Thus, the new contact structure supports the load with positive contact forces
as soon as $q>q_{max}$, and a new stable equilibrium state has been reached.

In the general case, we stressed the difference between the mechanical
response of the granular system without rearrangement, which can be deduced
from the geometry of the contact structure, \emph{via} matrix $G$, and the
mechanical response involving some rearrangement, the determination of which
requires some additional prescription (such as that of the GQSM) to move the
particles.

This difference is much less important within the ASD: as the matrices $G$
pertaining to either structure do not change in the motion, all displacement
coordinates will simply be found as follows:
\be
u_\mu = h_{l_1} G^{-1}_{\mu l_1}, \label{eqn:respasd}
\ee
where $h_{l_1}$ denotes the initial opening of contact $l_1$ and the matrix
$G$ is that of structure $S_1$. 

Equations~\ref{eqn:resp2} and~\ref{eqn:respasd} only differ by a scale factor,
interstice $h_{l_1}$. There is nothing especially singular in the
distribution of open interstices in dense granular systems at equilibrium. So,
it can be expected that macroscopic averages corresponding to both response
functions, \ref{eqn:resp2} and~\ref{eqn:respasd}, are proportional to one
another. Moreover, the response without rearrangement, expressed
by~\ref{eqn:resp2}, is the same with and without the ASD.

In the following subsection, we derive explicitly the form of the macroscopic
response function to small increments in applied external forces, in the case
of the triangular lattice model. These are large scale averages of
(combinations of) microscopic responses expressed by eqn.~\ref{eqn:respasd}. 

We shall therefore speculate that the results to be derived below, for the
form of such macroscopic Green's functions, are also valid for the average of
response functions without rearrangements in general.

\subsection{Macroscopic response of the triangular lattice model.}
In the model system studied in ref.~\cite{JNR97b}, the results of which are
recalled in paragraph VII.B.8, it is possible to find the form of macroscopic
equations to be solved when a small density of external forces $\delta {\bf
f}^{ext}$ is superimposed over an initial equilibrium state.

To do so, one just needs to translate the properties stated in paragraph
VII.B.8 in incremental form. 

First, let us impose, without loss of generality, a few conditions on
function $f$ defined in~\ref{eqn:defSigma}. It is convenient to choose a
symmetric function of $\epsilon_{\alpha \beta}$ and $\epsilon_{ \beta
\alpha}$, the derivation in~\ref{eqn:loimacro} being taken regarding both
strain components as independent variables.

Then, defining, in $\t2{\epsilon}$ space,
a norm $\Vert \t2{\epsilon}\Vert $ by
$$
\Vert \t2{\epsilon}\Vert ^2 = \t2{\epsilon} : \t2{\epsilon}
= \epsilon _{11}^2 + 2\epsilon _{12}^2 + \epsilon _{22}^2,
$$
one may enforce (replacing $f$ by $f/\Vert \nabla f \Vert$) the condition:
\be
\Vert \nabla f \Vert = 1,
\label{eqn:normf}
\ee
everywhere on $\Sigma$.

One starts from an equilibrium state in which the stress field,
$\t2{\sigma}$, is assumed to stay strictly inside the supported range,
defined by inequalities~\ref{eqn:suppsigma}, everywhere in the system. This
initial state is also characterized by a displacement field ${\bf u}_0$ and a
strain tensor field $\t2{\epsilon}$ (everywhere on $\Sigma$, and abiding
by~\ref{eqn:loimacro}), the origin being defined by the reference state (the
undisturbed regular lattice of spacing $a$). One then looks for the stress
increment field $\t2{\delta \sigma}$, displacement increment field ${\bf u}$
and strain increment field $\t2{\delta \epsilon}$ that result from the
application of $\delta {\bf f}^{ext}$. The problem is dealt with to first
order in any of these quantities, that are linear in $\delta {\bf f}^{ext}$,
assumed small.

Let us define
$$
A_{\alpha \beta \gamma \delta} = 
{\partial ^2 f\over  \partial \epsilon
_{\alpha \beta} \partial \epsilon _{\gamma \delta}},
$$
a fourth-order tensor that depends on $\t2{\epsilon}$. One has, upon
differentiating the macroscopic law
$$
\sigma _{\alpha \beta} =
\lambda {\partial f\over  \partial\epsilon _{\alpha \beta}},
$$
the decomposition of stress increments as
$$
\delta \sigma _{\alpha \beta } =
\delta \sigma ^{(1)}_{\alpha \beta } + \delta \sigma ^{(2)}_{\alpha \beta},
$$
with (summation over repeated indices)
$$
\delta \sigma ^{(1)}_{\alpha \beta } =
\lambda A_{\alpha \beta \gamma \delta} \delta\epsilon _{\gamma \delta},
$$
and $\delta \sigma ^{(2)}_{\alpha \beta}=
{\delta \lambda \over \lambda} \sigma_{\alpha \beta}$.

Condition~\ref{eqn:normf}, yields, by derivation,
$$
A_{\alpha \beta \gamma \delta}
{\partial f \over \partial \epsilon _{\alpha \beta}} =0,
$$
whence the orthogonality between $\sigma$ and
$\delta \sigma ^{(1)}$. Since $\t2{\epsilon}$ must remain 
on $\Sigma$, $\delta \epsilon$ is also orthogonal to $\sigma$. 

In view of the symmetry of the stress tensor and of the conditions imposed on
function $f$, tensor $A$ satisfies the following symmetries:
$$
A_{\alpha \beta \gamma \delta} =
A_{\beta \alpha \gamma \delta} = A_{\alpha \beta \delta \gamma}.
$$
Because it is a second-order derivative, one also has:
$$
A_{\alpha \beta \gamma \delta} = A_{\gamma \delta \alpha \beta}.
$$
Tensor $A$ is thus endowed with the same symmetry properties as a tensor of
elastic constants (or of viscosity coefficients).

We have seen that it might be viewed as a linear operator within the space of
symmetric second-order tensors that are orthogonal to $\t2{\sigma}$, or, in
other words, within the tangent plane to surface $\Sigma$ in strain space.
Because of the strict convexity of $\ddd$, this operator is \emph{positive
definite} (this is easily realized, as the curvature of $\Sigma$ is turned
inwards).

Transforming the equilibrium equation into one for the unknowns ${\bf u}$ and
$\delta \lambda$, using~\ref{eqn:epsu}, one obtains ($\partial _\alpha$
denoting a derivative with respect to coordinate $\alpha$)
\be
\partial _\beta 
\left[ \lambda  A_{\alpha \beta \gamma \delta}\partial _\delta
u_\gamma\right]- \partial_\beta \left({\delta \lambda \over\lambda}
\sigma_{\alpha \beta}\right) + \delta f^{ext}_\alpha =0, \label{eqn:ell1}
\ee
while the displacement field should satisfy
\be
\sigma_{\alpha \beta} \partial_\beta u_\alpha = 0.
\label{eqn:ell2}
\ee
Equations~\ref{eqn:ell1}-\ref{eqn:ell2}, supplemented by suitable boundary
conditions, define, because of the positive-definiteness of operator $A$, an
{\em elliptic boundary value problem}. The solution is unique provided 2
conditions (in 2D) involving ${\bf u}$  and/or its normal derivatives are
specified everywhere on the system boundary.

We now turn to the situation when the initial stress field is a uniform
hydrostatic pressure: 
$$
\sigma _{\alpha \beta} = P_0 \delta _{\alpha\beta},
$$
with a position-independent pressure $P_0$. In view of
condition~\ref{eqn:normf} on $f$, it should be noted 
that $\lambda$ coincides with  $P_0\sqrt{2}$ in this case. The corresponding
tangent space to $\Sigma$ is the space of traceless tensors.

In general, tensor $A$ reflects the common symmetries of the material (the
triangular lattice) and the stress tensor. In this particular case, it will
possess all the symmetries of the regular triangular lattice. The tensor of
elastic constants, in that case~\cite{LLel}, has the same symmetries as in an
isotropic medium. Because it operates within the space of traceless tensors,
tensor $A$ reduces to a scalar $K$: one has, for any traceless strain
increment,
$$
A_{\alpha \beta \gamma \delta} \delta \epsilon _{\gamma \delta}
= K  \delta \epsilon _{\alpha \beta}.
$$
\ref{eqn:ell1} has become
$$
KP_0\sqrt{2}\nabla ^2 {\bf u} - \nabla (\delta P) + \delta {\bf f}^{ext} =0,
$$
while~\ref{eqn:ell2} now states that the displacement field should be
divergenceless:
$$
{\bf \nabla}\cdot {\bf u} = 0.
$$
One recognizes {\em the Stokes problem for viscous incompressible flow}, in which the
displacement replaces the velocity field, the product $KP_0\sqrt{2}$
plays the role of the shear viscosity, and $\delta P$ is a pressure field
to be determined on solving the full boundary value problem.

Green's functions for the Stokes problem can be found, \emph{e.g.},
in~\cite{HS80}. In an infinite 2D medium, the velocity field varies
logarithmically with the distance to the point where a concentrated force is
applied. 
\subsection{Discussion.}
From the results just above, it can be
concluded that the form of the macroscopic equations ruling the displacement
field created by a small perturbation to a pre-stressed granular sample in
equilibrium should be elliptic, provided the microscopic rearrangements are
dealt with within the ASD.

From the discussion at the end of paragraph IX.C.6, we expect that operator
$G^{-1}$, in the general case, also averages macroscopically as the Green
function of an elliptic  second-order partial-differential operator.  One may
obtain a suitable macroscopic average on taking, \emph{e.g.,} the mean of all
matrix elements $G^{-1}_{\mu l}$ for which the vector pointing from bond $l$
to the center of the grain which coordinate $\mu$ belongs to is in some
prescribed small neighbourhood of a given vector.  

$G^{-1}$ rules the response without rearrangement. The general --and, in view
of the fragility property, most relevant-- case of mechanical response
involving rearrangements outside the ASD appears to involve more geometric
information than the one contained in matrix $G$: it could be
observed~\cite{CR99} that step 3) of the GQSM algorithm introduced in
paragraph IX.C.5 could involve a long sequence of  elementary rearrangements
replacing one contact by another. Unlike the distribution of open gaps between
adjacent particles, that of the magnitude of such complex rearrangements can be
quite wide and might significantly affect the macroscopic response in terms of
dispacements. This will be studied in a forthcoming publication. In the case
of a disordered granular assembly, no small parameter, like the level of
polydispersity of discs in the triangular lattice model, is available to
control the validity of the ASD. As found in section VIII, stable equilibrium
states of frictionless discs or spheres are especially scarce in configuration
space, as full rigidity is required. Outside the ASD, impenetrability
constraints do not limit a convex accessible domain of configuration space.
Whereas the route from one equilibrium state to another, within the ASD, can
be straight, it might have to follow a long and tortuous path outside the
approximation.  (The ASD amounts to simplify this complex geometry,
straightening up local curvatures, etc\dots )

Interestingly, Tkachenko and Witten~\cite{TW99}, following a suggestion by
Alexander~\cite{SA98}, speculated that, {\em as a consequence of the
isostaticity property}, the mechanics of frictionless sphere packings should
be described, at the continuum level, by laws of the type proposed in
refs.~\cite{WCC97,CWBC98}: the response to perturbating force fields satisfies
\emph{hyperbolic} partial differential equations. From considerations on the
floppy modes that appear within a subsystem that is isolated from the rest of
the sample, they derive a similar directional structure for matrix $G^{-1}$ as
for the macroscopic response in such theories: in a sample limited by a free
surface in the upwards direction, force perturbations are not felt above the
point where they are introduced.

Although we do not venture here to speculate on the form of macroscopic
equations that rule the mechanical response with rearrangements in a general,
disordered system for which the ASD might not be valid, our conclusions above
do go far enough as to clearly contradict the ones of~\cite{TW99}, since those
are concerned with the same object (operator $G^{-1}$). 

An explanation for this discrepancy could be that Tkachenko and Witten mainly
based their conclusions on the observation of packings (numerically) obtained
by sequential deposition algorithms under gravity.

When the stress tensor approaches the boundary of the region of supported
loads (\emph{i.e.,} when one of the conditions in ~\ref{eqn:suppsigma} is
almost an equality) one can observe~\cite{JNR97b}, for the triangular lattice
model, that the list of force-carrying contacts approaches a limit that
comprises all the bonds parallel to two of the three lattice directions,  and
none of the bonds parallel to the third. The topology of the backbone thus
approaches that of a square lattice. In this particular case~\cite{TW99}, it
is easy to check that a description in terms of \emph{force propagation},
involving hyperbolic equations, applies. The marginally supported stress
states of this model are the analog of the Coulomb condition for an isotropic
medium. When the Coulomb criterion is everywhere satisfied as an equality, the
material is everywhere on the verge of plastic failure, and it has long been
known (and exploited for the evaluation of critical loads~\cite{VS65}) that
the macroscopic equations are of the hyperbolic type. This situation has been
termed `incipient failure everywhere' (IFE) in~\cite{WCC97,CWBC98}. 

One may conjecture that deposition algorithms~\cite{VB72,MJ87} will
systematically produce internal states close to IFE. Specifically, we expect
sequential deposition under gravity to result in the `active' Rankine state,
in which the pressure on the lateral walls is barely sufficient to contain
macroscopic plastic flow of a horizontal granular layer sumitted to its own
weight. In the case of discs with a small or moderate polydispersity in 2D, 
the deposition algorithms do in fact produce networks of force-carrying
contacts that are very close to the limiting states of the triangular lattice
model (a deformed square lattice).

Therefore, we suspect that Tkachenko and Witten's arguments only apply to
those particular cases of limit states or IFE. 

There are, apart from the arguments put forward in~\cite{TW99}, other aspects
on which the general properties we have been discussing as well as the
numerical results obtained on the triangular lattice model appear at odds with
the assumption of a direct relationship between stress components, and related
theories. Leaving a more complete discussion to subsequent work, let us merely
point out that the nature of the boundary conditions has dramatic effects if
the macroscopic equations are hyperbolic. In fact, if a rigid boundary
transmitting a stress is replaced by a distribution of external forces imposed
independently on the grains that are close to the edge, such theories predict
this change to significantly affect the whole system (which has lost its
rigidity). In our experience~\cite{OR97b,Sofiane}, some rearrangement does
occur, but its effects are confined to a boundary layer of finite depth.

We also note that our results disagree with some of
Moukarzel's~\cite{MO98a,MO98b}, predicting perturbations due to a localized
force to increase \emph{exponentially} with distance. Although his results are
very accurate and were obtained on very large systems, the \emph{propagative}
nature of forces, which can be calculated from `top' to `bottom' in a single
sweep, is an explicit ingredient of his model, that was adapted from the one
of~\cite{HHR97}. Our results on the triangular lattice model disagree with his
because this very large effect of force perturbations (or, equivalently --
see~\ref{eqn:resp1} and~\ref{eqn:resp2}-- of bond length variations) would
cause the level of distortion of the regular lattice, due to the
polydispersity of discs, to increase very fast with the system size. Rather,
we observed it to approach a finite thermodynamic limit. Once again, we
suspect that the very peculiar properties obtained in these studies stem from
the consideration of a special case in which forces happen to possess a
propagative nature.

Finally, the (provisional) conclusion we propose here is, as already mentioned
in section VIIE, that the rigidity of the grains and the isostaticity
property do not \emph{necessarily} entail very special, critical or singular
macroscopic mechanical properties. Moreover, we expect -- as systems dealt
with within the ASD exhibit the same kind of elasticity as networks of rigid
cables -- that if unusual, exotic properties exist, then they are related to
the displacements (the rearrangements) rather than the network of forces (or
the operator $G$ attached to it).

\section{Conclusion and perspectives.}
Let us first briefly summarize the main results presented in this paper.

Specializing to frictionless grains, and assuming that granular packings,
under slowly varying sollicitations,  tend to stable equilibrium states, we
have shown that geometry determines, to a large extent, the mechanical
behaviour of such materials. 

Spatial arrangements of granular packings in equilibrium under a given load
are quite specific points in configuration space. Rigid grains that only
exert normal contact forces on one another, once submitted to a supported
load, will generically pack in such a way that the problem is isostatic,
\emph{i.e.,} there is no indeterminacy of forces. The value of all contact
forces is determined by equilibrium equations and the geometry of the contact
structure. This yields a rigourous upper bound on the contact coordination
number of any packing of rigid grains. These properties hold for compressive
or tensile contact forces. Contact structures, in equilibrium, are not always
rigid, especially (but not exclusively) in the case when contacts can sustain
tensions. Even if loose particles, that carry no force, are discarded from the
count, the upper bound on the coordination number might not be reached.

If the packing is such that the approximation of small displacements might be
well justified, in particular in the case of regular arrangements on lattices,
stronger properties were established, provided the problem can be coped with
in the framework of convex optimization theory (which requires the  definition
of a potential energy, thus excluding finite strength cohesion). Then
\begin{itemize}
\item
Not only the forces once the
contact structure is known, but the force-carrying structure itself
is entirely determined by the system geometry.
\item
Grain positions are also determined, apart from possible `floppy mode'
motions, of bounded amplitude, that do not affect the value of the potential
energy. 
\item
Displacements from the reference configuration on the one hand, and
\emph{contact forces on the other hand} are the solutions to two optimization
problems in duality.
\item
For rigid grains, force-carrying structures are the
exact analog of cost-minimizing directed paths in scalar transport problems. 
\end{itemize}
Such situations are thus very attractive from a theorist's point of view: the
reduction of the mechanical problem to one of random geometry is complete, and
analogies with other models of theoretical statistical physics (directed
percolation, directed polymer in a random environment) can be drawn and
exploited. However some important features of granular mechanics are absent:
such systems are devoid of plasticity and hysteresis.

Pursuing the stability analysis beyond the ASD in the case of discs or
spheres, we have shown that the force-carrying structure must be rigid if
contacts do not withstand tension, because any floppy mode would imply
instability. This entails that the force-carrying backbone in systems of rigid
spheres is, generically, an isostatic structure,  its coordination number is
equal to $2d$ in dimension $d$.

Analogous systems of cables (that resist tension, but no compression), on the
other hand, will generally keep some amount of floppiness, since mechanisms in
the equilibrium state are all stable. 

Assemblies of frictionless grains will, in general, exhibit internal friction,
due to the multiplicity of stable equilibrium states corresponding to the
same external load. This non-uniqueness might stem from the finite extent of
rearrangements or from bounded cohesion forces.

If submitted to slowly varying loads, packings of rigid grains will evolve via
a succession of jumps or crises separated by phases of rest. The isostaticity
property implies, for a system of rigid frictionless spheres, that the
concentration of forces is maximal during a phase of rest (forces cannot be
carried by a strictly smaller set of contacts), and that the concentration of
deformation is maximal at the beginning of a jump (there cannot exist a
strictly smaller list of interstices in which relative normal velocities are
not equal to zero). 

Although the motion in a rearranging event depends on the actual granular
dynamics, the forces during a phase of rest, and the direction of velocities
at the beginning of motion, are geometrically determined. 

Two kinds of response functions to force increments can be studied, depending
on whether the perturbation provokes a change in the contact list. Some
recent studies of response functions, without rearrangement of the grains,
were discussed and we argued that some of their conclusions might be specific
to sequential deposition models, in which  forces can be propagated along a
preferred direction. The fragility of frictionless granular assemblies in the
thermodynamic limit implies however that macroscopically meaningful
perturbations always involve some amount of rearrangement.

The results of the present article suggest both general perspectives and
specific problems, to be dealt with in future work.

An important feature of granular materials is the sparsity, in configuration
space, of equilibrium configurations. Those, especially for rigid grains,
have very specific characteristics. Moreover, they are generally suitable for
\emph{one} particular load. In such circumstances, it might not be adequate to
choose \emph{first} one specific geometric arrangement and contact structure,
built, \emph{e.g.,} by some convenient algorithm that respects impenetrability
conditions, and \emph{then} to apply external forces and see how they could be
balanced by contact forces. The list of active contacts is itself chosen
according to the external load. Many recent studies were devoted to the way
forces distribute among a fixed list of contacts, and to the ensuing
statistics of contact force values.  Although models along these lines might
capture \emph{some} of the physics, they ignore displacements.  Displacements,
as our results have amply shown here, are always part of the problem. The very
definition of a force requires the consideration of some amount of
displacement. A normal reaction force in the frictionless contact between two
rigid objects is a geometrically defined quantity, a Lagrange parameter
associated with an impenetrability constraint in configuration space. Large
assemblies of frictionless rigid grains are fragile: tiny load increments will
be associated with rearrangements of the contact structure. If one wishes to
understand the macroscopic mechanical behaviour of granular systems and its
relationship to grain-scale phenomena, the question of the \emph{magnitude} of
such rearrangements, in which the system moves from an equilibrium state to
another, is crucial.

Other, more specific questions, that are related to statistics and the
continuum limit, naturally follow from the mechanical properties we have been
presenting.  When is the ASD is a good approximation, apart from lattice
models ? Are the same states periodically revisited in cyclic sollicitations ?
What will be the density and the effect of floppy modes in systems of
non-spherical frictionless particles ? Will the staircase-like stress-strain
curve approach a smooth limit when the system size increases ? To what extent
are rearrangements sensitive to the actual dynamical rule ?  Such problems
would benefit from careful numerical simulations, and we shall address some of
these questions in forthcoming publications.

The treatment of granular systems with friction could be tackled with a
similar approach to the one developped here: one could investigate the range
of stability of a given contact structure, as the load gradually varies, by
purely static means. In the presence of friction, granular packings are also
observed, in experiments and dynamic numerical simulations, to evolve
by a succession of crises localized in time. We expect the geometry of the
assemblage to dictate, to a large extent, the way such sudden motions are
initiated.

It can be concluded that much of the promising prospects, as well as much of
the difficulties ahead, in the study of mechanical properties of granular
materials close to equilibrium, are in the understanding of the disordered,
yet quite peculiar, geometry of large systems that adapt their contact
network to sustain the load. 

\acknowledgments
The author wishes to thank J.-P. Bouchaud, X. Chateau,
E. Cl\'ement, G. Combe, P. Dangla, M. Jean, J. Jenkins,
J.-J. Moreau, S. Ouaguenouni, F. Radjai, J. Rajchenbach and
J. Socolar for stimulating contacts and conversations.

\bibliographystyle{prsty}

\end{document}